\documentclass[10pt]{article} 
\usepackage[preprint]{tmlr}

\usepackage{hyperref}
\usepackage{url}

\usepackage{algorithm}
\usepackage{algorithmic}
\let\classAND\AND
\let\AND\relax
\usepackage{algorithmic}

\let\AND\classAND
\AtBeginEnvironment{algorithmic}{\let\AND\algoAND}

\floatname{algorithm}{Algorithm}

\usepackage{star}

\newcommand{\e}{\mathbb{E}}

\newcommand{\mom}{\mathrm{\scriptstyle MoM}}
\newcommand{\pro}{\mathrm{\scriptstyle pro}}

\def\T{{ \mathrm{\scriptscriptstyle T} }}

\def\##1\#{\begin{align}#1\end{align}}
\def\$#1\${\begin{align*}#1\end{align*}}

\newcommand{\var}{\textrm{var}}
\newcommand{\cov}{\textrm{cov}}
\newcommand{\iid}{\textrm{i.i.d.\,}}

\newcommand{\Rom}[1]{\text{\uppercase\expandafter{\romannumeral #1\relax}}}

\renewcommand{\max}{\mathop{\mathrm{max}}}
\newcommand{\nn}{\nonumber}
\def\sn{\sum_{i=1}^n}

\newcommand{\s}{\mathrm{\scriptstyle p}}
\newcommand{\p}{\mathrm{\scriptstyle p}}

\newcommand{\taus}{\tau_*}
\newcommand{\tauvp}{\tau_\varpi}
\newcommand{\tauv}{\tau_{v_0}}

\newcommand{\tsigma}{\tau_{\sigma}}
\newcommand{\vs}{v_*}

\newcommand{\mus}{\mu^*}
\newcommand{\bartau}{\bar\tau}
\newcommand{\ultau}{\underline\tau}
\newcommand{\barv}{\bar v}

\newcommand{\wtau}{{\widehat\tau}}
\newcommand{\htau}{{\widehat\tau}}
\newcommand{\wv}{{\widehat v}}
\newcommand{\hmu}{{\hat\mu}}

\newcommand{\hv}{\hat v}
\newcommand{\ttau}{{\tilde\tau}}
\newcommand{\htheta}{{\hat\theta}}
\newcommand{\thetas}{{\theta^*}}

\newcommand{\tauvm}{\tau_{v_0}^2-1}

\newcommand{\init}{\mathrm{init}}
\newcommand{\med}{\mathrm{med}}

\newcommand{\llangle}{\left\langle}
\newcommand{\rrangle}{\right\rangle}

\newcommand{\scolor}[1]{{\color{magenta}#1}}

\newcommand{\scomment}[1]{\scolor{$\dagger$}\marginpar{\tiny\scolor{S:\ #1}}\hspace{-3pt}}

\usepackage{minitoc}

\title{Do we need to estimate the variance in robust mean estimation?}

\author{\name Qiang Sun \email qiang.sun@utoronto.ca \\
      \addr Department of Statistical Sciences\\
      University of Toronto
      }

\def\month{12}  
\def\year{2023} 
\def\openreview{\url{https://openreview.net/forum?id=CIv8NfvxsX}} 

\begin{document}

\maketitle

\begin{abstract}

In this paper, we propose self-tuned robust estimators for estimating the mean of heavy-tailed distributions, which refer to distributions with only finite variances. Our approach introduces a new loss function that considers both the mean parameter and a robustification parameter. By jointly optimizing the empirical loss function with respect to both parameters, the robustification parameter estimator can automatically adapt to the unknown data variance, and thus the self-tuned mean estimator can achieve optimal finite-sample performance. Our method outperforms previous approaches in terms of both computational and asymptotic efficiency. Specifically, it does not require cross-validation or Lepski's method to tune the robustification parameter, and the variance of our estimator achieves the Cram\'er-Rao lower bound. Project source code is available at \url{https://github.com/statsle/automean}.

\end{abstract}

\doparttoc 
\faketableofcontents 
\part{} 
\vspace{-40pt}

\section{Introduction}\label{sec:1}

The success of numerous statistical and learning methods 
heavily relies on the assumption of light-tailed or sub-Gaussian errors \citep{wainwright2019high}. A random variable $Z$ is considered to have sub-Gaussian tails if there exist constants $c_1$ and $c_2$ such that  $\PP(|Z-\EE Z|>t)\leq c_1\exp(-c_2t^2)$ for any $t \geq 0$. However, in many practical applications, data are often collected with a high degree of noise. For instance, in the context of gene expression data analysis, it has been observed that certain gene expression levels exhibit kurtoses  much larger than $3$, regardless of the normalization method used \citep{wang2015high}. Furthermore, a recent study on functional magnetic resonance imaging \citep{eklund2016cluster} demonstrates that the principal cause of invalid functional magnetic resonance imaging inferences is that the data do not follow the assumed Gaussian shape.  It is therefore important to develop robust statistical methods  with desirable statistical performance in the presence of heavy-tailed data.

In this paper, we focus on robust mean estimation problems, which serves as the foundation for tackling more general problems.   Specifically, we consider a generative model where data, $\{y_i, \, 1\leq i \leq n\}$, are generated according to
\#\label{model:main}
y_i=\mus + \varepsilon_i, ~1\leq i \leq n,
\#
where the random errors $\varepsilon_i \in \RR$ are independent and identically distributed (\iid\!) copies of   the random variable $\varepsilon$, following the law $F_0$. We assume that $\varepsilon$ is mean zero and posseses only a finite variance, without higher-order moments.  Let ${\EE}_{\varepsilon\sim F_0} \varepsilon=0$ and ${\EE}_{ \varepsilon \sim F_0} \varepsilon^2=\sigma^2$, where  the expectation ${\EE}_{\varepsilon \sim F_0} \varepsilon$ represents the expected value of $\varepsilon$ when it follows the distribution $F_0$.

When estimating the mean $\mus$, the sample mean estimator $\sum_{i=1}^n y_i/n$ is known to achieve, at best, a polynomial-type nonasymptotic confidence width \citep{catoni2012challenging}, when the errors have only finite variances. Specifically, there exists a distribution $F = F_{n, \delta}$ for $\varepsilon$ with a mean of $0$ and a variance of $\sigma^2$, such that the followings hold simultaneously: 
\$
\PP\left(\left|\sum_{i=1}^n \frac{y_i}{n} -\mu^*\right|\leq \sigma\sqrt{\frac{1}{2n}\cdot\frac{1}{\delta}}\right) 
&\geq 1-2\delta,~~~\forall~\delta\in  \big(0, 1/2 \big); \\
\PP\left(\left|\sum_{i=1}^n \frac{y_i}{n} -\mu^*\right|\leq  \sigma\sqrt{\frac{1}{2n}\cdot\frac{1}{\delta}}\left(1- \frac{2e\delta}{n}\right)^{{(n-1)}/{2}} \right) 
&\leq  1-2\delta,  ~~~\forall~\delta\in \big(0, (2e)^{-1}\big). 
\$ 
In simpler terms, this indicates that the sample mean does not converge quickly enough to the true mean when the errors have only finite variances.

\cite{catoni2012challenging} made an important step towards estimating the mean with faster concentration. Catoni introduced  a  robust mean estimator $\hat\mu(\tau)$, which depends on a tuning parameter $\tau$ and deviates from the true mean $\mu^*$ logarithmically in $1/\delta$. For sufficiently large $n$ and with $\tau$ properly tuned, $\hat\mu(\tau)$ satisfies the following concentration inequality: 
\# \label{catoni.bound}
	 \PP\left(   |  \hat{\mu}(\tau) - \mu^* | \leq  c\sigma \sqrt{\frac{1}{n}\cdot \log\left(\frac{1}{\delta}\right)}\right) \geq  1- 2\delta, ~~~\forall~\delta\in  \big(0, 1/2 \big),
\#
where $c$ is some constant. Estimators that satisfy this deviation property are referred to as sub-Gaussian mean estimators because they achieve the same performance as if the data were sub-Gaussian. Following Catoni's work, there has been a surge of research on sub-Gaussian estimators using the empirical risk minimization approach in various settings; see   
\cite{brownlees2015empirical, hsu2016loss, fan2017estimation, avella2018robust,  lugosi2019risk, lecue2020robust, wang2021new}  and \cite{sun2020adaptive}, among others.  For a recent review, we refer readers to  \cite{ke2019user}.

To implement Catoni's estimator \citep{catoni2012challenging}, the tuning parameter $\tau = \tau(\sigma)$, which depends on the unknown variance $\sigma^2$,  needs to be carefully tuned. However, in practice, this often involves computationally expensive methods such as cross-validation or Lepski's method \citep{catoni2012challenging}. For instance, when using the adaptive Huber estimator \citep{sun2020adaptive, avella2018robust} to estimate a $d\times d$ covariance matrix entrywise, as many as $d^2$ tuning parameters can be involved. If cross-validation or Lepski's method were employed, the computational burden would increase significantly as $d$ grows. Therefore, it is natural to ask the following question:
\begin{quote}
\textit{Is it possible to develop computationally efficient robust mean estimators for distributions with finite but unknown variances?}
\end{quote}

This paper addresses the previously mentioned challenge by introducing a self-tuned robust mean estimator for distributions with only two moments. We propose an empirical risk minimization (ERM) approach based on a new loss function. The proposed loss function is smooth with respect to both the mean parameter and the robustification parameter. By joint optimizing both parameters, we prove that the resulting robustification parameter can automatically adapt to the unknowns, allowing the resultant mean estimator to achieve sub-Gaussian performance up to logarithmic terms. Therefore, compared to prior methods, our approach eliminates the need for cross-validation or Lepski's method to tune the  robustification parameter. This significantly boots the computational efficiency of robust data analysis in practical applications. Furthermore, from an asymptotic perspective, we establish that our proposed estimator is asymptotically efficient, meaning that its variance achieves the Cramér-Rao lower bound asymptotically  \citep{van2000asymptotic}.

\paragraph{Related work} 
In addition to empirical risk minimization (ERM)-based methods, median-of-means (MoM) techniques \citep{devroye2016sub, lugosi2019risk, lecue2020robust} are commonly employed for constructing robust estimators in the presence of heavy-tailed distributions. A comprehensive survey on median-of-means can be found in the work by \cite{lugosi2019mean}. The MoM technique involves randomly dividing the complete dataset into $k$ subsamples and calculating the mean for each subsample. The MoM estimator is then determined as the median of these local mean estimators. The number of subsamples $k$ is the only user-defined parameter in MoM, and it can be chosen independently of the unknowns, rendering it tuning-free. However, based on our experience, MoM often exhibits undesirable numerical performance when compared to ERM-based estimators. To gain insight into this phenomenon, we adopt an asymptotic perspective and compare the asymptotic efficiencies of our estimator and the MoM estimator. We demonstrate that the relative efficiency of the MoM estimator with respect to ours is only $2/\pi \approx 0.64$.

\paragraph{Paper overview} 

Section \ref{sec:2} introduces a novel loss function and presents the empirical risk minimization (ERM) approach. The nonasymptotic theory is presented in Section \ref{sec:3}. In Section \ref{sec:3.5}, we compare our  estimator with the MoM mean estimator in terms of asymptotic performance. Section \ref{sec:5} provides numerical experiments. Finally, we conclude in Section \ref{sec:6}. The supplementary material contains  basic calculations,  algorithms, a comparison with Lepski's method, proofs of the main results, supporting lemmas, and additional results.

\paragraph{Notation}
We summarize here the notation that will be used throughout the paper. We  use $c$ and $C$ to denote generic constants which may change from line to line. For two sequences of real numbers $\{ a_n, {n\geq 1}\}$ and $\{ b_n, {n\geq 1}\}$, $a_n \lesssim b_n$ or $a_n = O(b_0)$ denotes $a_n \leq C b_n$ for some constant $C>0$, and  $a_n \gtrsim b_n$ if $b_n \lesssim a_n$. We use $a_n \propto b_n$ to denote that $a_n\gtrsim b_n$ and $a_n\lesssim b_n$.  The $\log$ operator is understood with respect to the base $e$. For a function $f(x, y)$, we use $\nabla_x f(x,y)$ or $\frac{\partial}{\partial x} f(x,y)$ to denote its partial derivative of $f(x,y)$ with respect to $x$. Let $\nabla f(x, y)$ denote the gradient of $f(x,y)$.  
For a vector $x\in \RR^d$, $\|x\|_2$ denotes its Euclidean norm. For a symmetric positive semi-definite matrix $\Sigma$, $\lambda_{\max}(\Sigma)$ denotes its largest eigenvalue.


\section{A new loss function for self-tuning}\label{sec:2}

 This section introduces a new loss function to  robustly estimate  the mean of distributions with only finite variances while automatically tuning the robustification parameter. We begin with the pseudo-Huber loss \citep{hastie2009elements}
\#\label{loss:sh}
\ell_\tau(x)  = \tau\sqrt{\tau^2+x^2}  -\tau^2=  \tau^2\sqrt{1+x^2/\tau^2}-\tau^2, 
\#
where $\tau$ serves as a tuning parameter. It exhibits behavior similar to the Huber loss \citep{huber1964robust}, approximating $x^2/2$ when $x^2\lesssim \tau^2$ and resembling a straight line with slope $\tau$  when $x^2\gtrsim\tau^2$. To see this, some algebra  yields  
\begin{gather*}
\left\{\begin{array}{ll}
\frac{\epsilon^2-2(1+\epsilon)}{2\epsilon^2}x^2\leq \ell_\tau (x)\leq  \frac{x^2}{2},   & \mbox{if } x^2\leq \tau^2\cdot 4(1+\epsilon)/\epsilon^2,  \\
	\frac{\tau |x|}{1+\epsilon}\leq \ell_\tau(x)\leq \tau|x|,   &  \mbox{if } x^2>  \tau^2\cdot 4(1+\epsilon)/\epsilon^2 .
	\end{array}  \right.
\end{gather*}
 We refer to $\tau$ as the robustification parameter because it controls the trade-off between the quadratic loss and the least absolute deviation loss, where the latter induces robustness. In practice, $\tau$ is typically tuned using computationally expensive methods such as Lepski's method \citep{catoni2012challenging} or cross-validation \citep{sun2020adaptive}.

To circumvent these computationally expensive procedures, our goal is to propose a new loss function of both the mean parameter $\mu$ and the robustification parameter  $\tau$ (or its equivalent) so that optimizing over them jointly yields an automatically tuned robustification parameter $\wtau$ and thus the correspondingly self-tuned mean estimator $\hmu(\wtau)$.   Unlike the Huber loss \citep{sun2020adaptive}, the pseudo-Huber loss is a smooth function of  $\tau$, making optimization with respect to $\tau$ possible. To motivate the new loss function,  let us first consider the estimator {$\hat\mu(\tau)$} with $\tau$ fixed {\it a priori}:
\#\label{loss:pseudo}
\hmu(\tau) = \argmin_{\mu} \left\{  \frac{1}{n}\sum_{i=1}^n  \ell_\tau(y_i-\mu)\right\}.
\#
Below, we provide an informal result, with its rigorous version presented as  Theorem \ref{thm:mean.fixed} in subsequent sections.

\begin{theorem}[Informal result]\label{thm:informal}
Take $\tau= \sigma\sqrt n/z$ with $z=\sqrt{\log (1/\delta)}$, and assume $n$ is sufficiently large. Then, for any $0<\delta< 1$, with probability at least $1-\delta$, we have
\$
|\widehat \mu(\tau)-\mu^*|
	&\lesssim \sigma \sqrt{\frac{\log(2/\delta)}{n}}. 
\$
\end{theorem}

The aforementioned  result indicates that when $\tau = \sigma\sqrt{n}/z$ with $z = \sqrt{\log(1/\delta)}$, the estimator $\widehat{\mu}(\tau)$ achieves the desired sub-Gaussian performance. The only unknown parameter in $\tau$ is the standard deviation $\sigma$.  In view of this, we treat $\sigma$ as an unknown parameter $v$, substitute $\tau = \sqrt{n}v/z$ into \eqref{loss:sh}, and obtain
\begin{equation}\label{loss:v}
\ell(x, v) := \ell_\tau(x) = \frac{nv^2}{z^2} \left(\sqrt{1+ \frac{x^2z^2}{nv^2}} -1 \right),
\end{equation}
where $z$ is a confidence parameter because it depends on $\delta$ as in the theorem above.

Instead of searching for the optimal $\tau$, we will seek the optimal $v$, which is expected to be close to the underlying standard deviation $\sigma$ intuitively. We will use the term ``robustification parameter" interchangeably for both $\tau$ and $v$, as they are equivalent up to a multiplier. However, directly minimizing $\ell(x,v)$ with respect to $v$ leads to meaningless solutions, specifically $v = 0$ and $v = +\infty$. To avoid these trivialities, we consider a new loss by dividing $\ell(x, v)$ by $v$ and then adding a linear penalty function $av$. This will be referred to as the penalized pseudo-Huber loss, formally defined below.

\begin{definition}[Penalized pseudo-Huber loss]\label{def:ssh}
The penalized pseudo-Huber loss $\ell^\p(x, v)$ is defined  as follows: 
\#\label{loss:pen}
\ell^\p(x,v) :=  \frac{\ell(x,v)+av^2}{v} = \frac{nv}{z^2} \left(\sqrt{1+ \frac{x^2z^2}{nv^2}} -1 \right) + av,
\#
where $n$ is the sample size, $z$ is a confidence parameter, and $a$ is an adjustment factor.
\end{definition}

We thus propose to  jointly optimize over $\mu$ and  $v$ by solving the following ERM problem: 
\# \label{opt:main}
\left\{\,\widehat\mu, \,\widehat v\,\right\} 
&=
\argmin_{\mu,\, v}\left\{L_n(\mu, v):= \frac{1}{n}\sn \ell^\p (y_i-\mu, v)\right\} \nn \\
&=\argmin_{\mu, \, v}\frac{1}{n}\sum_{i=1}^n\left\{\frac{nv}{z^2}\left(\sqrt{1+\frac{(y_i-\mu)^2z^2}{nv^2}} -1\right)+av \right\}.
\#


To gain insight into the loss function $L_n(\mu, v)$, let us  consider its population version:
\$
L(\mu, v)=\EE L_n(\mu, v)  = \frac{nv}{z^2} \EE\left(  \sqrt{1 +  \frac{(y-\mu)^2z^2}{nv^2} }   - 1\right) + a v. 
\$
Define the population oracle $v_*$ as the minimizer of  $L(\mu^*, v)$ with the true mean $\mu^*$ given {\it a priori}, that is
\$
\vs= \argmin_{\tau} L(\mus, v) = \argmin_{v} \left\{  \frac{nv}{z^2}  \EE\left(  \sqrt{1 +  \frac{(y-\mu)^2z^2}{nv^2} }   - 1\right) + a v   \right\}, 
\$
or equivalently, 
\$
\nabla_v L(\mu^*, v)\big|_{v =\vs} 
= \left\{\frac{n}{z^2} \left(\nabla_v \EE \sqrt{v^2 + \frac{\varepsilon^2z^2}{n}} -1 \right) + a \right\}\Bigg|_{v=v^*} = 0. 
\$
By interchanging the derivative with the expectation, we obtain 
\#\label{eq:tau.def}
\EE\frac{\vs}{\sqrt{\vs^2 + z^2\varepsilon^2/n}}
=1-\frac{az^2}{n}. 
\#

Let $\sigma^2_{x^2}:= \EE\{{\varepsilon^2} 1({\varepsilon^2}\leq x^2)\}$, where $1(A)$ is the indicator function of the set $A$. Our first key result utilizes the above characterization of $\vs$ to derive how $\vs$ is able to automatically adapt  to  the  standard deviation $\sigma$, promising  the effectiveness of our procedure.

\begin{theorem}[Self-tuning property of $\vs$]\label{thm:ada}
Suppose $n\geq az^2$.   Then, for any $\gamma \in[0,1)$, we have $\vs>0$ and  
\$
 \frac{(1-\gamma)\sigma^2_{\varphi\taus^2}}{2a}\leq \vs^2\leq \frac{\sigma^2}{2a}, 
\$
where $\varphi = \gamma/(1-\gamma)$ and $\taus = \vs\sqrt{n}/z.$
Moreover
\$
\lim_{n\rightarrow \infty} \vs^2 =  {\sigma^2}/({2a}).
\$
\end{theorem}

The above result indicates that for any $n \geq az^2$, the oracle $\vs^2$ can automatically adapt to the (truncated) variance. It is bounded between the scaled truncated variance $\sigma^2_{\varphi \taus}/(2a)$ and the scaled variance $\sigma^2/(2a)$. Since the second moment exists, we have $\sigma^2_{\varphi \taus^2}\rightarrow \sigma^2$ as $\varphi \taus^2 \rightarrow \infty$ by the dominated convergence theorem. For a large sample size $n$, $\sigma^2_{\varphi\taus^2}$ is close to $\sigma^2$, and therefore $\vs^2$ is approximately between $(1-\gamma )\sigma^2/(2a)$ and $\sigma^2/(2a)$. Furthermore, an asymptotic analysis reveals that $\lim_{n\rightarrow \infty} \vs^2 = {\sigma^2}/({2a})$. Taking $a=1/2$ yields $\lim_{n\rightarrow \infty} \vs^2 = \sigma^2$, indicating that the oracle $\vs^2$ with $a=1/2$ should approximate the true variance in the large sample limit. This also suggests the optimality of choosing $a=1/2$, which is assumed throughout the rest of the paper.

 
Our next result shows that the proposed empirical loss function is jointly convex in both $\mu$ and $v$, which allows us to employ standard first-order optimization algorithms to compute the global optima efficiently.

\begin{proposition}[Joint convexity]\label{prop:cvx}
The empirical loss function $L_n(\mu, v)$ in \eqref{opt:main} is jointly convex in both $\mu$ and $v$. Furthermore, if there exist at least two distinct data points, the empirical loss function is strictly convex in both $\mu$ and $v$ provided that $v>0$. 
\end{proposition}

Lastly, it was brought to our attention  that our formulation \eqref{opt:main} bears a resemblance to the concomitant estimator by \cite{huber2009robust}: 
\$
\argmin_{\mu, v} \left\{ \frac{1}{n} \sum_{i=1}^n \rho\left( \frac{y_i- \mu}{v}\right)v + av \right\},
\$
where $\rho$ is  any loss function, and $a$ is a user-specified constant. Notably, the selection of the appropriate constant $a$ is rarely addressed in the existing literature.  Our motivation, however, arises from a distinct perspective. We aspire to develop robust mean estimators that demonstrate improved finite-sample properties when dealing with heavy-tailed data. The empirical loss function $L_n$ we propose can be seen as an intricately adapted variant of theirs. Specifically, we employ the smooth pseudo-Huber loss, where we set the robustification parameter $\tau$ as $\tau = v\sqrt{n}/z$ to ensure the sub-Gaussian performance of the mean estimator. Here, $z$ serves as a carefully chosen confidence parameter. Simultaneously, we identify the optimal adjustment factor as $a=1/2$. To the best of our knowledge, all of these findings are the first among in the literature.


\section{Finite-sample theory}\label{sec:3}

This section presents the self-tuning property for estimated robustification parameter and then the finite-sample property of the self-tuned mean estimator. Recall $a=1/2$.

\subsection{Estimation with a fixed $v$}

With an abuse of notation, we use $\widehat{\mu}(v)$ to denote the minimizer of the empirical penalized pseudo-Huber loss in \eqref{opt:main} with $v$ fixed. Recall that we have used $\widehat{\mu}(\tau)$ to denote the minimizer of the empirical pseudo-Huber loss in \eqref{loss:pseudo}, and $\widehat{\mu}(v)$ is equivalent to $\widehat{\mu}(\tau)$ with $\tau = v\sqrt{n}/z$. We begin by examining the theoretical properties of $\widehat{\mu}(v)$. We require the following locally strong convexity assumption, which will be verified later in this subsection.

\begin{assumption}[Locally strong convexity in $\mu$]\label{ass:mean.sc}
The empirical Hessian matrix is locally strongly convex with respect to $\mu$ such that, for any $\mu \in \BB_r(\mu^*):=\{\mu: |\mu-\mu^*|\leq r\}$, 
\$
\inf_{\mu \in \BB_r (\mu^*)}\frac{ \llangle \nabla_\mu L_n(\mu,v) -  \nabla_\mu L_n(\mu^*,v),  \mu -\mu^*  \rrangle}{|\mu-\mu^*|^2} \geq \kappa_\ell>0 
\$
where $r>0$ is a local radius parameter.  
\end{assumption}

\begin{theorem}\label{thm:mean.fixed}
For any $0<\delta <  1$, let  $v> 0$ be fixed and $z^2 = \log (1/\delta)$. 
Assume Assumption \ref{ass:mean.sc} holds with {any} $r \geq r_0(\kappa_\ell):=\kappa_\ell^{-1}\left(\sigma/({\sqrt{2}v})+1\right)^2\,\sqrt{\log(2 /\delta)/{n}}$.  Then, with probability at least $1-\delta$, we have
\$
|\widehat \mu(v)-\mu^*|
&< \frac{1}{\kappa_\ell}\left(\frac{\sigma}{\sqrt{2}v}+1\right)^2\,\sqrt{\frac{\log(2 /\delta)}{n}} = \frac{C}{\kappa_\ell} \sqrt{\frac{\log(2 /\delta)}{n}},
\$
where $C=(\sigma/(\sqrt{2}v)+1)^2$ only depends on $v$ and $\sigma$.  
\end{theorem}

The above theorem states that under the assumption of locally strong convexity, $\widehat{\mu}(v)$ achieves a sub-Gaussian deviation bound when the data have only bounded variances. In particular, if we choose $v = \sigma$ in the theorem, we obtain
\$
|\widehat \mu(\sigma)-\mu^*|
	&\leq\frac{1}{\kappa_\ell}\left(\frac{\sigma}{\sigma}+1\right)^2\,\sqrt{\frac{\log(2/\delta)}{n}}
\leq \frac{4}{\kappa_\ell} \sqrt{\frac{\log(2/\delta)}{n}}. 
\$

Assumption \ref{ass:mean.sc} essentially requires the loss function to exhibit curvature in a small neighborhood ${\BB}_r(\mu^*)$, while the penalized loss \eqref{loss:pen} transitions from a quadratic function to a linear function roughly at $|x|=\tau \propto \sqrt{n}$. Quadratic functions always have curvature, so intuitively, Assumption \ref{ass:mean.sc} holds as long as
\$
  \sqrt{n} \gtrsim r \geq r_0(\kappa_\ell)\propto \sqrt{\frac{1}{n}}.
\$
The condition above is automatically guaranteed when $n$ is sufficiently large. Choosing $r$ to be the smallest $r_0(\kappa_\ell)$ results in Assumption \ref{ass:mean.lsc} being at its weakest. In other words, in this scenario, the empirical loss function only needs to exhibit curvature in a diminishing neighborhood of $\mus$, approximately with a radius of $\sqrt{1/n}$. The following lemma rigorously proves this claim.

 \begin{lemma}\label{lemma:mean.sc}
Suppose $v\geq v_0$. For any $0<\delta <  1$,  let $n\geq C\max\left\{z^2(\sigma^2 + {r^2})/v_0^2,\, \log(1/\delta)\right\}$ for some absolute constant $C$. Then,  with probability at least $1-\delta$, Assumption \ref{ass:mean.sc} with $\kappa_\ell = 1/(2v)$ and any local radius $r\geq r_0(\kappa_\ell)=r_0(1/(2v))$ holds uniformly over $v\geq v_0 > 0$. 
 \end{lemma}


The first sample complexity condition, $n\geq C z^2(\sigma^2+r^2)/v_0^2$, arises from the requirement that $\tauv^2:=v_0^2 n/z^2 \geq C(\sigma^2 + r^2)$. Because the robustification parameter $\tauv^2= v_0^2 n/z^2$ determines the size of the quadratic region, this requirement is minimal in the sense that Assumption \ref{ass:mean.lsc} can hold only when $\tau_v^2$ is larger than $r^2$ plus the noise variance $\sigma^2$. As argued before, Assumption \ref{ass:mean.lsc} holds with any $r$ such that $\sqrt{n} \gtrsim r \gtrsim \sqrt{1/n}$. For example, we can take $r\propto \sigma$ to be a constant, and this will not worsen the sample complexity condition. Finally, by combining Lemma \ref{lemma:mean.sc} and Theorem \ref{thm:mean.fixed}, we obtain the following result.

\begin{corollary}\label{coro:mean.fixed}
Suppose $v\geq v_0$.  For any $0<\delta <  1$, let $n \geq C\max\left\{ ({r^2} +\sigma^2)/v_0^2, 1\right\} \log(1/\delta)$ for some universal constant $C$, where  $r\geq 2r_0(1/(2v)) $. Take $z^2={\log(1/\delta)}$. Then, for any $v\geq v_0$, with probability at least $1-\delta$,  we have 
\$
|\hat\mu(v) -\mu^*|\leq 2v \left(\frac{\sigma}{\sqrt{2}v}+1\right)^2\,\sqrt{\frac{\log(4/\delta)}{n}} \lesssim v \sqrt{\frac{1+\log(1/\delta)}{n}}.
\$ 
\end{corollary}

\subsection{Self-tuned mean estimators}

We proceed to characterize the theoretical property of the self-tuned mean estimator.  
We need an additional constraint that $v_0\leq v\leq V_0$, and consider the constrained empirical risk minimization problem
\#\label{eq:const}
\left\{\,\widehat\mu, \,\widehat v\,\right\} 
 &=\argmin_{\mu,\, v_0\leq v \leq V_0}\left\{L_n(\mu, v):= \frac{1}{n}\sn \ell^\p (y_i-\mu, v)\right\}. 
\#
Indeed, when $v$ is either $0$ or $\infty$, the loss function is nonsmooth or  trivial, respectively. In other words, the loss function is not strongly convex in $\mu$ in either case, and the strong convexity is essential for our analysis.  Recall that $\tauv = v_0 \sqrt{n}/z$.


\begin{theorem}[Self-tuning property]\label{thm:v}
Assume that $n$ is sufficiently large. {Suppose 
$
v_0<  c_0 \left({\sigma_{\tauv^2/2-1}}/{\sigma_{\tauv^2/2}} \wedge 1 \right) \sigma_{\tauv^2/2-1} \leq  C_0 \sigma < V_0
$ 
where $c_0$ and $C_0$ are some universal constants.}  For any $0<\delta < 1$, let $z^2\geq \log(5/\delta)$.  Then, with probability at least $1-\delta$, we have 
\$
c_0  \left({\sigma_{\tauv^2/2-1}}/{\sigma_{\tauv^2/2}} \wedge 1 \right)  {\sigma_{\tauv^2/2-1}} \leq \hv \leq  C_0 \sigma. 
\$
\end{theorem}

The above theorem indicates that $\hv$ automatically adapts to the standard deviation, approximating $\sigma$, if $\sigma_{\tauv^2/2-1}$ approximates $\sigma$. This is expected to hold for a large sample size by the dominated convergence theorem. However, it is important to note that $\sigma_{\tauv^2}$ cannot be expected to be close to $\sigma$ at any predictable rate under the weak assumption that the data have bounded variances only. We will now proceed to characterize the finite-sample properties of the self-tuned mean estimator $\hmu (\hv)$.


\begin{theorem}[Self-tuned mean estimators]\label{thm:main_random}
Assume that $n$ is sufficiently large. {Suppose 
$
v_0<  c_0 \left({\sigma_{\tauv^2/2-1}}/{\sigma_{\tauv^2/2}} \wedge 1 \right) \sigma_{\tauv^2/2-1} \leq  C_0 \sigma < V_0
$ 
where $c_0$ and $C_0$ are some universal constants.}
{For any $0<\delta <  1$,} take $z^2=\log(n/\delta)$.  Then,  with probability at least $1-\delta$, we have
\$
|\widehat \mu(\hv)-\mu^*| \leq   C \cdot \sigma \, \sqrt{\frac{\log(n/\delta)}{n}}
\$
where $C$ is some constant. 

\end{theorem}


The above result asserts that the self-tuned mean estimator $\widehat{\mu} = \widehat{\mu}(\widehat{v})$ achieves the optimal deviation property up to a logarithmic factor. For practical applications, we suggest choosing $\delta = 0.05$, which corresponds to a failure probability of 0.05 or, equivalently, a confidence level of 0.95.

\section{Comparing with alternatives}\label{sec:3.5}

Other than the ERM-based approach, the median-of-means technique \citep{lugosi2019mean} {is another method to  construct robust estimators under heavy-tailed distributions.} The MoM mean estimator is constructed as follows:
\begin{enumerate}
\item Partition $ [n] = \{1, \ldots, n\}$ into $k$ blocks $\cB_1,\ldots, \cB_k$,  each with size $|\cB_i| \geq \lfloor  n/k \rfloor \geq 2$;
\item Compute the sample mean in each block
$
z_j = \sum_{i\in \cB_j}x_i/|\cB_j|;
$
\item Obtain the MoM mean estimator by taking the median of $z_j$'s:
\$
\hat\mu^{\mom} = {\med}(z_1, \ldots, z_k)
\$
where ${\med}(\cdot)$ represents the median operator. 
\end{enumerate}

The following theorem is taken from \cite{lugosi2019mean}. For simplicity and without loss of generality,  we assume throughout this section that $n$ is divisible by $k$ so that each block has exactly $m = n/k$ elements.

\begin{theorem}[Theorem 2 by \cite{lugosi2019mean} ]\label{thm:MoM} 
 For any $\delta \in (0, 1)$, if $k = \lceil 8 \log(1/\delta)\rceil$, then, with probability at least $1 -\delta$,
\$
\left|\hat\mu^{\mom} -\mu^* \right|\leq \sigma \sqrt{\frac{32\log(1/\delta)}{n}}. 
\$
\end{theorem}

The theorem above indicates that, to obtain a sub-Gaussian mean estimator, we only need to choose \(k = \lceil 8 \log(1/\delta)\rceil\) when constructing the MoM mean estimator. Thus, the MoM estimator is naturally tuning-free. However, in our numerical experiments, we have observed that the MoM estimator often exhibits inferior numerical performance compared to our proposed estimator. To shed light on this observation, we compare the asymptotic efficiencies of \(\hat{\mu}^{\text{MoM}}\) and our estimator \(\hat{\mu}(\hat{\tau})\) in the following two theorems. The first result is a direct consequence of \cite[Theorem 4]{minsker2019distributed} and we collect the proof in the appendix for completeness.

\begin{theorem}[Asymptotic inefficiency of MoM estimator]\label{thm:asym_mom}
Fix any $\iota\in(0, 1]$. Assume  $\EE|y_i-\mu^*|^{2+\iota}<\infty$.   
Suppose $k\rightarrow \infty$ and $k = o\big(n^{\iota/(1+\iota)}\big)$, then
\$
\sqrt{n} \left(\hat\mu^\mom - \mu^*\right) \rightsquigarrow \cN\left(0, \frac{\pi}{2}\sigma^2\right). 
\$
\end{theorem}

\begin{theorem}[Asymptotic efficiency of our estimator]\label{thm:asym}
Fix any $\iota\in(0, 1]$. Assume $\EE \varepsilon_i^{2+\iota}<\infty$ and the same assumptions as in Theorem \ref{thm:v}. {Take any $z^2 \geq 2\log(n)$.} Then
\$
\sqrt{n} \, (
\hmu(\hv) - \mus)
\rightsquigarrow \cN \left(0, \sigma^2\right). 
\$
\end{theorem}


We emphasize that the MoM mean estimator shares the same asymptotic property as the median estimator \citep{van2000asymptotic} due to taking the median operation in the last step, and thus it is asymptotically inefficient. In stark contrast, our proposed estimator achieves full asymptotic efficiency. The relative efficiency $e_{\rm r}$ of the MoM estimator with respect to our estimator is 
\$
e_{\rm r}\left(\hat\mu^\mom, \hat \mu(\hv)\right) = \frac{2}{\pi}\approx 0.64. 
\$
This indicates that our proposed estimator outperforms the MoM estimator in terms of asymptotic performance, partially explaining the empirical success of our method; see the numerical results in Section \ref{sec:5} for details.

\begin{figure}[t]
  \centering
  \includegraphics[width=0.60\textwidth]{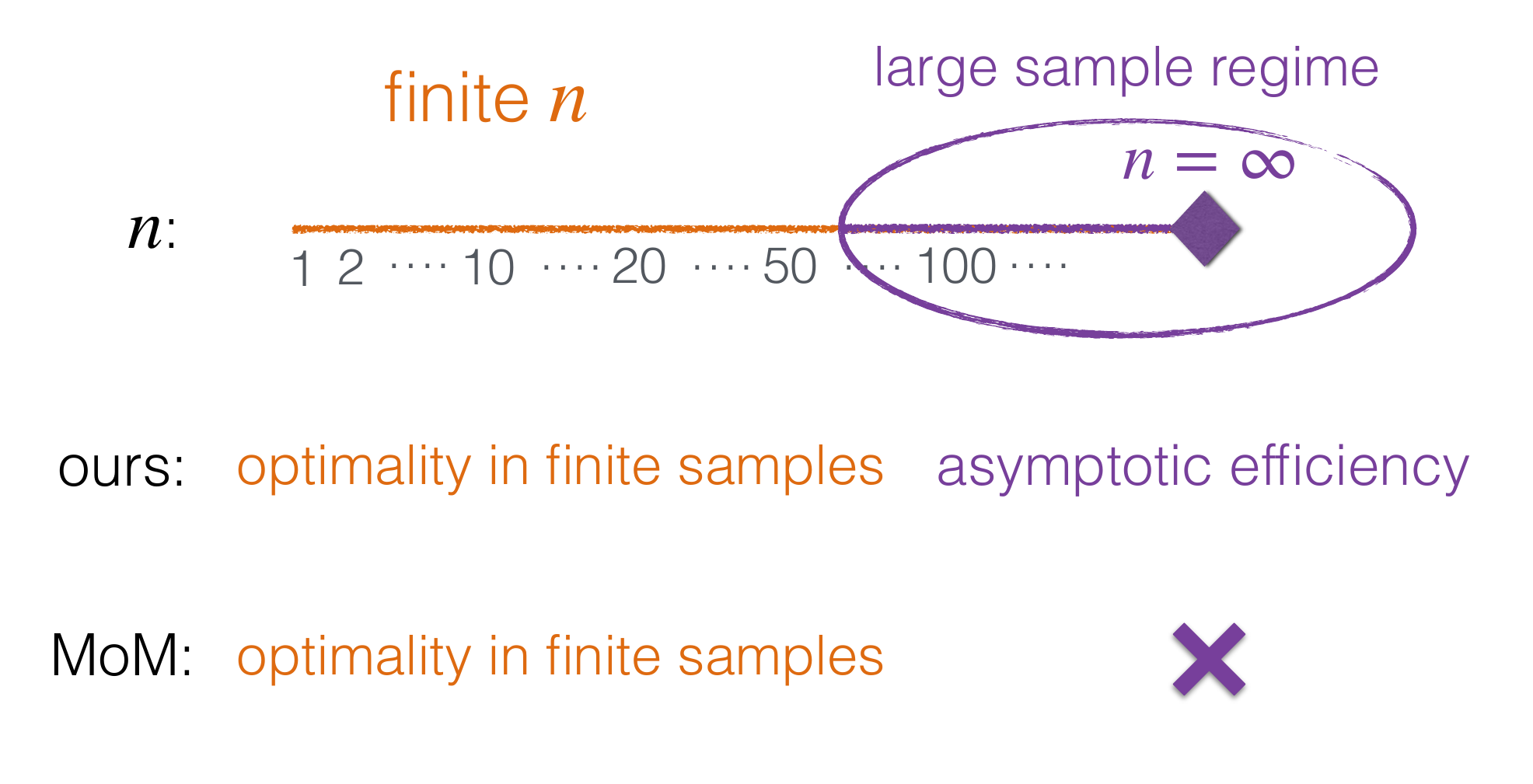}
  \caption{Comparing our self-tuned estimator with the MoM estimator in terms of adaptivity.}
  \label{fig:0}
  \end{figure}

We provide an intuitive explanation for the optimal performance of our self-tuned estimator in both finite-sample and asymptotic regimes. Because this estimator  is a self-tuned version of the pseudo-Huber estimator in \eqref{loss:pseudo},  we discuss the latter for simplicity. As per Theorem \ref{thm:informal},  taking $\tau = \sigma \sqrt{n/\log(1/\delta)}$ guarantees the sub-Gaussian performance of $\hat{\mu}(\tau)$ in the finite-sample regime. Meanwhile, as $n$ approaches infinity, $\tau = \sigma \sqrt{n/\log(1/\delta)}$  also grows to infinity, causing the pseudo-Huber loss to approach the least squares loss. This loss corresponds to the negative log-likelihood of Gaussian distributions, and its minimization yields an asymptotically efficient mean estimator.

For MoM estimators, the situation differes. To achieve the desired  sub-Gaussian performance in the finite-sample regime, the number of blocks $k$ must be at least $\lceil 8 \log(1/\delta)\rceil$, as proven in Theorem \ref{thm:MoM} by \cite{lugosi2019mean}. Conversely, for attaining asymptotic efficiency and approximating the sample mean estimator in large samples, the number of blocks should reduce to $1$ as the sample size $n$ increases. Consequently, MoM estimators exhibit a dichotomy between optimal finite-sample and asymptotic properties. This disparity likely stems from the discontinuous nature of the MoM estimator, which can not smoothly transition from requiring at least $k=3$ blocks for median calculation to functioning as an empirical mean estimator.

In summary, our self-tuned estimator can achieve optimal performance in both finite-sample and large-sample regimes. We point out that the large-sample regime is used to approximate the regime when the sample size is relatively large instead of describing the case of $n=\infty$.  We will refer to this ability as {\it adaptivity to both finite-sample and large-sample regimes}, or simply {\it adaptivity}. The MoM estimator does not naturally possess this adaptivity due to its discontinuous nature. Figure \ref{fig:0} provides a comparison between our self-tuned estimator and the MoM estimator in terms of adaptivity.

Another popular estimator is the trimmed mean estimator \citep{lugosi2021robust}. The univariate trimmed mean estimator operates as follows: (i) Split the data points into two subsamples with equal size, (ii) use the first subsample to determine the trimming parameters,  and (iii) employ the second subsample to construct the trimmed mean estimator. Due to this sample splitting scheme,  the trimmed mean estimator lacks sample efficiency.


\section{Numerical studies}\label{sec:5}

This section examines numerically the finite-sample performance of our proposed robust mean estimator when dealing with heavy-tailed data. Throughout our numerical examples, we take $z= \sqrt{\log(n/\delta)}$ with $\delta=0.05$ as recommended by Theorem \ref{thm:main_random}. This choice guarantees that the result stated in the theorem holds with a probability of at least $0.95$.

We investigate the robustness and efficiency of our proposed estimator under two distinct distribution settings for the random variable $y$:
\begin{enumerate}
\item[1.] Normal distribution $\cN(\mu,\sigma^2)$ with mean $\mu = 0$ and  variance $\sigma^2 \geq 1$.
\item[2.] Skewed generalized $t$ distribution ${\sf sgt}(\mu, \sigma, \lambda, p, q)$, where mean $\mu=0$, skewness $\lambda = 0.75$, standard deviation $\sigma = \sqrt{q/(q-2)}$, shape parameter $p = 2$, and  shape parameter $q>2$.
\end{enumerate}
For each of the above settings, we generate an independent samples of size $n = 100$ and compute four mean estimators: our proposed estimator (ours), the sample mean estimator (sample mean), the MoM mean estimator (MoM), and the trimmed mean estimator (trimmed mean).

\begin{figure}[t]
    \centering
        \includegraphics[width=0.49\linewidth]{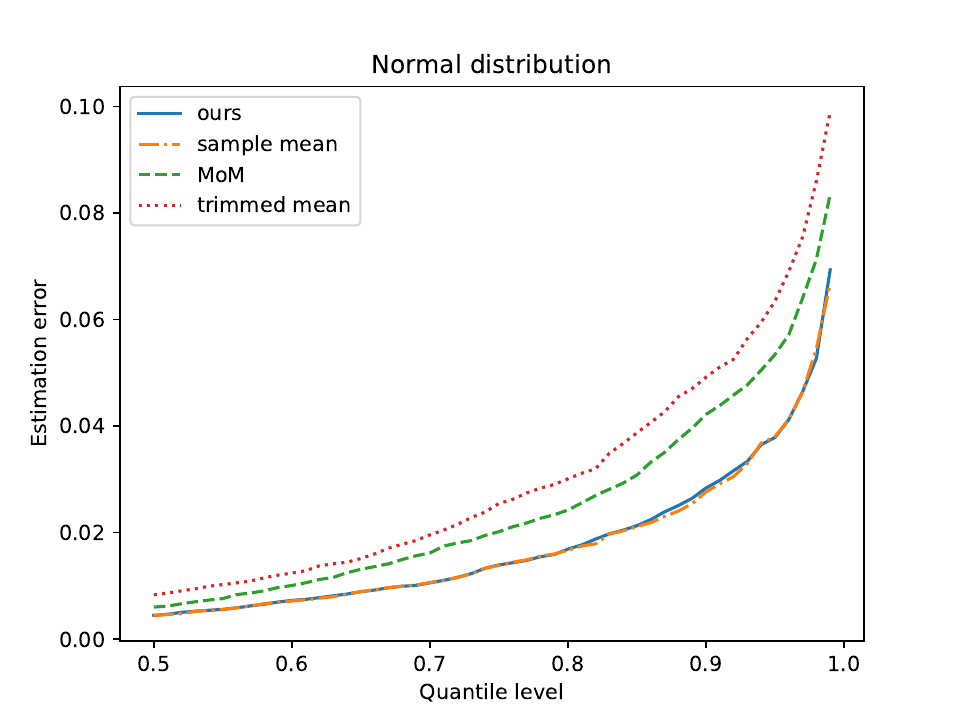}
  \includegraphics[width=0.49\linewidth]{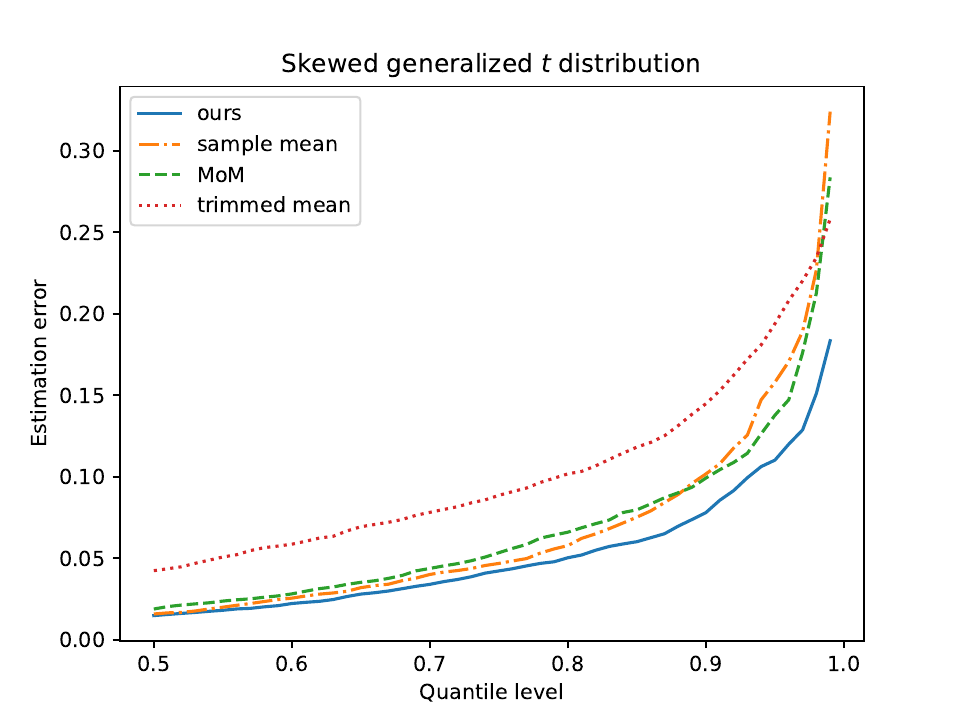}
    \caption{ The $\alpha$-quantile of the estimation error (estimation error, $y$-axis) versus $\alpha$ (quantile level, $x$-axis) for  our estimator, the sample mean estimator, the MoM estimator, and the trimmed mean estimator. 
    }
    \label{fig:1}
\end{figure}

\begin{figure}[t]
    \centering
        \includegraphics[width=0.49\linewidth]{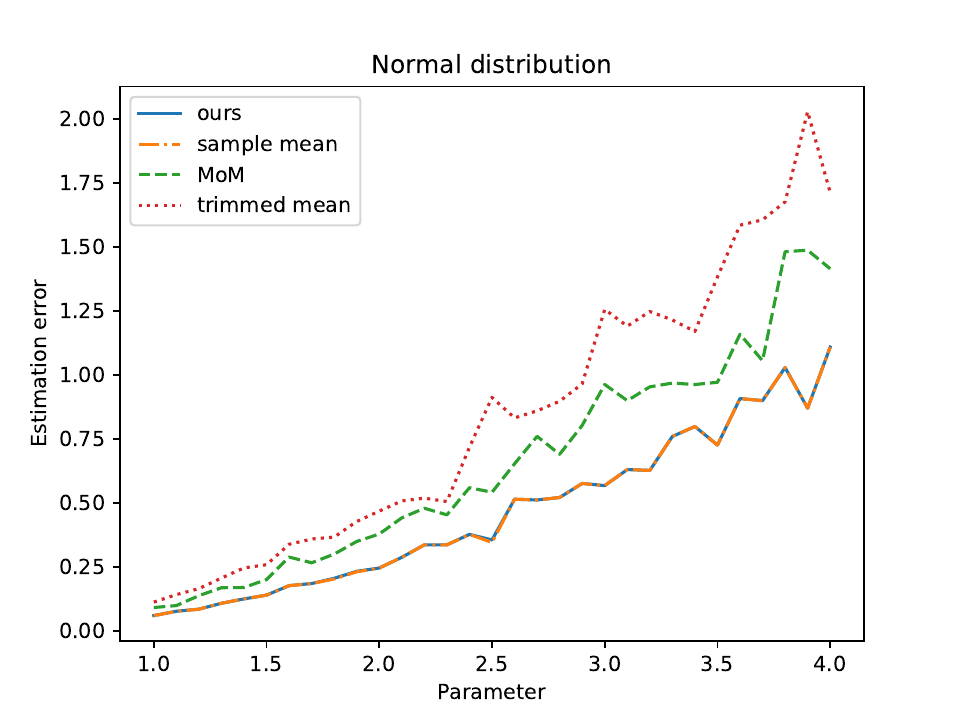}
  \includegraphics[width=0.49\linewidth]{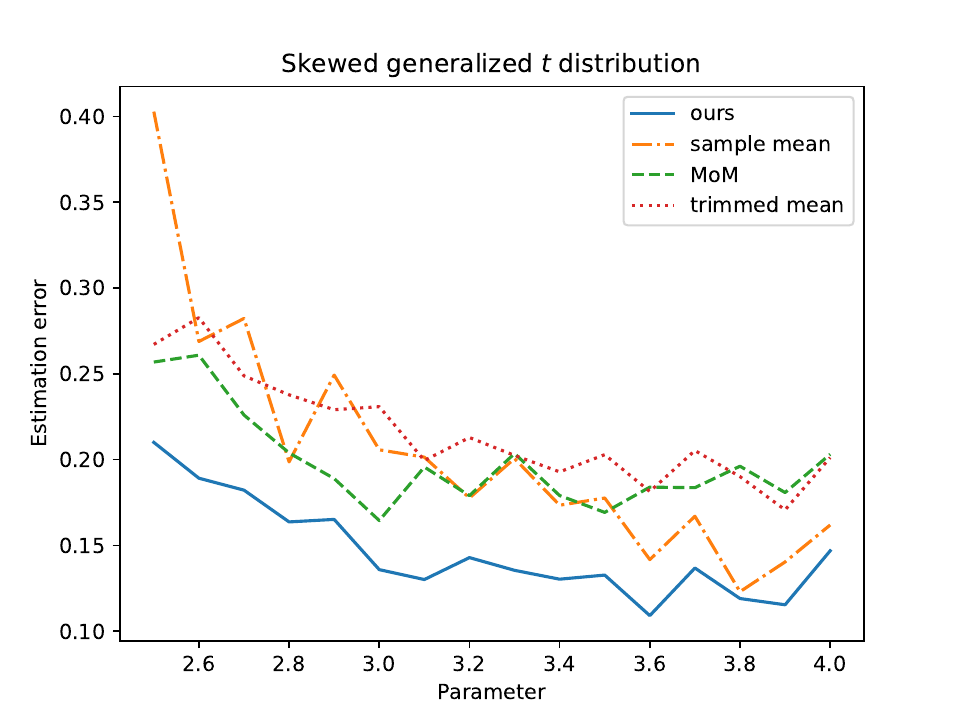}
    \caption{ Empirical 99\%-quantile of the estimation error (estimation error, $y$-axis) versus a distributution parameter (parameter, $x$-axis) for  our estimator, the sample mean estimator, the MoM estimator and the trimmed mean estimator. The distribution parameter is $\sigma$ for normal distribution and $q$ for skewed generalized $t$ distribution. 
    }
    \label{fig:2}
\end{figure}

Figure \ref{fig:1} displays the $\alpha$-quantile of the estimation error $\|\hmu -\mu\|_2^2$, with $\alpha$ ranging from 0.5 to 0.99, based on 1000 simulations for both distributional settings. 
For Settings 1 (normal distribution) and 2 (skewed generalized $t$ distribution), we set $\sigma^2=1$ and $q=2.5$, respectively.  In the case of normal distributions, our proposed estimator performs almost identically to the sample mean estimator, both of which outperform the MoM and trimmed mean estimator. Since the sample mean estimator is optimal for Gaussian data, this suggests that our estimator does not sacrifice statistical efficiency when applied to Gaussian data. In the case of heavy-tailed skewed generalized $t$ distributions, the estimation error of the sample mean estimator grows rapidly with increasing $\alpha$. This contrasts with the three robust estimators: our estimator, the MoM mean estimator, and the trimmed mean estimator.  Our estimator consistently outperforms the others in both settings.

Figure \ref{fig:2} examines the 99\%-quantile of the estimation error versus a distribution parameter, based on 1000 simulations. For Gaussian data, the  distribution parameter  is $\sigma$, and we vary $\sigma$ from 1 to 4 in increments of 0.1. For skewed generalized $t$ distributions, the  distribution parameter  is $q$, and we vary $q$ from 2.5 to 4  in increments of 0.1. For Gaussian data, our estimator performs identically to the optimal sample mean estimator, with both outperforming the MoM and trimmed mean estimators. 
In the case of skewed generalized $t$ distributions with $q\leq 3$, all three robust mean estimators either outperform or are as competitive as the sample mean estimator. However, when $q > 3$, the sample mean estimator starts to outperform both the MoM and trimmed mean estimators. Our proposed estimator, on the other hand, consistently outperforms all other methods across the entire range of parameter values.

\begin{figure}[t]
    \centering
        \includegraphics[width=0.48\linewidth]{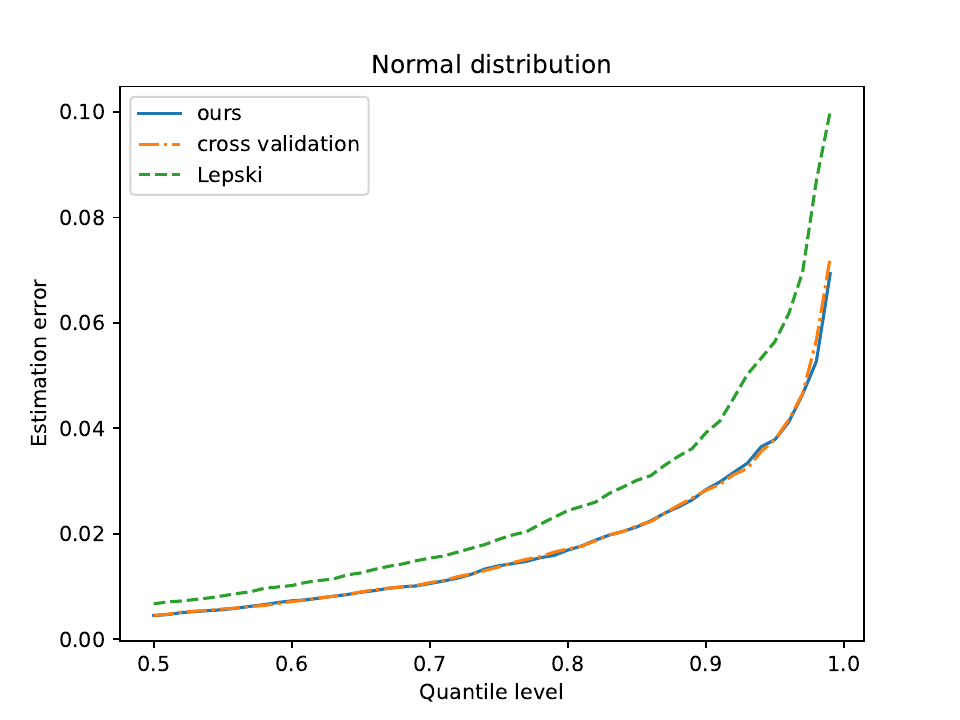}
  \includegraphics[width=0.48\linewidth]{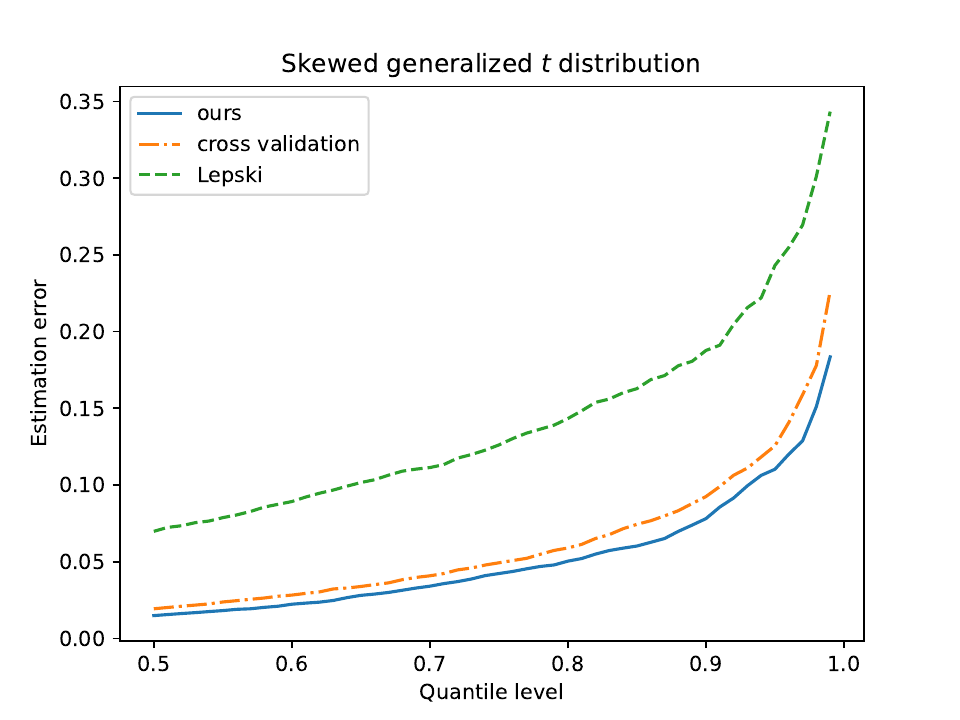}
    \caption{The $\alpha$-quantile of the estimation error (estimation error, $y$-axis) versus $\alpha$ (quantile level, $x$-axis) for our estimator, cross validation and Lepski's method. 
    }
    \label{fig:3}
\end{figure}

\begin{figure}[t]
    \centering
        \includegraphics[width=0.49\linewidth]{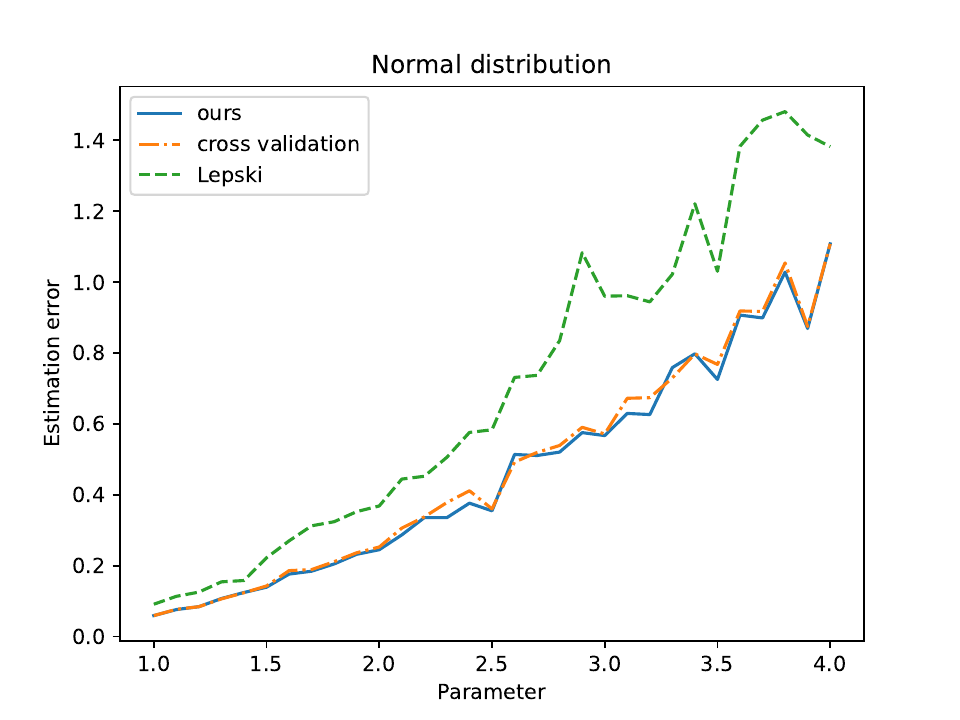}
  \includegraphics[width=0.49\linewidth]{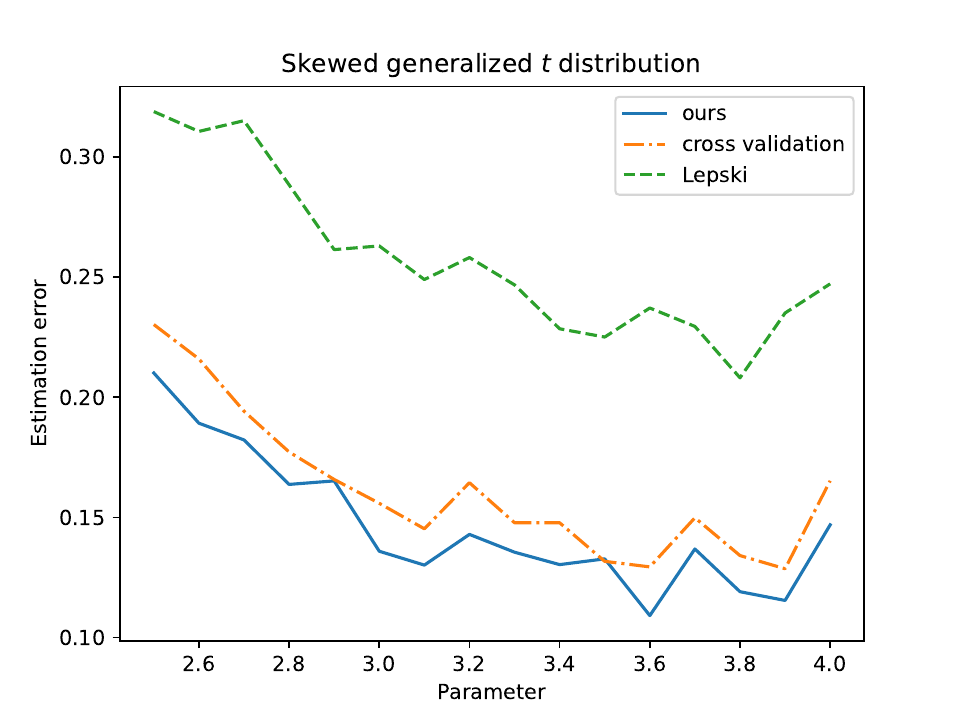}
    \caption{ The empirical 99\%-quantile of the estimation error (estimation error, $y$-axis) versus a distributution parameter (parameter, $x$-axis) for our estimator, cross validation and Lepski's method. 
    }
    \label{fig:4}
\end{figure}

We also conduct a computational performance comparison of our self-tuned method with pseudo-Huber loss + cross-validation,  and pseudo-Huber loss + Lepski's method. For cross-validation, we pick the best $\tau$ from a list of candidates $\{1,2,\ldots, 100\}$ using 10-fold cross-validation. In the case of Lepski's method, we follow the appendix and choose $V = 2$, $\rho = 1.2$, and $s=50$. We run 1000 simulations for the mean estimation problem in Setting 1 with $\sigma^2=1$ and a sample size of $n=100$. All computations are performed on a MacBook Pro with an Apple M1 Max processor and 64 GB of memory. The runtimes  are summarized in Table \ref{table1}.  Our proposed method is approximately $90\times$ faster than cross-validation and about $10\times$ faster than Lepski's method. The runtimes for the sample mean, MoM, and trimmed mean estimators in the same scenario are $0.018$, $0.111$, and $0.057$ seconds, respectively. Additionally, we compare the runtimes of our estimator with increasing sample sizes. Specifically, for $n=100, 1000, 10000, 100000$, the runtimes are $1.54$, $1.58$, $3.02$, and $25.04$ seconds, respectively.

Finally, we compare their statistical performance in both settings while varying the distribution parameter in the same manner as in Figure \ref{fig:2}. The results are summarized in Figure \ref{fig:3} and Figure \ref{fig:4}. In both figures, our method and cross-validation exhibit similar performance, with both outperforming Lepski's method. We suspect this is because Lepski's method depends on three additional hyperparameters: $V$, $\rho$, and $s$, and our chosen values may not be optimally tuned. This also indicates that Lepski's method does not consistently achieve good empirical performance, despite its elegant theoretical justifications.

\begin{table}[!t]
\begin{center}
\caption{{Comparing different tuning methods:} Run time (in seconds) for 1000 simulations  in Setting 1 with $\sigma^2=1$ and $n=100$.}
\begin{tabular}{ccc}
  \hline
{ours} & Lepski's method &  cross validation  \\
 \hline
   1.5 & 16.7 & 133.5\\
 \hline
\end{tabular}
\label{table1}
\end{center}
\end{table}

In summary, the most attractive feature of our method is its self-tuning property: (i) It is as efficient as the sample mean estimator for normal distributions and more efficient than popular robust alternatives for asymmetric and/or heavy-tailed distributions; (ii) It incurs much lower computational cost than cross-validation and Lepski's method. The latter property is particularly important for large-scale inference with a myriad of parameters to be tuned.

\section{Conclusions and discussions}\label{sec:6}

This paper proposes self-tuned estimators for estimating the means of heavy-tailed distributions, specifically those with only finite variances. Our approach introduces a new loss function that depends on both the mean parameter and a robustification parameter. By jointly optimizing these parameters, we demonstrate that the robustification parameter estimator can automatically adapt to the data variance. Consequently, the corresponding self-tuned mean estimator achieves optimal sub-Gaussian performance in finite samples, even when the data exhibit only second moments. This self-tuning property makes our approach computationally efficient, setting it apart from prior methods that necessitate cross-validation or Lepski's method for tuning the robustification parameters. In terms of asymptotic performance, the proposed estimator achieves asymptotic efficiency, distinguishing it from the widely used median-of-means estimator. We refer to the ability of performing optimally in both finite-sample and large-sample regimes as ``adaptivity", a feature our estimator possesses and the MoM estimator lacks.

In what follows, we extend our estimator to the multivariate case and discuss the limitations of our current work.

\paragraph{An extension to the multidimensional case}

We briefly discuss how to extend the proposed estimator to the multivariate case. Assume model \eqref{model:main} but with $y_i, \mus, \varepsilon_i \in \mathbb{R}^d$ being \iid such that $\mathbb{E}[\varepsilon_i]=0$ and $\text{cov}(\varepsilon_i)=\Sigma$. A simple strategy, as recommended by one of the referees, is to apply the univariate estimator coordinate-wise and then combine them to form the final estimator $\hat{\mu}$. Let $\sigma_{kk}^2$ be the $k$-th diagonal term of $\Sigma$, and $\sigma^2_{kk, x^2}= \EE[\varepsilon_{ik}^2 1(\varepsilon_{ik}^2 \leq x^2)]$, where $\varepsilon_{ik}$ is the $k$-th coordinate of $\varepsilon_i$. Then the following proposition holds. 

\begin{proposition}[Finite-sample property of the multivariate self-tuned mean estimator]\label{prop:multi}
Assume that $n$ is sufficiently large. Suppose $v_0<  c_0  {\sigma_{kk, \tauv^2-1}} \leq  C_0 \sigma_{kk} < V_0$ for all $1\leq k \leq d$, where $c_0$ and $C_0$ are some constants.
For any $\delta\in (0,1)$, take $z^2=\log(n/\delta)$. Then with probability at least $1-\delta$, we have
\$
\|\widehat \mu -\mu^*\|_2 \leq   C  \, \sqrt{\frac{\tr(\Sigma)\, \log(nd/\delta)}{n}},
\$
where $C$ is some constant. 
\end{proposition}

We also have the following asymptotic result, which states that the multivariate mean estimator also achieves asymptotic efficiency.

\begin{proposition}[Asymptotic efficiency of the multivariate self-tuned mean estimator]\label{prop:asym_multi}
Fix any $\iota\in(0, 1]$.  Assume $\max_{1\leq k\leq d}\EE [\varepsilon_{ik}^{2+\iota}] < \infty$  and the same assumptions as in Proposition \ref{prop:multi}.
Take any $z^2 \geq 2\log(n)$. Then
\$
\sqrt{n} \, (
\hmu(\hv) - \mus)
\rightsquigarrow \cN \left(0, \Sigma\right). 
\$
\end{proposition}

\paragraph{Limitations}  

One limitation of our self-tuned estimator is that its finite-sample performance depends on unknown constants, making it challenging to compute the sample complexity in advance for a fixed confidence level. Additionally, the proposed estimator achieves optimality only up to a logarithmic term. It remains an open question whether this logarithmic factor can be eliminated. Another limitation of this study is its scope. We primarily focus on robust mean estimators as they represent the simplest case, and the proofs are already quite complex. Nevertheless, our approach can potentially extend to more general settings, such as regression and matrix estimation problems. While we have provided an extension to the multivariate case, it is worth noting that this extension is not optimal in terms of finite-sample properties; for the optimal finite-sample property in the multivariate case, one can refer to the paper by \cite{lugosi2019mean}. Additionally, exploring the asymptotic properties of multivariate median-of-means estimators could be an intriguing avenue for future research.

\vspace{10pt}
\paragraph{Acknowledgement} The author would like to thank Stanislav Minsker for helpful discussions on the asymptotic properties of median-of-means estimators.


\bibliographystyle{tmlr}
\bibliography{ref}

\newpage
\appendix

\addcontentsline{toc}{section}{Appendix} 
\part{Appendix} 
\parttoc 

\section{Basic facts}\label{sec:basic}

This section collects some basic facts such as first-order derivatives and the Hessian matrix for the empirical loss function. 
{Let $\tau = v\sqrt{n}/z$ throughout the appendix.} Recall that our loss function is
\$
L_n(\mu, v) 
&= \frac{1}{n}\sum_{i=1}^n \ell(y_i -\mu, v) 
= \frac{1}{n}\sum_{i=1}^n\left\{\frac{\sqrt{n}}{z}\sqrt{\frac{nv^2}{z^2} + {(y_i-\mu)^2}}  - \left(\frac{n}{z^2} - a \right) v \right\} \\
&=  \frac{1}{n}\sum_{i=1}^n\left\{\frac{\sqrt{n}}{z} \left(\sqrt{\tau^2 + {(y_i-\mu)^2}}  - \tau \right) + a\cdot \frac{\tau}{z\sqrt{n}} \right\} . 
\$
The first-order and second-order derivatives of $L_n(\mu,v)$ are  
\begin{align*}
\nabla_\mu L_n(\mu,v)
&= -\frac{1}{n}\sum_{i=1}^n \frac{y_i-\mu}{v\sqrt{1+z^2(y_i-\mu)^2/(nv^2)}}
= -  \frac{\sqrt{n}}{z}\cdot \frac{1}{n} \sum_{i=1}^n  \frac{y_i - \mu}{\sqrt{\tau^2 + (y_i -\mu)^2}},\\
\nabla_v L_n(\mu,v)
&= \frac{1}{n}\sum_{i=1}^n \frac{n/z^2}{\sqrt{1 + {z^2(y_i-\mu)^2}/({nv^2})}}-\left(\frac{n}{z^2} - a\right)
= \frac{n}{z^2} \cdot \frac{1}{n}\sum_{i=1}^n \left(\frac{\tau}{\sqrt{\tau^2 + {(y_i-\mu)^2}}} - 1\right)  + a
\end{align*}
where $a=1/2$. 
The Hessian matrix is
\$
H(\mu, v)
= \begin{bmatrix}
\frac{\sqrt n}{z} \frac{1}{n}  \sum_{i=1}^n \frac{\tau^2}{\big({\tau^2+(y_i-\mu)^2}\big)^{3/2}} & \frac{n}{z^2} \frac{1}{n}\sum_{i=1}^n \frac{\tau(y_i-\mu)}{(\tau^2+(y_i-\mu)^2)^{3/2}} \\
 \frac{n}{z^2} \frac{1}{n}\sum_{i=1}^n \frac{\tau(y_i-\mu)}{(\tau^2+(y_i-\mu)^2)^{3/2}} & \frac{n^{3/2}}{z^{3}} \frac{1}{n}\sum_{i=1}^n \frac{(y_i-\mu)^2}{(\tau^2+(y_i-\mu)^2)^{3/2}} 
\end{bmatrix}. 
\$

\section{Population bias}

Let  $\mu^*(v)$ be the population version of the pseudo-Huber regression coefficient with $v$ fixed {\it a priori} 
\$
\mu^*(v) =\argmin_{\mu} \left\{ L(\mu, v):= \EE L_n(\mu,v)\right\}. 
\$
Recall that $\tau = v\sqrt{n}/z$. Let
\$
\psi_v(x)&:=\nabla_x{\ell}^\s(x, v)=\frac{ \sqrt{n}x}{z\sqrt{\tau^2+x^2}},\\
h_v(x)&:=\nabla^2_{x}{\ell}^\s(x, v)= \frac{ \sqrt{n}}{z\sqrt{\tau^2+x^2}} -\frac{ \sqrt{n}x^2}{z(\tau^2+x^2)^{3/2}}=\frac{\sqrt{n}\tau^2}{z(\tau^2+x^2)^{3/2}}. 
\$

\begin{assumption}\label{ass:mean.lsc}
The second-order derivative of  $L(\mu, v)$ satisfies
\$
\nabla_\mu^2 L(\mu, v) \geq \kappa_\ell > 0
\$
for any $\mu\in \BB(r, \mu^*):=\{\mu: |\mu-\mu^*|\leq r\}$, where $r>0$ is some local radius parameter and we use the same $\kappa_\ell$ as in Assumption \ref{ass:mean.sc} without loss of generality. 
\end{assumption}


Our next proposition shows that the population bias is of order $\sqrt{n}/(z\tau^2)$.

\begin{proposition}[Population bias]\label{prop:bias}
Assume that Assumption \ref{ass:mean.lsc} holds with some $r>{\sqrt{n}\sigma^2}/(2{\kappa_\ell\tau^2}).$ Then
\$
|\mu^*(v)-\mu^*|\leq  \frac{\sigma^2}{2z\kappa_\ell}\cdot\frac{\sqrt{n}}{\tau^2}\lesssim \frac{\sqrt{n}}{z\tau^2}. 
\$
\end{proposition}

\begin{proof}[Proof of Proposition \ref{prop:bias}]
Let ${\Delta} = {\mu^*}- {\mu^*(v)}$ and  $h_v({\mu}) = n^{-1} \sn \e \{ \ell^\s (y_i - {\mu}, v ) \}$. We first assume that $|\Delta|\leq r$. 
By the first order optimality of ${\mu^*(v)}$, we have 
\$
\nabla h_v({\mu^*(v)}) = {0},
\$
and thus
\#
\langle {\Delta}, \nabla^2 h_v (\tilde{{\mu}}) {\Delta} \rangle = \langle \nabla h_v({\mu^*}) - \nabla h_v({\mu^*}(v)),  {\Delta} \rangle  = \langle \nabla h_v({\mu^*}) , {\Delta} \rangle = -\frac{1}{n} \sn \EE[\psi_v({{\varepsilon_i}})] {\Delta}   \label{eq:mean.be}
\#
where $\tilde{\mu}= \lambda {\mu^*}+ (1-\lambda) {\mu^*(v)}$ for some $0\leq \lambda \leq 1$.

Since $\EE ( {{\varepsilon_i}} ) =0$, we have
\#
\left|\EE \{ -\psi_v({{\varepsilon_i}} ) \}\right| 
&=   \frac{\sqrt{n}}{z}\cdot\left| \EE \left\{ \frac{-{{\varepsilon_i}}/\tau}{\sqrt{1+{\varepsilon_i^2}/\tau^2}}\right\}\right|
=\frac{\sqrt{n}}{z}\cdot\left|\EE \left\{ \frac{{{\tau^{-1}}{{\varepsilon_i}}} \left(\sqrt{1+{\varepsilon_i^2}/\tau^2}-1\right)}{\sqrt{1+{\varepsilon_i^2}/\tau^2}}\right\}\right| \nn \\
&\leq \frac{\sqrt{n}}{2z}\cdot  \EE \left| \frac{({\varepsilon_i})^3/\tau^3}{\sqrt{1+{\varepsilon_i^2}/\tau^2}}\right|  
 \leq \frac{\sqrt{n}\sigma^2}{2z\tau^2}, \label{eq:mean.gb}
\#
where the first inequality uses  the inequality $\sqrt{1+x^2}\leq 1+x^2/2$, and the last inequality uses the fact that $\sqrt{1+{\varepsilon_i^2}/\tau^2}\geq 1\vee |{\varepsilon_i}|/\tau.$

Using equality \eqref{eq:mean.be} together with Assumption \ref{ass:mean.lsc} and inequality \eqref{eq:mean.gb}, and canceling the term $|\Delta|$ on both sides,  we obtain 
\$
| \Delta|\leq \frac{\sqrt{n}\sigma^2}{2z\kappa_\ell\tau^2}. 
\$

Lastly, we prove that it must hold that $|\Delta|\leq r$. If not, then we shall construct an intermediate solution between $\mu^*$ and $\mu^*(v)$, denoted by $\mus_{\eta}(v) =\mu^*+ \eta (\mu^*(v)-\mu^*)$, such that $|\mus_{\eta}(v) -\mu^* |= r$. Specifically, we can  choose some $\eta \in(0,1)$ such that  $|  \mus_{\eta}(v) -\mu^*|=r$. We then proceed the above calculation and obtain 
\$
|  \mu^*_\eta(v) -\mu^* |\leq \frac{\sqrt{n}\sigma^2}{2z\rho_\ell\tau^2} <r. 
\$
This is a contradiction. 
\end{proof}

\section{An alternating gradient descent algorithm}

This section presents an alternating gradient descent algorithm to optimize \eqref{eq:const}. The algorithm generates the solution sequence $\{(\mu_k, v_k): k\geq 0\}$ with the initialization $(\mu_0, v_0)=(\mu_\init, v_\init)$. At the working solution $(\mu_{k}, v_{k})$ for any $k \geq 0$, the $(k+1)$-th iteration involves the following two steps:
\begin{enumerate}
\item $\mu_{k+1} = \mu_k  -\eta_1\nabla_\mu  L_n(\mu_k, v_{k})$,
\item $\tilde v_{k+1} = v_k  -\eta_2\nabla_\tau  L_n(\mu_{k+1}, v_k)$  and $v_{k+1}= \min\{\max\{\tilde v_{k+1}, v_0\}, V_0\}$,
\end{enumerate}
where $\eta_1$ and  $\eta_2$ are the learning rates and 
\begin{align*}
\nabla_\mu L_n(\mu, v) &= -\frac{1}{n}\sum_{i=1}^n \frac{y_i-\mu}{v\sqrt{1+z^2(y_i-\mu)^2/(nv^2)}}, \\
\nabla_v L_n(\mu, v) &= \frac{1}{n}\sum_{i=1}^n \frac{n/z^2}{\sqrt{1 + {z^2(y_i-\mu)^2}/({nv^2})}}-\left(\frac{n}{z^2} - a\right).
\end{align*}
The above two steps are repeated until convergence. The algorithm routine is summarized in Algorithm \ref{alg:1}. The learning rates $\eta_1$ and $\eta_2$ can be chosen adaptively in practice. In our experiments, we utilize alternating gradient descent with the Barzilai and Borwein method and backtracking line search.  

\begin{algorithm}[t]
  \begin{algorithmic}
  \STATE{{\bf Input}:} \(\mu_\init, v_\init, v_0, V_0, \eta_1,  \eta_2, \, (y_1,\ldots, y_n)\)
 \FOR{\(k= 0, 1, \dots\) until convergence}
  \STATE{\(\mu_{k+1} = \mu_k  -\eta_1\nabla_\mu  L_n(\mu_k,v_{k})\)}
   \STATE{$\tilde v_{k+1} = v_k  -\eta_2\nabla_\tau  L_n(\mu_{k+1},v_k) $  and $v_{k+1}= \min\{ \max\{\tilde v_{k+1}, v_0\}, V_0\}$  }
 \ENDFOR
 \STATE{{\bf Output}:} $\widehat  \mu= \mu_{k+1}$, $\widehat v = v_{k+1}$
 \end{algorithmic}
\caption{An alternating gradient descent algorithm.}
\label{alg:1}
\end{algorithm}

\section{Comparing with Lepski's method}\label{sec:lepski}

We compare our method with Lepski's method. Specifically, we employ Lepski's method to tune the robustification parameter $v$ and, consequently $\tau=v\sqrt{n}/z$, in the empirical pseudo-Huber loss:
\$
L^{h}_n(\mu, v) := \frac{1}{n} \sum_{i=1}^n \left(\tau\sqrt{\tau^2+(y_i-\mu)^2}  -\tau^2\right). 
\$

Lepski's method proceeds as follows. Let $v_{\max}$ be an upper bound for $\sigma$, and $\tau_{\max} = v_{\max}\sqrt{n}/z$ with $z=\sqrt{\log(1/\delta)}$.  Let $n$ be sufficiently large.  Then with probability at least $1-\delta$, we have
\$
|\widetilde\mu(v_{\max}) -\mus|\leq 6 v_{\max}\sqrt{\frac{\log(4/\delta)}{n} } =: \epsilon(v_{\max}, \delta ),
\$
where $\widetilde\mu(v_{\max}) = \argmin_{\mu} L_n(\mu, v_{\max})$.  Let us by convention set $\epsilon(v_{\max}, 0 )= +\infty$.  Clearly, $\epsilon(v_{\max}, \delta )$ is homogeneous in the sense that
\$
\epsilon(v_{\max}, \delta ) = B(\delta) v_{\max}, ~~ \text{where}~B(
\delta) =  6\sqrt{\frac{\log(4/\delta)}{n}}. 
\$

For some parameters $V\in\RR$, $\rho>1$,  and $s\in \NN$,  we choose the following probability measure  $\cV$ for $v_{\max}$
\$
\cV(v_{\max}) 
=
\begin{cases}
{1}/(2s+1), &\text{if}~v_{\max}= V\rho^{k}, \,  k\in \ZZ, \, |k|\leq s, \\
0, &\text{otherwise}. 
\end{cases}
\$
Let us consider for any $v_{\max}$ such that $\epsilon(v_{\max}, \delta  \cV(v_{\max}))<\infty$ the confidence interval 
\$
I(v_{\max})= \tilde \mu(v_{\max}) + \epsilon(v_{\max}, \delta \,\cV(v_{\max}))\times [-1, 1], 
\$
where 
\$
\epsilon(v_{\max}, \delta\,\cV(v_{\max})) = 6v_{\max}\sqrt{\frac{\log(4/\delta)+\log(2s+1)}{n}}
\$ 
if $v_{\max}= V\rho^k$ for any $k\in \ZZ$ and $|k|\leq s$. We set $I(v_{\max})=\RR $ when $ \epsilon(v_{\max}, \delta \cV(v_{\max}))=+\infty$.

Let us consider the non-decreasing family of closed intervals
\$
J(v_1) =\bigcap \left\{I(v_{\max}): v_{\max}\geq v_1  \right\}, \, v_1 \in \RR_+. 
\$
In this definition, we can restrict the intersection to the support of $\cV$ , since otherwise $I(v_{\max}) = \RR$. Lepski's method picks the center point of the intersection
\$
\bigcap \left\{J(v_1): v_1\in \RR_+, \, J(v_1)\ne \emptyset   \right\}
\$
to be the final estimator $\widehat \mu_{\rm Lepski}$. Then the following result is due  to \cite{catoni2012challenging}. 

\begin{proposition}\label{prop:lepski}
Suppose $|\log(\sigma/V)|\leq 2 s \log (\rho)$. Then with probability at least $1-\delta$
\$
|\widehat\mu_{\rm Lepski}-\mu^*|\leq 12\rho \sigma \sqrt{\frac{\log(4/\delta)+\log(2s+1)}{n}}. 
\$
\end{proposition}

If we take the grid fine enough such that $s=n$, then the upper  bound above reduces to  
\$
12 \rho \sigma \sqrt{\frac{\log(4/\delta)+\log(2n+1)}{n}},
\$
which agrees with deviation bound for our proposed estimator, up to a constant multiplier. Therefore, our proposed estimator is comparable to Lepski's method in terms of the deviation upper ound. Computationally, our estimator is self-tuned  and  thus computationally more efficient  than Lepski's method; detailed numerical results can be found in Section \ref{sec:5}.


\section{Proofs for Section \ref{sec:2}}

\subsection{Proofs for Theorem \ref{thm:ada}}
\begin{proof}[Proof of Theorem \ref{thm:ada}]
We prove first the finite-sample result and then the asymptotic result. Recall that $\taus = \vs \sqrt{n}/z$. 

\paragraph{Proving the finite-sample result.} 
On one side, if $\vs=0$ and by the definition of $\vs$,  $\vs$ satisfies
\$
1-\frac{az^2}{n}
=\EE\frac{\sqrt{n}\vs}{\sqrt{n\vs^2+{z^2{\varepsilon^2}}}} = 0,
\$
which is a contradiction. Thus $\vs>0$. 
Using the convexity of $1/\sqrt{1+x}$ for $x>-1$ and Jensen's inequality acquires
\$
1-\frac{az^2}{n}
=\EE\frac{\sqrt{n}\vs}{\sqrt{n\vs^2+{z^2{\varepsilon^2}}}}
= \EE\frac{1}{\sqrt{1+z^2{\varepsilon^2}/(n\vs^2)}}
\geq \frac{1}{\sqrt{1 + z^2\sigma^2/(n\vs^2)}}\geq 1- \frac{z^2\sigma^2}{2n\vs^2},
\$
where the last inequality uses the inequality $(1+x)^{-1/2}\geq 1-x/2$, i.e., Lemma \ref{lemma:ine} (i) with $r = -1/2$. This implies
\$
\vs^2 \leq \frac{\sigma^2}{2a}.
\$

On the other side, using the concavity of $\sqrt{x}$, we obtain, for any $\gamma \in [0, 1)$, that 
\#
1-\frac{az^2}{n}
&=\EE\frac{\sqrt{n}\vs}{\sqrt{n\vs^2+z^2\varepsilon^2}}
= \EE\frac{1}{\sqrt{1+ \sigma^2 z^2\varepsilon^2/(n\vs^2)}}  \nn \\
&\leq  \sqrt{\EE \left(\frac{1}{{1+ z^2\varepsilon^2/(n\vs^2)}}\right)}  \nn \\
&\leq \sqrt{\EE\left\{\left(1 - (1-\gamma) \frac{{z^2\varepsilon^2}}{n\vs^2}\right)1\left(\frac{z^2\varepsilon^2}{n\vs^2}\leq \frac{\gamma}{1-\gamma}\right)+\frac{1}{{1+z^2\varepsilon^2/(n\vs^2)}}1\left(\frac{z^2\varepsilon^2}{n\vs^2}> \frac{\gamma}{1-\gamma}\right)\right\}}  \nn \\
&\leq \sqrt{{1} -  (1-\gamma)\, \EE\left\{\frac{z^2\varepsilon^2}{n\vs^2} 1\left(\frac{z^2\varepsilon^2}{n\vs^2}\leq \frac{\gamma}{1-\gamma}\right)\right\}} \nn \\
&\leq \sqrt{1  -  (1-\gamma) \, \frac{\EE\left\{{\varepsilon^2} 1\left({\varepsilon^2}\leq  {\gamma \taus^2}/({1-\gamma})\right)\right\}}{ {n\vs^2}/{z^2}  }},  \label{eq:oracle_1}
\#
where the second inequality uses Lemma \ref{lemma:ine_2}, that is, 
\$
(1+x)^{-1}\leq 1- (1-\gamma)x, ~\textnormal{for any}~ x\in \left[ 0, \frac{\gamma}{1-\gamma} \right]. 
\$
Taking square on both sides of inequality \eqref{eq:oracle_1} and using the fact that $n\geq az^2$ together with Lemma \ref{lemma:ine} (i) with $r=2$, aka $(1+x)^2\geq 1+2x$ for $x\geq -1$, we obtain
\$
1- \frac{2az^2}{n}\leq \left(1-\frac{az^2}{n} \right)^2\leq 1- (1-\gamma)\, \frac{\EE\{{\varepsilon^2} 1({\varepsilon^2}\leq   \gamma \taus^2/(1-\gamma))\}}{ n\vs^2/z^2},
\$
or equivalently
\$
\vs^2\geq \frac{\sigma^2_{\varphi \taus^2}}{2a},
\$ 
where $\varphi = \gamma/(1-\gamma)$. 
Combining the upper bound and the lower bound for $\vs^2$ completes the proof for the finite-sample result.


\paragraph{Proving the asymptotic result.} 
The above derivation implies that $\vs < \infty$ for any $a> 0$. By the definition of $\vs$, we obtain
\#\label{eq:oracle_2}
\frac{az^2}{n} 
&= 1-\EE\frac{1}{\sqrt{1+{z^2{\varepsilon^2}}/(n\vs^2)}}. 
\# 
We must have $n\vs^2/z^2\rightarrow \infty$. Otherwise assume 
\$
\limsup_{n\rightarrow \infty} n\vs^2 / z^2 \leq M <\infty.
\$ 
Taking $n\rightarrow \infty$, the left hand side of the above equality  goes to $0$ while the right hand is lower bounded as
\$
1-\EE\frac{1}{\sqrt{1+ \varepsilon^2/M}}     
&\geq 1-  \sqrt{\EE \left(\frac{1}{{1+ \varepsilon^2/M}}\right)}  \\
&\geq 1- \sqrt{1- \frac{\EE\left\{\varepsilon^21(\varepsilon^2\leq M)\right\}}{2M}}\\
&\geq 1- \sqrt{\frac{1}{2}} >0, 
\$
where the first two inequalities follow from the same arguments in deriving \eqref{eq:oracle_1} but with $\gamma=1/2$, and the third inequality uses the fact that 
\$
\EE\{\varepsilon^2 1(\varepsilon^2 \leq M)\} \leq M. 
\$ 
This is a contradiction. Thus $n\vs^2/z^2\rightarrow \infty$. 
Multiplying both sides of the above equality by $n$, taking $n\rightarrow \infty$, and using the dominated convergence theorem, we obtain
\$
az^2 &= \lim_{n\rightarrow \infty} \EE\left( n \cdot \frac{ \sqrt{1+{z^2{\varepsilon^2}}/(n\vs^2)} - 1}{\sqrt{1+{z^2{\varepsilon^2}}/(n\vs^2)}}\right)\\
&= \lim_{n\rightarrow \infty} \EE\left( n \cdot \frac{1}{\sqrt{1+{z^2{\varepsilon^2}}/(n\vs^2)}} \cdot \frac{ \sqrt{1+{z^2{\varepsilon^2}}/(n\vs^2)} - 1}{z^2\varepsilon^2 / (2n\vs^2)}\cdot \frac{z^2\varepsilon^2}{2n\vs^2}\right)\\
&= \frac{\EE z^2\varepsilon^2}{2 \lim_{n\rightarrow \infty} \vs^2},
\$
and thus $\lim_{n\rightarrow \infty}\vs^2 = \sigma^2/(2a)$.  This proves the asymptotic result. 

\end{proof}

\subsection{Proof of Proposition \ref{prop:cvx}}
\begin{proof}[Proof of Proposition \ref{prop:cvx}]
The convexity proof consists of two steps: (1) proving that $L_n(\mu, v)$ is jointly convex in $\mu$ and $v$; (2) proving that $L_n(\mu, v)$ is strictly convex, provided that there are at least two distinct data points. 

To show that $L_n(\mu, v)=n^{-1}\sum_{i=1}^n\ell^\p(y_i - \mu, v)$ in \eqref{opt:main} is jointly convex in $\mu$ and $v$, it suffices to show that each $\ell^\p(y_i-\mu, v)$ is jointly  convex in $\mu$ and $v$.  Recall that $\tau = v\sqrt{n}/z.$ The Hessian matrix of $\ell^\p(y_i-\mu,v)$ is
\$
H_i(\mu, v) 
&=\frac{\sqrt n}{z}\cdot \frac{1}{\big(\tau^2 + (y_i-\mu)^2\big)^{3/2}}\begin{bmatrix}
    \tau^2      & (\sqrt{n}/{z}) \ \tau(y_i-\mu)  \\
     (\sqrt{n}/{z}) \   \tau(y_i-\mu)   &  (\sqrt{n}/{z})^2 \  (y_i-\mu)^2 
\end{bmatrix}
 \succeq 0,
\$
and thus positive semi-definite. Therefore,  $L_n(\mu,v)$ is jointly convex in $\mu$ and $v$. 

We proceed to show (2). 
Because  the Hessian matrix $H(\mu, v)$ of $L_n(\mu, v)$ satisfies $H(\mu,v)=n^{-1}\sum_{i=1}^n H_i(\mu,v)$ and each $H_i(\mu,v)$ is positive semi-definite, we only need to show that $H(\mu,v)$ is of full rank. Without generality, assume that $y_1\ne y_2$. Then
\$
H_1(\mu,v)+H_2(\mu,v)
&=\frac{\sqrt n}{z}\cdot\sum_{i=1}^2  \frac{1}{\big(\tau^2 + (y_i-\mu)^2\big)^{3/2}}
\begin{bmatrix}
    \tau^2      &(\sqrt{n}/{z}) \  \tau(y_i-\mu) \\
      (\sqrt{n}/{z}) \  \tau(y_i-\mu)   & (\sqrt{n}/{z})^2 \ (y_i-\mu)^2 
\end{bmatrix}.
\$
Some algebra yields 
\$
\det\left(H_1(\mu,v)+H_2(\mu,v)\right)
&= \frac{n^2\tau^2}{z^4} \cdot\frac{(y_1-y_2)^2}{(\tau^2+(y_1-\mu)^2)^{3/2}(\tau^2+(y_2-\mu)^2)^{3/2}} \ne 0
\$
for any $\tau >0$ ($v > 0$),  and $\mu\in\RR$, provided that $y_1\ne y_2$. Therefore,  $H_1(\mu, v)+H_2(\mu, v)$ is of  full rank and thus is $H(\mu,\tau)$, provided $v >0$, $\mu\in\RR$, and $y_1\ne y_2$. 
\end{proof}

\subsection{Supporting lemmas}

\begin{lemma}\label{lemma:ine_2}
Let $0\leq \gamma <1$. For any $0 \leq x \leq \gamma/(1-\gamma)$, we have
\$
(1+x)^{-1}\leq 1- (1-\gamma)x. 
\$
\end{lemma}

\begin{proof}[Proof of Lemma \ref{lemma:ine_2}]
To prove the lemma,  it suffices to show, for any $\gamma \in [0, 1)$, that 
\$
1 
&\leq (1+x)- (1-\gamma)x(1+x), ~~~\forall~0\leq x\leq \frac{\gamma}{1-\gamma},
\$
which is equivalently to
\$
x\left(x -   \frac{\gamma}{1-\gamma}   \right) \leq 0, ~~~\forall~0\leq x\leq \frac{\gamma}{1-\gamma}.
\$ 
The above inequality always holds, and this completes the proof. 

\end{proof}


\section{Proofs for the fixed $v$ case}
This section collects proofs for Theorem \ref{thm:mean.fixed}, Lemma \ref{lemma:mean.sc}, and Corollary \ref{coro:mean.fixed}. 
Recall that $\tau = v\sqrt{n}/z$, and the gradients with respect to $\mu$ and $v$ are
\$
\nabla_\mu L_n(\mu, v)&= -\frac{1}{n}\sum_{i=1}^n \frac{y_i-\mu}{v\sqrt{1+z^2(y_i-\mu)^2/(nv^2)}}
= -\frac{\sqrt{n}}{z} \cdot \frac{1}{n}\sum_{i=1}^n \frac{y_i - \mu}{\sqrt{\tau^2 + (y_i -\mu)^2}} ,\\
\nabla_v L_n(\mu, v)&= \frac{1}{n}\sum_{i=1}^n \frac{n/z^2}{\sqrt{1 + {z^2(y_i-\mu)^2}/({nv^2})}}-\left(\frac{n}{z^2} - a\right)
= \frac{n}{z^2} \cdot \frac{1}{n}\sum_{i=1}^n \left(\frac{\tau}{\sqrt{\tau^2 + {(y_i-\mu)^2}}} - 1\right)  + a. 
\$


\subsection{Proof of Theorem \ref{thm:mean.fixed}}
\begin{proof}[Proof of Theorem \ref{thm:mean.fixed}]
Because $\widehat\mu (v)$ is the stationary point of $L_n(\mu,v)$, we have 
\$
\frac{\partial}{\partial \mu} L_n(\widehat \mu (v),v)
&= -\frac{1}{n}\sum_{i=1}^n \frac{y_i-\widehat\mu(v)}{v\sqrt{1 + z^2(y_i-\widehat\mu (v))^2/(nv^2)}}
= -   \frac{\sqrt{n}}{z}\cdot \frac{1}{n}  \sum_{i=1}^n \frac{y_i - \hmu(v)}{\sqrt{\tau^2 + (y_i -\hmu(v))^2}}=0. 
\$
Let $\Delta = \hmu(v) -\mu$. We first assume that $|\Delta|:=|\widehat \mu(v)-\mu^*|\leq r_0 \leq r$. Using Assumption \ref{ass:mean.sc} obtains
\$
\kappa_\ell|\widehat\mu(v)-\mu^*|^2
&\leq \llangle \frac{\partial}{\partial\mu} L_n(\widehat \mu(v),v) - \frac{\partial}{\partial\mu} L_n(\mu^*,v),  \widehat\mu(v) -\mu^*   \rrangle \\
&\leq \left|\frac{1}{\sqrt{n}}\sum_{i=1}^n \frac{{\varepsilon_i}}{z\sqrt{\tau^2+ {\varepsilon_i^2}}}\right| \left|\widehat\mu(v) -\mu^* \right|,
\$
or equivalently
\$
\kappa_\ell|\widehat\mu(v)-\mu^*|
&\leq \left|\frac{1}{\sqrt n}\sum_{i=1}^n \frac{{\varepsilon_i}}{z\sqrt{\tau^2+ {\varepsilon_i^2}}}\right|. 
\$
Applying Lemma \ref{lemma:con.1} with the fact that $\left|\EE \left( {\tau{\varepsilon_i}}/({\tau^2+{\varepsilon_i^2}})^{1/2} \right)\right|  \leq  {\sigma^2}/{(2\tau)}$, we obtain with probability at least $1-2\delta$ that
\$
\kappa_\ell|\widehat\mu(v)-\mu^*|
&\leq \left|\frac{\sqrt{n}}{\tau}\frac{1}{n}\sum_{i=1}^n \frac{\tau{\varepsilon_i}}{z\sqrt{\tau^2+ {\varepsilon_i^2}}}\right|
\leq \frac{\sqrt{n}}{z\tau}\left(\sigma\sqrt{\frac{2\log(1/\delta)}{n}} + \frac{\tau\log(1/\delta)}{3n} +\frac{\sigma^2}{2\tau}\right) ,
\$
or equivalently
\$
\kappa_\ell|\widehat\mu(v)-\mu^*|
&\leq \sqrt{\frac{2\log(1/\delta)}{z^2\tau^2/{\sigma^2}}} +  \frac{\log(1/\delta)}{3z\sqrt n} +\frac{\sqrt{n}\sigma^2}{2z\tau^2}.  
\$
Since  $\tau = v\sqrt{n}/z$, we have 
\$
\kappa_\ell|\widehat\mu(v)-\mu^*|
\leq  \left(\frac{\sqrt{2}\sigma}{v}+\frac{\sqrt{\log (1/\delta)}}{3z}\right)\sqrt{\frac{\log(1/\delta)}{n}} + \frac{1}{2}\cdot\frac{\sigma^2}{v^2}\cdot\frac{z}{\sqrt n}. 
\$
Taking $z = \sqrt{\log({1}/\delta)}$ then yields
\$
\kappa_\ell|\widehat\mu(v)-\mu^*|
&\leq  \left(\frac{\sqrt{2}\sigma}{v}+\frac{\sqrt{\log(1/\delta)}}{3\sqrt{\log({1}/\delta)}}\right)\sqrt{\frac{\log(1/\delta)}{n}} + \frac{1}{2}\cdot\frac{\sigma^2}{v^2}\cdot \sqrt{\frac{\log({1}/\delta)}{n}} \\
&\leq   \left(\frac{\sqrt{2}\sigma}{v}+\frac{1}{3} + \frac{1}{2}\cdot \frac{\sigma^2}{v^2} \right)\sqrt{\frac{\log({1}/\delta)}{n}}  \\
&< \left(1 + \frac{\sigma}{\sqrt{2}v}\right)^2\,\sqrt{\frac{\log({1} /\delta)}{n}}
\$
for any $\delta \in (0, 1/2)$.  Moving $\kappa_\ell$ to the right hand side and using a change of variable $2\delta \rightarrow \delta$, we obtain  
\$
|\widehat\mu(v)-\mu^*| 
&< \frac{1}{\kappa_\ell} \cdot \left(1 + \frac{\sigma}{\sqrt{2}v}\right)^2\,\sqrt{\frac{\log(2 /\delta)}{n}}\\
&=  {r}_0 \leq {r}. 
\$
This completes the proof, provided that  $|\Delta|\leq r_0$.

Lasty, we  show that  $|\Delta|\leq r_0$ must hold. If not, we shall construct an intermediate solution between $\mu^*$ and $\widehat\mu(v)$, denoted by $\mu_{\eta} =\mu^*+ \eta (\widehat\mu(v)-\mu^*)$, such that $| \mu_{\eta} -\mu^* |= r_0$. Specifically, we can  choose some $\eta \in(0,1)$ such that  $|  \mu_{\eta} -\mu^* |=r_0$. We then repeat  the above calculation and  obtain 
\$
|\widehat\mu(v)-\mu^*|
&\leq \frac{1}{\kappa_\ell} \cdot \left(\frac{\sqrt{2}\sigma}{v}+\frac{1}{3} + \frac{1}{2}\cdot \frac{\sigma^2}{v^2} \right)\sqrt{\frac{\log({2}/\delta)}{n}} \\
&< r_0= \frac{1}{\kappa_\ell} \cdot\left(1 + \frac{\sigma}{\sqrt{2}v}\right)^2\,\sqrt{\frac{\log({2} /\delta)}{n}} 
\$
which  is a  contradiction. Therefore, it must  hold that $|\Delta|\leq r_0$. 
\end{proof}


\subsection{Proof of Lemma \ref{lemma:mean.sc}}

\begin{proof}[Proof of Lemma \ref{lemma:mean.sc}]

We first prove that, with probability at least $1-\delta$,  Assumption \ref{ass:mean.sc} with $\kappa_\ell = 1/(2v)$ and radius $r$ holds for any fixed $v\geq v_0$. 
Recall that $\tau = v\sqrt{n}/z$. For notational simplicity, let $\Delta = \mu- \mu^*$ and  $\tauv = v_0 \sqrt{n}/z$. It follows that
\#
 \langle \nabla_\mu  L_n(\mu, v) -  \nabla_\mu L_n(\mu^*, v),\, \Delta \rangle
 &=\llangle \frac{1}{\sqrt n}\sum_{i=1}^n \frac{{\varepsilon_i}}{z\sqrt{\tau^2+{\varepsilon_i^2}}}  -\frac{1}{\sqrt n}\sum_{i=1}^n \frac{y_i-\mu}{z\sqrt{\tau^2+(y_i-\mu)^2}},\,  \Delta \rrangle\nn \\
 &= \frac{1}{\sqrt n}\sum_{i=1}^n \frac{\tau^2}{z(\tau^2+(y_i-\tilde\mu)^2)^{3/2}} \, \Delta^2,  \nn
 \#
 where $\tilde\mu$ is some convex combination of $\mu^*$ and $\mu$, that is,  $\tilde\mu =(1- \lambda)\mu^* + \lambda\mu$ for some $\lambda \in [0,1]$. Obviously, we have $|\tilde \mu -\mu^*|= \lambda |\Delta|\leq |\Delta| \leq r$.  Since 
 $
 (y_i -\tilde \mu )^2\leq 2{\varepsilon_i^2} + 2\lambda^2\Delta^2\leq 2{\varepsilon_i^2} + 2\Delta^2 \leq 2{\varepsilon_i^2} + 2r^2 
$
the above displayed equality implies that, with probability at least $1-\delta$, 
\#
 &\inf_{\mu\in \BB_{r}(\mus)} \frac{\langle \nabla_\mu  L_n(\mu, v ) -  \nabla_\mu L_n(\mu^*, v)  , \mu -  \mu^* \rangle}{|\mu-\mu^*|^2} \nn \\
 &\geq \frac{\sqrt{n}}{z} \cdot \frac{1}{n}\sum_{i=1}^n \frac{\tau^2}{(\tau^2+2r^2+2{\varepsilon_i^2})^{3/2}}  \nn \\
 &= \frac{\sqrt{n}}{z} \cdot \frac{\tau^2}{ (\tau^2+2r^2)^{3/2}} \cdot \frac{1}{n}\sum_{i=1}^n \frac{(\tau^2+2r^2)^{3/2}}{(\tau^2+2r^2+2{\varepsilon_i^2})^{3/2}} \nn  \\
 &\geq  \frac{\sqrt{n}}{z} \cdot \frac{\tau^2}{ (\tau^2+2r^2)^{3/2}} 
 	\cdot \left( \EE  \frac{(\tauv^2+2r^2)^{3/2}}{(\tauv^2+2r^2+2{\varepsilon_i^2})^{3/2}}- \sqrt{\frac{\log(1/\delta)}{2n}} \right)   \nn \\ 
 &= \frac{\sqrt{n}}{z} \cdot \frac{ \tau^2}{(\tau^2+2r^2)^{3/2}} \cdot  \left(\Rom{1} - \sqrt{\frac{\log(1/\delta)}{2n}}  \right), \label{eq:mean.sc.1}
 \#
 where the last inequality uses Lemma \ref{lemma:c4}.

It remains to lower bound \Rom{1}. Using the convexity of $1/(1+x)^{3/2}$ and Jensen's inequality, we obtain
 \$
\frac{1}{n}\sum_{i=1}^n \EE \frac{(\tauv^2+2{r^2})^{3/2}}{(\tauv^2+2{r^2}+ 2{\varepsilon_i^2})^{3/2}}
&= \EE \frac{(\tauv^2+2{r^2})^{3/2}}{(\tauv^2+ 2{r^2} + 2{\varepsilon_i^2})^{3/2}}\\
&=  \EE \frac{1}{(1+ 2{\varepsilon_i^2}/(\tauv^2+ 2{r^2}))^{3/2}}\\
&\geq  \frac{1}{(1+ 2\sigma^2/({\tauv^2+ 2{r^2}}))^{3/2}}\\
&= \frac{(\tauv^2+2{r^2})^{3/2}}{(\tauv^2+ 2{r^2}+ 2\sigma^2)^{3/2}}. 
 \$
 Plugging the above lower bound   into \eqref{eq:mean.sc.1} and using the facts
 \$
 \frac{\tau^3}{(\tau^2+ 2{r^2})^{3/2}} &\geq \frac{\tauv^3}{(\tauv^2 + 2{r^2})^{3/2}}~~\text{for}~\tauv\geq\tau~~~\text{and}~~~\frac{\tau^3}{(\tau^2 + 2{r^2})^{3/2}} \leq 1, 
\$
we obtain with probability at least $1-\delta$
 \$
&\inf_{\mu\in \BB_r(\mus)}\frac{\langle \nabla_\mu  L_n(\mu) -  \nabla_\mu L_n(\mu^*)  , \mu -  \mu^* \rangle}{|\mu-\mu^*|^2} \\
&\geq \frac{\sqrt{n}}{z}\cdot \frac{\tau^2}{(\tau^2+2{r^2})^{3/2}}\cdot \left( \frac{(\tauv^2+2{r^2})^{3/2}}{(\tauv^2+ 2{r^2}+ 2\sigma^2)^{3/2}} 
       -  \sqrt{\frac{ \log(1/\delta)}{2n}}    \right)\\
&\geq \frac{\sqrt{n}}{z\tau}\cdot \frac{\tau^3}{(\tau^2+2{r^2})^{3/2}}\cdot \left( \frac{(\tauv^2+2{r^2})^{3/2}}{(\tauv^2+ 2{r^2}+ 2\sigma^2)^{3/2}} 
       -  \sqrt{\frac{ \log(1/\delta)}{2n}}    \right)\\
 &= \frac{\sqrt{n}}{z\tau} \, \left(\frac{\tau^3}{(\tau^2+ 2{r^2})^{3/2}}\cdot\frac{(\tauv^2+ 2{r^2})^{3/2}}{(\tauv^2+ 2{r^2} + 2\sigma^2)^{3/2}} -\frac{\tau^3}{(\tau^2+ 2{r^2})^{3/2}}\cdot \sqrt{\frac{\log(1/\delta)}{2n}}\right)\\
 &\geq \frac{\sqrt n}{z\tau} \, \left(\frac{1}{(1+(2{r^2}+2\sigma^2)/\tauv^2)^{3/2}} - \sqrt{\frac{\log(1/\delta)}{2n}}  \right)\\
 &= 	\frac{1}{v} \, \left(\frac{1}{(1+(2{r^2}+2\sigma^2)/\tauv^2)^{3/2}} - \sqrt{\frac{\log(1/\delta)}{2n}}  \right)\\
 &\geq \frac{1}{2v}
 \$
 provided $\tauv^2 \geq 4{r^2}+4\sigma^2$ and $n \geq C\log(1/\delta)$ for some large enough absolute  constant $C$.

 Lastly, the above result holds uniformly over $v\geq v_0$ with probability at least $1-\delta$ since the probability event does not depend on $v$. 
 
 \end{proof}

\subsection{Proof of Corollary \ref{coro:mean.fixed}}

\begin{proof}[Proof of Corollary \ref{coro:mean.fixed}]
Recall $z=\sqrt{\log(1/\delta)}$ and
\$
r \geq  2v \left(\frac{\sigma}{\sqrt{2}v}+1\right)^2\,\sqrt{\frac{\log(2/\delta)}{n}}. 
\$ 
If
$
n \geq C\max\left\{ ({r^2} +\sigma^2)/v_0^2, 1\right\} \log(1/\delta), 
$ 
which is guaranteed by the conditions of the corollary, 
then Lemma \ref{lemma:mean.sc} implies that, with probability at least $1-\delta$, Assumption \ref{ass:mean.sc} holds with $\kappa_\ell = 1/(2v)$ and radius $r$ uniformly over $v\geq v_0$. Denote this probability event by $\cE$.  
If Assumption \ref{ass:mean.sc} holds, then by Theorem \ref{thm:mean.fixed}, we have 
\$
\PP\left( 
|\widehat \mu(v)-\mu^*|
\leq 2v \left(\frac{\sigma}{\sqrt{2}v}+1\right)^2\,\sqrt{\frac{\log(2/\delta)}{n}} \,\bigg|\,\cE\right)\geq 1-\delta.
\$
Thus 
\$
&\PP\left( 
|\widehat \mu(v)-\mu^*|
> 2v \left(\frac{\sigma}{\sqrt{2}v}+1\right)^2\,\sqrt{\frac{\log(2/\delta)}{n}}\right)\\
&= \PP\left( 
|\widehat \mu(v)-\mu^*|
> 2v \left(\frac{\sigma}{\sqrt{2}v}+1\right)^2\,\sqrt{\frac{\log(2/\delta)}{n}}, \,\cE\right) \\
&\qquad + \PP\left( 
|\widehat \mu(v)-\mu^*|
> 2v \left(\frac{\sigma}{\sqrt{2}v}+1\right)^2\,\sqrt{\frac{\log(2/\delta)}{n}},\,  \cE^c\right) \\
&\leq  \PP\left( 
|\widehat \mu(v)-\mu^*|
> 2v \left(\frac{\sigma}{\sqrt{2}v}+1\right)^2\,\sqrt{\frac{\log(2/\delta)}{n}} \,\bigg|\,\cE\right) + \PP\left( \cE^c\right)\\
&\leq 2\delta. 
\$
Then with probability at least $1-2\delta$, we have 
\$
|\widehat \mu(v)-\mu^*|
\leq 2v \left(\frac{\sigma}{\sqrt{2}v}+1\right)^2\,\sqrt{\frac{\log(2/\delta)}{n}}. 
\$
Using a change of variable $2\delta \rightarrow \delta$ finishes the proof. 
\end{proof}

\subsection{Supporting lemmas}

This subsection collects two supporting lemmas that are used earlier in this section. 

\begin{lemma}\label{lemma:con.1}
Let $\varepsilon_i$ be i.i.d. random variables such that $\EE\varepsilon_i=0$ and $\EE \varepsilon_i^2=1$.  For any $0 < \delta < 1$, with probability at least $1-2\delta$, we have  
\$
\left|\frac{1}{n}\sum_{i=1}^n \frac{\tau{\varepsilon_i}}{\sqrt{\tau^2+{\varepsilon_i^2}}}-\EE\frac{\tau{\varepsilon_i}}{\sqrt{\tau^2+{\varepsilon_i^2}}} \right|
 \leq  \sigma\sqrt{\frac{2\log(1/\delta)}{n}}+ \frac{\tau\log(1/\delta)}{3n}.
\$
\end{lemma}
\begin{proof}[Proof of Lemma \ref{lemma:con.1}]
The random  variables $Z_i:=\tau \psi_\tau({\varepsilon_i})=\tau{{\varepsilon_i}}/({\tau^2+{\varepsilon_i^2}})^{1/2}$ with $\mu_z=\EE Z_i$ and $\sigma_z^2=\var (Z_i)$ are bounded i.i.d. random variables such that 
\$
|Z_i|&=\left|{\tau {\varepsilon_i}}/({\tau^2+{\varepsilon_i^2}})^{1/2}\right| \leq |{\varepsilon_i}|\wedge \tau\leq \tau,\\
|\mu_z|&=|\EE Z_i|=\left|\EE \left( {\tau{\varepsilon_i}}/({\tau^2+{\varepsilon_i^2}})^{1/2} \right)\right| \leq  \frac{\sigma^2}{2\tau}, \\
\EE Z_i^2 &=\EE\left(\frac{\tau^2{\varepsilon_i^2}}{\tau^2+{\varepsilon_i^2}}\right) \leq {\sigma^2},\\
\sigma^2_z&:=\var (Z_i)=\EE\big({\tau{\varepsilon_i}}/({\tau^2+{\varepsilon_i^2}})^{1/2}-\mu_z\big)^2 \\
&=\EE\left(\frac{\tau^2{\varepsilon_i^2}}{\tau^2+{\varepsilon_i^2}}\right)-\mu_z^2
\leq {\sigma^2}.
\$
For third and higher order absolute moments, we have 
\$
\EE |Z_i|^k&=\EE\left|\frac{\tau{\varepsilon_i}}{\sqrt{\tau^2+{\varepsilon_i^2}}}\right|^k\leq {\sigma^2}\tau^{k-2}\leq \frac{k!}{2}{\sigma^2} (\tau/3)^{k-2}, ~\text{for all integers}~ k\geq 3. 
\$

Using Lemma \ref{lemma:bernstein.ine} with $v=n{\sigma^2}$ and $c=\tau/3$, we have for any $t> 0$
\$
\PP\left(\left|\sum_{i=1}^n \frac{\tau{\varepsilon_i}}{\sqrt{\tau^2+{\varepsilon_i^2}}}-\sum_{i=1}^n\EE\frac{\tau{\varepsilon_i}}{\sqrt{\tau^2+{\varepsilon_i^2}}} \right|\geq \sqrt{2n{\sigma^2}t}+ \frac{\tau t}{3}\right)\leq 2\exp\left(-t\right).
\$
Taking $t= \log (1/\delta)$ acquires  that for any $0 < \delta < 1$
\$
\PP\left(\left|\frac{1}{n}\sum_{i=1}^n \frac{\tau{\varepsilon_i}}{\sqrt{\tau^2+{\varepsilon_i^2}}}-\frac{1}{n}\sum_{i=1}^n\EE\frac{\tau{\varepsilon_i}}{\sqrt{\tau^2+{\varepsilon_i^2}}} \right|\leq \sigma\sqrt{\frac{2\log(1/\delta)}{n}}+ \frac{\tau\log(1/\delta)}{3n}\right)\geq 1- 2\delta.
\$
This completes the proof. 

\end{proof}

\begin{lemma}\label{lemma:c4}
For any $0  < \delta< 1$, with probability at least $1-\delta$,
\$
\frac{1}{n}\sum_{i=1}^n \frac{\tau^3}{(\tau^2+{\varepsilon_i^2})^{3/2}}   -  \EE\frac{\tau^3}{({\tau^2+{\varepsilon_i^2}})^{3/2}}
&\geq  - \sqrt{\frac{\log(1/\delta)}{2n}}. 
\$ 
{Moreover, with probability at least $1-\delta$, it  holds uniformly over $\tau \geq \tauv \geq 0$ that }
\$
\frac{1}{n}\sum_{i=1}^n \frac{\tau^3}{(\tau^2+{\varepsilon_i^2})^{3/2}} 
&\geq   \EE\frac{\tauv^3}{({\tauv^2+{\varepsilon_i^2}})^{3/2}} - \sqrt{\frac{\log(1/\delta)}{2n}}. 
\$
\end{lemma}

\begin{proof}[Proof of Lemma \ref{lemma:c4}]
The random  variables $Z_i = Z_i(\tau):=\tau^3/(\tau^2+{\varepsilon_i^2})^{3/2}$ with $\mu_z=\EE Z_i$ and $\sigma_z^2=\var (Z_i)$ are bounded i.i.d. random variables such that 
\$
0\leq Z_i&={\tau^3}/({\tau^2+{\varepsilon_i^2}})^{3/2}\leq 1. 
\$
Therefore, using Lemma \ref{lemma:hoeffding} with $v=n$  acquires that  for any $t>0$
\$
\PP\left( \sum_{i=1}^n  \frac{\tau^3}{(\tau^2+{\varepsilon_i^2})^{3/2}} - \sum_{i=1}^n \EE  \left(  \frac{\tau^3}{(\tau^2+{\varepsilon_i^2})^{3/2}} \right) \leq - \sqrt{\frac{nt}{2}}  \right)\leq  \exp(-t). 
\$
Taking $t= \log (1/\delta)$ acquires  that for any $0< \delta < 1$
\$
\PP\left( \frac{1}{n}\sum_{i=1}^n  \frac{\tau^3}{(\tau^2+{\varepsilon_i^2})^{3/2}} - \frac{1}{n}\sum_{i=1}^n  \EE  \left(  \frac{\tau^3}{(\tau^2+{\varepsilon_i^2})^{3/2}} \right) > - \sqrt{\frac{ \log(1/\delta)}{2n}}  \right)>  1-\delta. 
\$

The second result follows from the fact that $Z_i(\tau)$ is an increasing function of $\tau$. Specifically, we have with probability at least $1-\delta$
\$
\frac{1}{n}\sum_{i=1}^n  \frac{\tau^3}{(\tau^2+{\varepsilon_i^2})^{3/2}} 
&\geq \frac{1}{n}\sum_{i=1}^n  \frac{\tauv^3}{(\tauv^2+{\varepsilon_i^2})^{3/2}}   \\
&\geq  \EE  \left(  \frac{\tauv^3}{(\tauv^2+{\varepsilon_i^2})^{3/2}} \right)  + \frac{1}{n}\sum_{i=1}^n  \frac{\tauv^3}{(\tauv^2+{\varepsilon_i^2})^{3/2}}     -  \EE  \left(  \frac{\tauv^3}{(\tauv^2+{\varepsilon_i^2})^{3/2}} \right)\\
&\geq \EE  \left(  \frac{\tauv^3}{(\tauv^2+{\varepsilon_i^2})^{3/2}} \right)  - \sqrt{\frac{\log(1/\delta)}{2n}}. 
\$
This finishes the proof. 

\end{proof}

\section{Proofs for the self-tuned case}

This section collects the proofs for Theorems \ref{thm:v} and \ref{thm:main_random}.

\subsection{Proof of Theorem of \ref{thm:v}}

\begin{proof}[Proof of Theorem of \ref{thm:v}]


Recall that $\tau = v\sqrt{n}/z$. For simplicity, let $\htau= \hv\sqrt{n}/z$. 
Define the profile loss $L_n^\pro(v)$  as
\$
L_n^\pro(v): = L_n(\widehat\mu(v),v) =\min_{\mu} L_n(\mu, v). 
\$
Then it is convex and its first-order gradient is
\#\label{thm1:eq.1}
\nabla L_n^\pro(v) = \nabla  L_n(\widehat\mu(v), v)
&=\frac{\partial}{\partial v} \widehat \mu(v) \cdot \frac{\partial}{\partial v} L_n (\mu, v)\Big|_{\mu=\widehat\mu(v)}+\frac{\partial}{\partial v} L_n (\mu, v)\Big|_{\mu=\widehat\mu(v)}=\frac{\partial}{\partial v} L_n (\widehat\mu(v), v),
\#
where we use the fact that $\partial/\partial\mu\, L_n(\mu,v)|_{\mu=\widehat\mu(v)}=0$, implied by  the stationarity of $\widehat\mu (v)$.   

\paragraph{Assuming that the constraint is inactive.} We first assume that the constraint is not active for any stationary point $\hv$,  that  is, any stationary point $\widehat v$ is an interior point of $[v_0, V_0]$, aka  $\widehat v \in(v_0, V_0)$.   By the joint convexity of $L_n(\mu, v)$ and the convexity of $L_n^\pro(v)$,   $(\widehat\mu (\widehat v), \widehat v)$  and  $\widehat v$ are stationary points of $L_n(\mu,v)$ and $L_n(\widehat\mu (v), v )$, respectively. Thus  we have
\$
\frac{\partial}{\partial \mu} L_n( \mu , v) \Big|_{(\mu, v)=(\widehat\mu(\hv), \hv)}
&= -\frac{\sqrt{n}}{z}\cdot \frac{1}{n}\sum_{i=1}^n \frac{y_i-\widehat\mu (\hv)}{\sqrt{\wtau^2+(y_i-\widehat\mu (\hv))^2}}=0,\\
\frac{\partial}{\partial v}  L_n(\mu, v)\Big|_{(\mu, v)=(\widehat\mu(\hv), \hv)}
&= \frac{n}{z^2}\cdot \frac{1}{n}\sum_{i=1}^n \frac{\widehat{\tau}}{\sqrt{\widehat\tau^2+(y_i-\widehat\mu(\hv))^2}}-\left(\frac{n}{z^2}- a\right)=0,\\
\nabla L_n^\pro(v) \Big|_{v = \hv} 
&=  \nabla L_n(\widehat \mu (\hv),\hv) \Big|_{v = \hv}  
=\frac{\partial}{\partial v}  L_n(\widehat \mu (v),v)\Big|_{v=\hv} = \frac{\partial}{\partial v}  L_n(\mu, v)\Big|_{(\mu, v)=(\widehat\mu(\hv), \hv)}=0, 
\$
where the first two equalities are on partial derivatives of $L_n(\mu,v)$ and the last one is on the derivative of the profile loss $L_n^\pro(v)\equiv L_n(\widehat \mu (v), v)$. 

Recall that $\tau = \sqrt{n} v/z$. 
Let $f(\tau) = {z^2} \nabla L_n^\pro(v) /n$, that is,
\$
f(\tau)  
& = \frac{1}{n} \sum_{i=1}^n \frac{\tau}{\sqrt{\tau^2  + (y_i -\widehat\mu(v))^2 }} - \left(1- \frac{az^2}{n}\right). 
\$
In other words, $\widehat \tau= \sqrt{n}\hv /z$ satisfies $f(\wtau )= 0$. Assuming that the conststraint is inactive, we split the proof  into two steps.

\paragraph{Step 1: Proving $\hv\leq C_0\sigma$ for some universal constant $C_0$.} 
We will employ the method of proof by  contradiction. Assume there exists some $v$ such that 
\$
v> (1+\epsilon)\sqrt{{r^2}+\sigma^2} ~~~\text{and}~~~\nabla L^\pro_v(v)=0; 
\$  
or equivalently, there exists some $\tau$ such that
\#\label{eq:tau_bigger}
\tau> (1+\epsilon)\sqrt{{r^2}+\sigma^2}\sqrt{n}/z =: \bar\tau ~~\text{and}~~  f(\tau) = 0, 
\#
where $\epsilon$ and $r$ are to be determined later. Let $\tau_{v_0}= v_0 \sqrt{n}/z.$
Then, provided $n$ is large enough, Lemma \ref{lemma:mean.sc} implies that  Assumption \ref{ass:mean.sc} with 
$
\kappa_\ell  = {1}/{(2v)}
$
and {local radius $r\geq r_0(\kappa_\ell)$}
holds uniformly over $v\geq v_0$ conditional on the following event 
\$
\cE_1 :=  \left\{  \frac{1}{n}\sum_{i=1}^n \frac{(\tauv^2+2{r^2})^{3/2}}{(\tauv^2+2{r^2}+ 2{\varepsilon_i^2})^{3/2}}- \frac{1}{n}\sum_{i=1}^n \EE \frac{(\tauv^2+2{r^2})^{3/2}}{(\tauv^2+2{r^2}+ 2{\varepsilon_i^2})^{3/2}} \geq  - \sqrt{\frac{\log(1/\delta)}{2n}}\right\}. 
\$
 Conditional on the intersection of  event $\cE_1$ and the following event 
\$
\cE_2
:=\left\{ \sup_{v\in[v_0,V_0]} \left|\frac{1}{n}\sum_{i=1}^n \frac{{\varepsilon_i}}{\sqrt{\tau^2+ {\varepsilon_i^2}}}\right| \leq C \cdot\frac{V_0}{v_0} \cdot  \frac{\log (n/\delta)}{n }\right\}, 
\$
where $z\lesssim \sqrt{\log(n/\delta)}$ and $C$ is some constant, and following the proof of Theorem \ref{thm:mean.fixed}, for any fixed $v$ and thus fixed $\tau = v\sqrt{n}/z$, we have 
\$
\kappa_\ell|\widehat\mu(v)-\mu^*|
&\leq \left|\frac{1}{\sqrt n}\sum_{i=1}^n \frac{{\varepsilon_i}}{z\sqrt{\tau^2+ {\varepsilon_i^2}}}\right|. 
\$
Thus, for any $v$ such that  $ v_0 \vee \bar v_0 := v_0\vee (1+\epsilon)\sqrt{{r^2}+\sigma^2} < v < V_0$, we have on $\cE_2$ that 
\$
 \sup_{ v_0\vee \bar v_0< v < V_0} \kappa_\ell(v) \, |\widehat\mu(v)-\mu^*|
 &\leq   \sup_{v\in[v_0,V_0]} \kappa_\ell(v) \, |\widehat\mu(v)-\mu^*| \\
&\leq \sup_{v\in[v_0,V_0]} \left|\frac{1}{\sqrt n}\sum_{i=1}^n \frac{{\varepsilon_i}}{z\sqrt{\tau^2+ {\varepsilon_i^2}}}\right|\\
&\leq C \cdot\frac{V_0}{v_0} \cdot \frac{\log(n/\delta)}{z\sqrt{n}}, 
\$
which, by Lemma \ref{lemma:mean.sc}, yields 
\#\label{def:r}
\sup_{v\in [v_0, V_0]}|\widehat \mu(v)-\mu^*|
\leq  2 C \cdot \frac{V_0^2}{v_0} \cdot \frac{\log(n/\delta)}{z \sqrt n} =: r.
\#
The above $r$ can be further refined by using the finer lower bound $\barv_0$ of $v$ instead of $v_0$, but we use $v_0$ for simplicity. Let $\Delta = \mus -\hmu(v)$, and we have $|\Delta|\leq r$. Let the event $\cE_3$ be
\$
\cE_3
:=\left\{\frac{1}{n}\sum_{i=1}^n \frac{\sqrt{\bartau^2 + 2({r^2}+ \varepsilon_i^2)} - \bartau }{\sqrt{\bartau^2 + 2({r^2}+ \varepsilon_i^2)}}  - \EE\left(  \frac{\sqrt{\bartau^2 + 2({r^2}+ \varepsilon_i^2)} - \bartau }{\sqrt{\bartau^2 + 2({r^2}+ \varepsilon_i^2)}} 
 \right)   
 \leq \sqrt{\frac{\log(1/\delta) 2({r^2}+ \sigma^2)}{n\bartau^2}} +\frac{\log(1/\delta)}{3n}\right\}. 
\$
Thus  on the event $\cE_1 \cap \cE_2 \cap \cE_3$ and using the fact that $1- 1/\sqrt{1+x}$ is an increasing function, we have 
\$
f(\tau)
&= \frac{az^2}{n}  - \frac{1}{n}\sum_{i=1}^n \frac{\sqrt{\tau^2 + (\Delta + \varepsilon_i)^2} -\tau }{\sqrt{\tau^2 + (\Delta + \varepsilon_i)^2}}
\geq  \frac{az^2}{n}  -  \frac{1}{n}\sum_{i=1}^n \frac{\sqrt{\tau^2 + 2({r^2} + \varepsilon_i^2)} -\tau }{\sqrt{\tau^2 + 2({r^2} + \varepsilon_i^2)}} \\
&>  \frac{az^2}{n}  -  \frac{1}{n}\sum_{i=1}^n \frac{\sqrt{\bartau^2 + 2({r^2} + \varepsilon_i^2)} -\bartau }{\sqrt{\bartau^2 + 2({r^2} + \varepsilon_i^2)}}  \tag{$\tau< \bar\tau$}\\
&\geq  \frac{az^2}{n}  - \left\{ \EE \left( \frac{\sqrt{\bartau^2 + 2({r^2} + \varepsilon_i^2)} -\bartau }{\sqrt{\bartau^2 + 2({r^2} + \varepsilon_i^2)}}\right) +  \frac{1}{n}\sum_{i=1}^n \frac{\sqrt{\bartau^2 + 2({r^2} + \varepsilon_i^2)} -\bartau }{\sqrt{\bartau^2 +  2({r^2} + \varepsilon_i^2)}} - \EE \left( \frac{\sqrt{\bartau^2 + 2({r^2} + \varepsilon_i^2)} -\bartau }{\sqrt{\bartau^2 + 2({r^2} + \varepsilon_i^2)}}\right) \right\}\\
&\geq  \frac{az^2}{n} - \left(\frac{{r^2}+\sigma^2}{\bartau^2} + \sqrt{\frac{\log(1/\delta)\cdot 2({r^2}+ \sigma^2)}{n\bartau^2}} +\frac{\log(1/\delta)}{3n} \right)\\
&= \frac{z^2}{n}\left(a - \frac{\log(1/\delta)}{3z^2}\right) -  \left( \frac{{r^2}+ \sigma^2}{{r^2}+\sigma^2} \frac{z^2}{(1+\epsilon)^2 n} + \sqrt{\frac{{r^2}+\sigma^2}{{r^2}+\sigma^2} \frac{2z^2\log(1/\delta)}{(1+\epsilon)^2n^2}} \right)  \tag{Definition of $\bar\tau$} \\
&\geq \frac{(a-1/3)z^2}{n} -  \left( \frac{{r^2}+ \sigma^2}{{r^2}+\sigma^2} \frac{z^2}{(1+\epsilon)^2 n} + \sqrt{\frac{{r^2}+\sigma^2}{{r^2}+\sigma^2} \frac{2z^4}{(1+\epsilon)^2n^2}} \right)    \tag{$z^2\geq \log(1/\delta)$}\\
&\geq \frac{(a-1/3)z^2}{n} - \frac{z^2}{n}\cdot\left( \frac{1}{(1+\epsilon)^2} + \sqrt{\frac{2}{(1+\epsilon)^2}} \right)\\
&= \frac{z^2}{n} \left( a -\frac{1}{3} -  \frac{1}{(1+\epsilon)^2} -  \sqrt{\frac{2}{(1+\epsilon)^2} }\right)\\
&\geq  0, 
\$
provided that 
\$
\frac{1}{1+ \epsilon} \leq   \frac{\sqrt{1+2(a-1/3)}-1}{\sqrt{2}}, 
\$
or equivalently
\$
\epsilon \geq \frac{\sqrt{4a + 2/3} + 2/3 + \sqrt{2} - 2a}{2(a-1/3)} =: \epsilon(a). 
\$
In other words, conditional on the event $\cE_1\cap\cE_2\cap\cE_3$ and taking $\epsilon \geq \epsilon(a)$,    $f(\tau)>0$ for $\tau > \bartau:= (1+\epsilon) \sqrt{{r^2}+\sigma^2} \sqrt{n}/z$. This contradicts with \eqref{eq:tau_bigger}, and thus  
\$
\htau \leq (1+ \epsilon)\sqrt{{r^2}+\sigma^2}\sqrt{n}/z. 
\$
If $a=1/2$ and conditional on the same event, the above holds with
\$
\epsilon 
= 9 
\geq  \epsilon\left(1/2\right).
\$
If $n$ is large enough such that 
$ 
12\sigma \geq   10\sqrt{{r^2}+\sigma^2}, 
$ 
then  conditional on the  event $\cE_1\cap \cE_2 \cap \cE_3$, we have 
\$
v_0 \leq \hv \leq C_0 \sigma,
\$
where $C_0 =12$.

\paragraph{{Step 2:  Proving  $\widehat v\geq c_0  \left( \frac{\sigma_{\tauv^2/2-1}}{\sigma_{\tauv^2/2}} \wedge 1   \right) \sigma_{\tauv^2-1}$ for some universal constant $c_0$.}}  
We will again employ the method of proof by contradiction. Let
\$
g(\tau ) := \left(\frac{1}{n}\sum_{i=1}^n \frac{\tau^2}{\sqrt{\tau^2+ (\Delta + \varepsilon_i )^2}}\right)^2 - \left(1-\frac{az^2}{n}\right)^2. 
\$
Assume there exists some $v$ such that   
\$
v< c~~~\text{and}~~~\frac{\partial}{\partial v} L_n (\widehat\mu(v), v)=0;
\$
or equivalently,  assume there exists some $\tau$ such that
\#\label{eq:tau_lower}
\tau< c\sqrt{n}/z =: \underline \tau ~~\text{and}~~ g(\tau)= 0. 
\#
It is  impossible that $c\leq  v_0$ because any stationary point $v$ is in $(v_0, V_0)$. Thus $c> v_0$.   
Let $\Delta=\hmu(v) - \mus$. Then on the event $\cE_1 \cap \cE_2$, using the facts that $\sqrt{x}$ is a concave function and $1/\sqrt{1+ y/x}$ is an increasing function of $x$, we have 
\$
\frac{1}{n}\sum_{i=1}^n \frac{\tau^2}{\sqrt{\tau^2+ (\Delta + \varepsilon_i )^2}}
&=   \frac{1}{n}\sum_{i=1}^n \frac{1}{\sqrt{1+ (\Delta + \varepsilon_i )^2/\tau^2}} \\
&\leq    \frac{1}{n}\sum_{i=1}^n \frac{1}{\sqrt{1+ (\Delta + \varepsilon_i )^2/\ultau^2}} \\
&\leq \sqrt{\frac{1}{n}\sum_{i=1}^n \frac{1}{1+ (\Delta+\varepsilon_i)^2/\ultau^2}} \\
&\leq  \sqrt{\frac{1}{n}\sum_{i=1}^n \frac{1}{1+ \ultau^{-2}(\Delta+\varepsilon_i)^2\cdot 1\left((\Delta + \varepsilon_i)^2\leq \ultau^2\right)}} \\
&\leq \sqrt{1 - \frac{1}{n}\cdot\frac{1}{2\ultau^2}\sum_{i=1}^n (\Delta + \varepsilon_i)^2 \cdot 1\left((\Delta +\varepsilon_i)^2\leq \ultau^2 \right)}. 
\$
By the proof from step 1,  we have  on the event $\cE_1\cap \cE_2$ that
\$
\sup_{v\in [v_0, V_0]}|\widehat \mu(v)-\mu^*| \leq r,
\$
where $r$ is defined in \eqref{def:r}. 
Then
\$
g(\tau)
&\leq 1 - \frac{1}{n}\cdot\frac{1}{2\ultau^2}\sum_{i=1}^n (\Delta + \varepsilon_i)^2 \cdot 1\left((\Delta +\varepsilon_i)^2\leq \ultau^2\right) -  \left(1-\frac{az^2}{n}\right)^2\\
&< \frac{2az^2}{n} - \frac{1}{n}\cdot\frac{1}{2\ultau^2}\sum_{i=1}^n (\Delta + \varepsilon_i)^2 \cdot 1\left((\Delta +\varepsilon_i)^2\leq \ultau^2\right)  \tag{as long as $az^2/n>0$}\\
&\leq \frac{2az^2}{n} - \frac{1}{n}\cdot\frac{1}{2\ultau^2}\sum_{i=1}^n \left(  \varepsilon_i^2 + 2\Delta\varepsilon_i  \right) \cdot 1\left(\varepsilon_i^2\leq \frac{\ultau^2}{2} - {r^2} \right) \\
&\leq  \frac{2az^2}{n} - \frac{1}{2\ultau^2}\left(  \frac{1}{n}\sum_{i=1}^n  \varepsilon_i^2  1\left(\varepsilon_i^2\leq \frac{\ultau^2}{2} - {r^2} \right)    - \frac{2}{n} \sum_{i=1}^n r|\varepsilon_i| 1\left( \varepsilon_i^2\leq \frac{\ultau^2}{2} - {r^2}\right)  \right)  \\
&=  \frac{2az^2}{n} - \frac{1}{2\ultau^2}\left(  \Rom{1} - 2r\cdot \Rom{2}  \right). 
\$

Define the probability event  $\cE_4$ as 
\$
\cE_4 := \cE_{41} \cap\cE_{42}, 
\$
where 
\$
 \cE_{41}&=:\left\{  \frac{1}{n}\sum_{i=1}^n  \varepsilon_i^2  1\left(\varepsilon_i^2\leq \frac{\ultau^2}{2} - {r^2} \right) \geq \EE  \varepsilon_i^2  1\left(\varepsilon_i^2\leq \frac{\ultau^2}{2} - {r^2} \right)  -  \sigma_{\frac{\ultau^2}{2}}\sqrt{\frac{\ultau^2\log(1/\delta)}{n}} -  \frac{\ultau^2\log(1/\delta)}{6n}\right\} ~~~\text{and} \\
  \cE_{42}&=:\left\{ \frac{1}{n }\sum_{i=1}^n |\varepsilon_i| 1\left( \varepsilon_i^2\leq \frac{\ultau^2}{2} - {r^2}\right)
\leq  \EE |\varepsilon_i|  1\left(\varepsilon_i^2\leq \frac{\ultau^2}{2} - {r^2} \right) +  \sqrt{\frac{2 \sigma^2_{\ultau^2/2}\log(1/\delta)}{n}} + \frac{\ultau\log(1/\delta)}{3\sqrt{2}n} \right\}.
\$
If $n$ is sufficiently large such that 
\begin{align*}
&{r^2}\leq  \epsilon_0 \lesssim \left(\frac{\log n + \log (1/\delta)}{z\sqrt{n}}\right)^2\leq 1 ~~~\text{and} \\
&\frac{r}{\ultau^2} \left(\sigma^2_{\ultau^2/2} + \sqrt{\frac{2\sigma^2_{\ultau^2/2} \log(1/\delta)}{n}}+ \frac{\ultau\log(1/\delta)}{3\sqrt{2}n} \right) 
\leq    \frac{1}{12} \frac{\log (1/\delta)}{n}, 
\end{align*}
then conditional on  $\cE_4$,  we have 
\$
\Rom{1} &\geq \EE  \varepsilon_i^2  1\left(\varepsilon_i^2\leq \frac{\ultau^2}{2} - {r^2} \right)  -  \sigma_{\frac{\ultau^2}{2}}\sqrt{\frac{\ultau^2\log(1/\delta)}{n}} -  \frac{\ultau^2\log(1/\delta)}{6n} ~~~\text{and}\\
\Rom{2} 
&\leq  \EE |\varepsilon_i|  1\left(\varepsilon_i^2\leq \frac{\ultau^2}{2} - {r^2} \right) +  \sqrt{\frac{2 \sigma^2_{\ultau^2/2}\log(1/\delta)}{n}} + \frac{\ultau\log(1/\delta)}{3\sqrt{2}n}.
\$
Thus conditional on $\cE_4$ we have 
\$
g(\tau)
&<  \frac{2az^2}{n} - \frac{1}{2\ultau^2}\left(  \Rom{1} - 2r\cdot \Rom{2}  \right)\\
&\leq  \frac{2az^2}{n} - \frac{1}{2\ultau^2} \left( \EE  \varepsilon_i^2  1\left(\varepsilon_i^2\leq \frac{\ultau^2}{2} - {r^2} \right)  -  \sigma_{{\ultau^2}/{2}}\sqrt{\frac{\ultau^2\log(1/\delta)}{n}} -  \frac{\ultau^2\log(1/\delta)}{6n}\right) \\
 & \qquad + \frac{r}{\ultau^2}  \left(   \EE |\varepsilon_i|  1\left(\varepsilon_i^2\leq \frac{\ultau^2}{2} - {r^2} \right) +  \sqrt{\frac{2 \sigma^2_{\ultau^2/2}\log(1/\delta)}{n}} + \frac{\ultau\log(1/\delta)}{3\sqrt{2}n} \right)\\
 &\leq \frac{2az^2}{n} - \frac{\sigma^2_{\ultau^2/2- \epsilon_0}}{2\ultau^2}  + \frac{\sigma_{\ultau^2/2}{\sqrt{\log(1/\delta)}}}{2 \ultau\sqrt{n}} + \frac{\log(1/\delta)}{12n} 
 + \frac{r}{\ultau^2} \left(\sigma^2_{\ultau^2/2} + \sqrt{\frac{2\sigma^2_{\ultau^2/2} \log(1/\delta)}{n}}+ \frac{\ultau\log(1/\delta)}{3\sqrt{2}n} \right) \\
 &\leq \frac{z^2}{n} \left(2a + \frac{\log(1/\delta)}{z^2}\cdot \frac{1}{6}\right) - \frac{\sigma^2_{\ultau^2/2-\epsilon_0}}{2\ultau^2}  + \frac{\sigma_{\ultau^2/2}{\sqrt{\log(1/\delta)}}}{2 \ultau\sqrt{n}}\\
  &= \frac{z^2}{2n} \left(4a + \frac{\log(1/\delta)}{z^2}\cdot \frac{1}{3} - \frac{\sigma^2_{\ultau^2/2-\epsilon_0}}{c^2}  + \frac{\sigma_{\ultau^2/2}}{c}\cdot\frac{\sqrt{\log(1/\delta)}}{z} \right) \tag{$\underline\tau = c\sqrt{n}/z$}\\
 &\leq \frac{z^2}{2n} \left( 4a + \frac{1}{3} - \frac{\sigma^2_{\ultau^2/2 -\epsilon_0}}{c^2} + \frac{\sigma_{\ultau^2/2}}{c}    \right)  \tag{ $z^2\geq \log(1/\delta)$}\\
 &\leq 0, 
\$
for any  $c$ such that    
\$
c 
\leq  \frac{\sigma_{\ultau^2/2}}{2(4a+{1}/{3})}\left( \sqrt{1+ \frac{4(4a+{1}/{3}) \sigma^2_{\ultau^2/2  - \epsilon_0}}{\sigma^2_{\ultau^2/2}}}  - 1 \right),
\$
In other words, conditional  on the event $\cE_1\cap \cE_2\cap\cE_4$ and taking  any $c$  satisfying the above inequality, we have  
\$
g(\tau)<0 ~\text{for any}~ \tau< \ultau = c\sqrt{n}/z. 
\$
This is a contradiction.  Thus,  $\htau \geq  \ultau = c \sqrt{n}/z$, or equivalently $\hv \geq c> v_0$. 
{
Using the inequality
\$
\sqrt{1+x} -1 
&\geq  1(x\geq 3) + \frac{x}{3} 1(0\leq x<3) \geq  \frac{x}{3} \wedge 1~~~\forall~x\geq 0, 
\$ 
we obtain
\$
&\frac{\sigma_{\ultau^2/2}}{2(4a+1/3)}\left( \sqrt{1+ \frac{4(4a+{1}/{3}) \sigma^2_{\ultau^2/2 - \epsilon_0}}{\sigma^2_{\ultau^2/2}}}  - 1 \right)\\
&= \frac{3\sigma_{\tau_{v_0}^2/2}}{14}\left( \sqrt{1+ \frac{28 \sigma^2_{\ultau^2/2 - \epsilon_0}}{3\sigma^2_{\ultau^2/2}}}  - 1 \right) \tag{$a=1/2$} \\
&\geq  \frac{3\sigma_{\ultau^2/2}}{14} \left( \frac{28 \sigma^2_{\ultau^2/2-\epsilon_0}}{9\sigma^2_{\ultau^2/2}} \wedge 1  \right)\\
&= \frac{2 \sigma^2_{\ultau^2/2-\epsilon_0}}{3\sigma_{\ultau^2/2}} \wedge \frac{3 \sigma_{\ultau^2/2}}{14} \\
&\geq \frac{1}{5}\left(\frac{\sigma_{\ultau^2/2-1}}{\sigma_{\ultau^2/2}} \wedge 1 \right) \sigma_{\ultau^2/2-1} \\
&\geq \frac{1}{5}\left(\frac{\sigma_{\tauv^2/2-1}}{\sigma_{\tauv^2/2}} \wedge 1 \right) \sigma_{\tauv^2/2-1}. 
\$
Therefore we can take $c=  5^{-1}(\sigma_{\tauv^2/2-1}/\sigma_{\tauv^2/2}\wedge 1)\sigma_{\tauv^2/2-1}$. 
}
Thus on the event $\cE_1\cap\cE_2\cap\cE_4 $, we have 
\$
\hv  \geq c := c_0  \left( \frac{\sigma_{\tauv^2/2-1}}{\sigma_{\tauv^2/2}} \wedge 1   \right) \sigma_{\tauv^2/2-1},
\$
where $c_0=1/5$ is a universal constant. 
This finishes the proof of step 2.

\paragraph{Proving that the constraint is inactive.}

If $\hv\not\in (v_0, V_0)$, then $\hv\in \{v_0, V_0\}$. Suppose $\hv= v_0$, then $\hv = v_0 < c$. Recall that  $\tau_{v_0}= v_0 \sqrt{n}/z.$ Then we must have 
$
f(\tau_{v_0}) \geq 0, 
$
and thus $g(\tau_{v_0}) \geq 0.$ However, conditional on the probability event $\cE_1\cap \cE_2 \cap \cE_4$, repeating the above analysis in step 2 obtains $g(\tau_{v_0})< 0$. This is a contradiction. Therefore $\hat v \ne v_0$. Similarly, conditional on probability event $\cE_1\cap \cE_2 \cap \cE_3$, we can obtain  $\hv \ne V_0$. Therefore, conditional on the probability event $\cE_1\cap \cE_2 \cap \cE_3 \cap \cE_4$,  the constraint must be inactive, aka $\hv\in (v_0, V_0)$.

Using the first result of Lemma \ref{lemma:c4} with $\tau^2$ and ${\varepsilon_i^2}$ replaced by $\tauv^2+ 2{r^2}$ and $2{\varepsilon_i^2}$ respectively, Lemma \ref{lemma:uniform}, Lemma \ref{lemma:c6} with $\tau^2$ and $w_i^2$ replaced by $\bartau^2$ and $2({r^2}+{\varepsilon_i^2})$ respectively, and Lemma \ref{lemma:lb_tech}, we obtain
\$
\PP(\cE_1)\geq 1-\delta, ~\PP(\cE_2)\geq 1-\delta, ~\PP(\cE_3)\geq 1-\delta, ~\PP(\cE_4)\geq 1-2\delta,
\$
and thus
\$
\PP(\cE_1\cap \cE_2\cap \cE_3 \cap \cE_4 ) \geq 1-5\delta. 
\$
Putting the above results together,  and using Lemmas \ref{lemma:uniform} and \ref{lemma:lb_tech}, we obtain with probability at least $1-5\delta$ that
\$
{c_0(\sigma_{\tauv^2/2-1}/\sigma_{\tauv^2/2}\wedge 1)\sigma_{\tauv^2/2-1}}\leq \hv \leq C_0 \sigma. 
\$
Using a change of variable $5\delta \rightarrow \delta$ completes the proof. 
\end{proof}

\subsection{Proof of Theorem \ref{thm:main_random}}

\begin{proof}[Proof of Theorem \ref{thm:main_random}]

On the probability event $\cE_1\cap \cE_2 \cap \cE_3 \cap \cE_4$ where $\cE_k$'s are defined the same as in the proof of Theorem \ref{thm:v}, we have  
\$
c_0 (\sigma_{\tauv^2/2-1}/\sigma_{\tauv^2/2}\wedge 1) {\sigma_{\tauv^2/2-1}}\leq \hv \leq  C_0\sigma. 
\$
Following the proof of Theorem \ref{thm:mean.fixed}, for any fixed $v$ and thus $\tau$, we have 
\$
\kappa_\ell|\widehat\mu(v)-\mu^*|
&\leq \left|\frac{1}{\sqrt n}\sum_{i=1}^n \frac{{\varepsilon_i}}{z\sqrt{\tau^2+ {\varepsilon_i^2}}}\right|. 
\$
For any  $v$ such that  $ c_0' \sigma_{\tauv^2/2-1} \leq  v \leq  C_0 \sigma$ where $c_0' = c_0(\sigma_{\tauv^2/2-1}/\sigma_{\tauv^2/2}\wedge 1)  $ and any $z >0$, using Lemma \ref{lemma:uniform} but with $v_0$ and $V_0$ replaced by $c_0'\sigma_{\tauv^2/2-1}$ and $C_0\sigma$ respectively,  we obtain with probability at least $1-\delta$
\$
 \sup_{ v\in[ c_0' \sigma_{\tauv^2/2-1},\  C_0 \sigma]} \kappa_\ell(v) \, |\widehat\mu(v)-\mu^*|
 &\leq   \sup_{v\in[ c_0' \sigma_{\tauv^2/2-1}, \   C_0 \sigma]} \kappa_\ell(v) \, |\widehat\mu(v)-\mu^*| \\
&\leq \sup_{v\in[ c_0'\sigma_{\tauv^2/2-1}, \  C_0 \sigma]} \left|\frac{1}{\sqrt n}\sum_{i=1}^n \frac{{\varepsilon_i}}{z\sqrt{\tau^2+ {\varepsilon_i^2}}}\right|\\
&\leq \frac{\sigma}{c_0'\sigma_{\tauv^2/2-1}} \sqrt{\frac{2\log(n/\delta)}{n}} + \frac{1}{z} \frac{\log(n/\delta)}{\sqrt{n}} \\
&\qquad \qquad 
      + \frac{\sigma^2}{2c_0'^2\sigma^2_{\tauv^2/2-1}}\frac{z}{\sqrt{n}} 
       + \frac{3(C_0\sigma - c_0'\sigma_{\tauv^2/2-1})}{\sigma_{\tauv^2/2-1}} \frac{1}{z\sqrt{n}}, 
\$
which yields
\$
\sup_{ v\in[c_0'\sigma_{\tauv^2/2-1},\ C_0\sigma]} |\widehat \mu(v)-\mu^*|
\leq  C \sigma \, \frac{\log(n/\delta)\vee z^2\vee 1}{z \sqrt n}, 
\$
where $C$ is some constant only depending on $\sigma/\sigma_{\tauv^2/2-1}$, $c_0'$, and $C_0$. Putting the above pieces together and if $\log(1/\delta)\leq z^2\leq \log(n/\delta)$, we obtain with probability at least $1-6\delta$ that 
\$
|\widehat \mu(\hv)-\mu^*| \leq \sup_{ v\in[c_0'\sigma_{\tauv^2/2-1}, \ C_0\sigma]}|\widehat \mu(v)-\mu^*|
\leq  C \cdot \sigma \, \frac{\log(n/\delta)\vee 1}{z \sqrt n}. 
\$ 
Using a change of variable  $6\delta \rightarrow \delta$ and then setting $z=\log(n/\delta)$ gives
\$
|\widehat \mu(\hv)-\mu^*| \leq \sup_{ v\in[c_0'\sigma_{\tauv^2/2-1}, \ C_0\sigma]}|\widehat \mu(v)-\mu^*|
\leq  C \cdot \sigma \, \sqrt{\frac{\log(n/\delta)}{n}}
\$ 
with a lightly different constant $C$, 
provided that $\log(n/\delta)\geq 1$, aka $n\geq e\delta$. This completes the proof. 
\end{proof}


\subsection{Supporting lemmas}

We collect supporting lemmas, aka Lemmas \ref{lemma:uniform}, \ref{lemma:c6},  and \ref{lemma:lb_tech}, in this subsection. 

\begin{lemma}\label{lemma:uniform}
Let  $0< \delta < 1$. Suppose $\sigma \lesssim V_0$ and  $z\lesssim \sqrt{\log(n/\delta)}$. Then, with probability at least $1-\delta$, we have 
\$
\sup_{v\in[v_0,V_0]} \left|\frac{1}{n}\sum_{i=1}^n \frac{{\varepsilon_i}}{\sqrt{\tau^2+ {\varepsilon_i^2}}}\right| \leq C \cdot \frac{V_0}{v_0} \cdot \frac{\log (n/\delta)}{n }
\$
where $C$ is some constant. 
\end{lemma}

\begin{proof}[Proof of Lemma \ref{lemma:uniform}]

To prove the uniform bound over $[v_0, V_0]$, we adopt a covering argument. 
For any $0< \epsilon\leq 1$, there exists an $\epsilon$-cover  $\cN$  of $ [v_0, V_0]$ such that 
$
|\cN|\leq {3(V_0 - v_0)}/{\epsilon}. 
$
Let $\tau_w = w\sqrt{n}/z$. Then for every $v \in [v_0, V_0]$, there exists a $w\in \cN \subset [v_0, V_0]$ such that $|w-\tau|\leq \epsilon$ and
\$
 \left|\frac{1}{\sqrt n}\sum_{i=1}^n \frac{{\varepsilon_i}}{z\sqrt{\tau^2+ {\varepsilon_i^2}}}\right|
&\leq \left|\frac{1}{\sqrt n}\sum_{i=1}^n \frac{{\varepsilon_i}}{z\sqrt{\tau_w^2+ {\varepsilon_i^2}}}\right| \\
&\qquad + \left|\frac{1}{\sqrt n}\sum_{i=1}^n \frac{{\varepsilon_i}}{z\sqrt{\tau_w^2+ {\varepsilon_i^2}}}- \frac{1}{\sqrt n}\sum_{i=1}^n \frac{{\varepsilon_i}}{z\sqrt{\tau^2+ {\varepsilon_i^2}}}   \right| \\
&\leq \left|\frac{1}{\sqrt n}\sum_{i=1}^n \frac{{\varepsilon_i}}{z\sqrt{\tau_w^2+ {\varepsilon_i^2}}}-\EE\left[\frac{1}{\sqrt n}\sum_{i=1}^n \frac{{\varepsilon_i}}{z\sqrt{\tau_w^2+ {\varepsilon_i^2}}}\right]  \right| \\
&\qquad + \left|\EE\left[\frac{1}{\sqrt n}\sum_{i=1}^n \frac{{\varepsilon_i}}{z\sqrt{\tau_w^2+ {\varepsilon_i^2}}}\right]\right|\\
& \qquad + \left|\frac{1}{\sqrt n}\sum_{i=1}^n \frac{{\varepsilon_i}}{z\sqrt{\tau_w^2+ {\varepsilon_i^2}}}-\frac{1}{\sqrt n}\sum_{i=1}^n \frac{{\varepsilon_i}}{z\sqrt{\tau^2+ {\varepsilon_i^2}}} \right|\\
&=   \Rom{1} + \Rom{2} + \Rom{3}. 
\$
For \Rom{2}, we have
\$
\Rom{2}\leq \frac{\sqrt n}{z} \cdot \frac{\sigma^2}{2\tau_w^2}\leq \frac{z\sigma^2}{2v_0^2 \sqrt{n}}. 
\$   For \Rom{3}, using the inequality 
\$
\left|\frac{x}{\sqrt{\tau_w^2+x^2}}-\frac{x}{\sqrt{\tau^2+x^2}}\right|\leq \frac{|\tau_w-\tau|}{2\,|\tau_w|\wedge |\tau|},
\$
we obtain
\$
\Rom{3}\leq  \frac{\sqrt{n}}{z}\cdot\frac{\epsilon}{2(w\wedge v)}\leq   \frac{\sqrt{n}}{z}\cdot\frac{\epsilon}{2v_0}. 
\$
We then bound \Rom{1}. For any fixed $\tau_w$, applying  Lemma \ref{lemma:con.1} with the fact that $\left|\EE \left( {\tau_w{\varepsilon_i}}/({\tau_w^2+{\varepsilon_i^2}})^{1/2} \right)\right|  \leq  {\sigma^2}/{(2\tau_w)}$, we obtain with probability at least $1-2\delta$
\$
\left|\frac{1}{\sqrt n}\sum_{i=1}^n \frac{{\varepsilon_i}}{z\sqrt{\tau_w^2+ {\varepsilon_i^2}}}-\EE\left[\frac{1}{\sqrt n}\sum_{i=1}^n \frac{{\varepsilon_i}}{z\sqrt{\tau_w^2+ {\varepsilon_i^2}}}\right]  \right|
&\leq \frac{\sqrt{n}}{z\tau_w}\left(\sigma\sqrt{\frac{2\log(1/\delta)}{n}} + \frac{\tau_w\log(1/\delta)}{n} \right) \\
&\leq \frac{\sigma}{z \tauv}\sqrt{2\log(1/\delta)} + \frac{1}{z} \frac{\log(1/\delta)}{\sqrt{n}}
\$
where $\tauv= v_0\sqrt{n}/z$. Therefore, putting above pieces together and  using the union bound, we obtain with probability at least $1 - 6\epsilon^{-1}(V_0 -v_0)\delta$
\$
\sup_{v\in[v_0,V_0]} \left|\frac{1}{\sqrt n}\sum_{i=1}^n \frac{{\varepsilon_i}}{z\sqrt{\tau^2+ {\varepsilon_i^2}}}\right|
&\leq \sup_{w\in \cN}\left|\frac{1}{\sqrt n}\sum_{i=1}^n \frac{{\varepsilon_i}}{z\sqrt{\tau_w^2+ {\varepsilon_i^2}}}-\EE\left[\frac{1}{\sqrt n}\sum_{i=1}^n \frac{{\varepsilon_i}}{z\sqrt{\tau_w^2+ {\varepsilon_i^2}}}\right]  \right| \\
&\qquad  + \frac{z\sigma^2}{2v_0^2\sqrt{n}} + \frac{\sqrt{n}}{z}\cdot\frac{\epsilon}{2v_0}\\
&\leq \frac{\sigma}{v_0}  \sqrt{\frac{2\log(1/\delta)}{n}} + \frac{1}{z} \frac{\log(1/\delta)}{\sqrt{n}} 
      + \frac{\sigma^2}{2v_0^2}\frac{z}{\sqrt{n}} 
       + \frac{\sqrt{n}}{z}\cdot\frac{\epsilon}{2v_0}. 
\$
Taking $\epsilon= 6(V_0-v_0)/ n$, we obtain with probability at least $1-n\delta $
\$
\sup_{v\in[v_0,V_0]} \left|\frac{1}{\sqrt n}\sum_{i=1}^n \frac{{\varepsilon_i}}{z\sqrt{\tau^2+ {\varepsilon_i^2}}}\right|
\leq \frac{\sigma}{v_0}  \sqrt{\frac{2\log(1/\delta)}{n}} + \frac{1}{z} \frac{\log(1/\delta)}{\sqrt{n}} 
      + \frac{\sigma^2}{2v_0^2}\frac{z}{\sqrt{n}} 
       + \frac{3(V_0 -v_0)}{v_0} \frac{1}{z\sqrt{n}}. 
\$
Thus with probability at least $1-\delta$, we have
\$
\sup_{v\in[v_0,V_0]} \left|\frac{1}{\sqrt n}\sum_{i=1}^n \frac{{\varepsilon_i}}{z\sqrt{\tau^2+ {\varepsilon_i^2}}}\right|
&\leq \frac{\sigma}{v_0}  \sqrt{\frac{2\log(n/\delta)}{n}} + \frac{1}{z} \frac{\log(n/\delta)}{\sqrt{n}} 
      + \frac{\sigma^2}{2v_0^2}\frac{z}{\sqrt{n}} 
       + \frac{3(V_0 -v_0)}{v_0} \frac{1}{z\sqrt{n}}\\
&\leq C \cdot \frac{V_0}{v_0} \cdot \frac{\log(n/\delta)}{ z\sqrt{n}}
\$
provided $ z\lesssim \sqrt{\log (n/\delta)}$, where $C$ is a constant only depending on $\sigma^2/(v_0V_0)$. When $v_0$ and $V_0$ are taken symmetrically around $1$, $v_0V_0$ is close to $1$.  Multiplying both sides by $z/\sqrt{n}$ finishes the proof. 
\end{proof}

\begin{lemma}\label{lemma:c6}
Let $w_i$ be i.i.d. copies of $w$.  For any $0< \delta < 1$, with probability at least $1-\delta$    
\$
\frac{1}{n}\sum_{i=1}^n \frac{\sqrt{\tau^2+ w_i^2} - \tau }{\sqrt{\tau^2+w_i^2}}   -  \EE\left(\frac{\sqrt{\tau^2 + w_i^2} - \tau}{\sqrt{\tau^2 +w_i^2}} \right)
&\leq   \sqrt{\frac{\log(1/\delta) \, \EE w_i^2 }{n\tau^2}} + \frac{\log(1/\delta)}{3n}. 
\$
\end{lemma}
\begin{proof}[Proof of Lemma \ref{lemma:c6}]
The random  variables 
\$
Z_i = Z_i(\tau):=  \frac{\sqrt{\tau^2+ w_i^2} - \tau }{\sqrt{\tau^2+w_i^2}}  
=\frac{\sqrt{1+ w_i^2/\tau^2} - 1}{\sqrt{1+w_i^2/\tau^2}} 
\$ with $\mu_z=\EE Z_i$ and $\sigma_z^2=\var (Z_i)$ are bounded i.i.d. random variables such that 
\$
0\leq Z_i\leq 1\wedge \frac{w_i^2}{2\tau^2}. 
\$
Moreover we have
\$
\EE Z_i^2 &\leq \frac{\EE w_i^2}{2\tau^2},~\sigma^2_z:=\var (Z_i)\leq \frac{\EE w_i^2}{2\tau^2}. 
\$
For third and higher order absolute moments, we have 
\$
\EE |Z_i|^k&\leq   \frac{ \EE w_i^2}{2\tau^2} \leq \frac{k!}{2}\cdot  \frac{\EE w_i^2}{2\tau^2} \cdot \left(\frac{1}{3}\right)^{k-2}, ~\text{for all integers}~ k\geq 3. 
\$
Therefore, using Lemma \ref{lemma:bernstein.ine} with $v=n \, \EE w_i^2 /(2\tau^2)$ and $c=1/3$ acquires that  for any $t >  0$
\$
\PP\left(\sum_{i=1}^n \frac{(1+ w_i^2/\tau^2)^{1/2} - 1}{(1+ w_i^2/\tau^2)^{1/2}} - \sum_{i=1}^n  \EE  \left( \frac{(1+ w_i^2/\tau^2)^{1/2} - 1}{(1+ w_i^2/\tau^2)^{1/2}}  \right)
\geq -\sqrt{\frac{tn \, \EE w_i^2 }{\tau^2}} - \frac{t}{3} \right)\leq  \exp(-t). 
\$
Taking $t= \log (1/\delta)$ acquires  that for any $0< \delta < 1$
\$
\PP\left( \frac{1}{n}\sum_{i=1}^n \frac{(1+ w_i^2/\tau^2)^{1/2} - 1}{(1+ w_i^2/\tau^2)^{1/2}} -  \EE  \left( \frac{(1+ w_i^2/\tau^2)^{1/2} - 1}{(1+ w_i^2/\tau^2)^{1/2}} \right) 
> - \sqrt{\frac{\log(1/\delta) \, \EE w_i^2 }{n\tau^2}} - \frac{\log(1/\delta)}{3n} \right)
>  1-\delta. 
\$
This finishes the proof. 

\end{proof}

\begin{lemma}\label{lemma:lb_tech}
For any $0 <\delta < 1$,   we have with probability at least $1-\delta$ that  
\$
\frac{1}{n}\sum_{i=1}^n \varepsilon_i^2  1\left(\varepsilon_i^2\leq \frac{\ultau^2}{2} - {r^2} \right) 
&\geq \frac{1}{n}\sum_{i=1}^n\EE \varepsilon_i^2  1\left(\varepsilon_i^2\leq \frac{\ultau^2}{2} - {r^2} \right)
-  \sigma_{\ultau^2/2}\sqrt{\frac{\ultau^2\log(1/\delta)}{n}} - \frac{\ultau^2\log(1/\delta)}{6n}. 
\$
For any $0 <\delta < 1$, we have with probability at least $1-\delta$ that 
\$
\frac{1}{n}\sum_{i=1}^n |\varepsilon_i|  1\left(\varepsilon_i^2\leq \frac{\ultau^2}{2} - {r^2} \right)
\leq  \frac{1}{n}\sum_{i=1}^n\EE |\varepsilon_i|  1\left(\varepsilon_i^2\leq \frac{\ultau^2}{2} - {r^2} \right)  +\sqrt{\frac{2 \sigma^2_{\ultau^2/2}\log(1/\delta)}{n}}+ \frac{\ultau\log(1/\delta)}{3\sqrt{2}n}. 
\$
Consequently, we have, with probability at least $1-2\delta$, the above two inequalities hold simultaneously. 
\end{lemma}
\begin{proof}[Proof of Lemma \ref{lemma:lb_tech}]
We prove the first two results and the last result directly follows from  first two. 
\paragraph{First result.}
Let $Z_i=  \varepsilon_i^2  1\left(\varepsilon_i^2\leq {\ultau^2}/{2} - {r^2} \right).$ The random  variables $Z_i$ with $\mu_z=\EE Z_i$ and $\sigma_z^2=\var (Z_i)$ are bounded i.i.d. random variables such that 
\$
|Z_i|
&=\left| \varepsilon_i^2  1\left(\varepsilon_i^2\leq {\ultau^2}/{2} - {r^2} \right) \right| 
\leq \ultau^2/2,\\
|\mu_z|&=|\EE Z_i|=\left|\EE \left( \varepsilon_i^2  1\left(\varepsilon_i^2\leq {\ultau^2}/{2} - {r^2} \right)\right)\right| \leq  \sigma^2_{\ultau^2/2}, \\
\EE Z_i^2 &=\EE\left( \varepsilon_i^4 1\left( \varepsilon_i^2\leq {\ultau^2}/{2} - {r^2} \right)\right) 
 \leq  {\ultau^2\sigma^2_{\ultau^2/2}}/{2},\\
\sigma^2_z&:=\var (Z_i)=\EE\big(    Z_i-\mu_z\big)^2 \leq {\ultau^2\sigma^2_{\ultau^2/2}}/{2}. 
\$
For third and higher order absolute moments, we have 
\$
\EE |Z_i|^k
&=\EE\left| \varepsilon_i^2  1\left(\varepsilon_i^2\leq {\ultau^2}/{2} - {r^2} \right) \right|^k
\leq \frac{ \ultau^2 \sigma^2_{\ultau^2/2} }{2}\left(\frac{\ultau^2}{2}\right)^{k-2}\leq \frac{k!}{2}\frac{\ultau^2\sigma^2_{\ultau^2/2}}{2} \left( \frac{\ultau^2}{6}\right)^{k-2}, ~\text{for all integers}~ k\geq 3. 
\$

Using Lemma \ref{lemma:bernstein.ine} with $v=n\ultau^2{\sigma^2_{\ultau^2/2}}/2$ and $c=\ultau^2/6$, we have for any $t> 0$
\$
\PP\left(\sum_{i=1}^n \varepsilon_i^2  1\left(\varepsilon_i^2\leq \frac{\ultau^2}{2} - {r^2} \right) -\sum_{i=1}^n\EE \varepsilon_i^2  1\left(\varepsilon_i^2\leq \frac{\ultau^2}{2}- {r^2} \right) \leq  
-\sqrt{{n\ultau^2\sigma^2_{\ultau^2/2}}{t}} -  \frac{\ultau^2 t}{6}\right)\leq \exp\left(-t\right).
\$
Taking $t= \log (1/\delta)$ acquires  the desired result. 

\paragraph{Second result.} 
With an abuse of notation, let $Z_i=  |\varepsilon_i| 1\left( \varepsilon_i^2\leq {\ultau^2}/{2} - {r^2}\right).$ The random  variables $Z_i$ with $\mu_z=\EE Z_i$ and $\sigma_z^2=\var (Z_i)$ are bounded i.i.d. random variables such that 
\$
|Z_i|
&=\left| \varepsilon_i 1\left(\varepsilon_i^2\leq {\ultau^2}/{2} - {r^2} \right) \right| 
\leq {\ultau}/{\sqrt{2}},\\
|\mu_z|
&=|\EE Z_i|=\left|\EE \left( |\varepsilon_i|  1\left(\varepsilon_i^2\leq {\ultau^2}/{2} - {r^2} \right)\right)\right| 
\leq {\sqrt{2} \sigma^2_{\ultau^2/2}}/{\ultau}, \\
\EE Z_i^2 
&=\EE\left( \varepsilon_i^2 1\left( \varepsilon_i^2\leq {\ultau^2}/{2} - {r^2} \right)\right) 
 \leq {\sigma^2_{\ultau^2/2}},\\
\sigma^2_z&:=\var (Z_i)=\EE\big(    Z_i-\mu_z\big)^2  \leq {\sigma^2_{\ultau^2/2}}. 
\$
For third and higher order absolute moments, we have 
\$
\EE |Z_i|^k
&=\EE\left| |\varepsilon_i|  1\left(\varepsilon_i^2\leq {\ultau^2}/{2} - {r^2} \right) \right|^k
\leq  \sigma^2_{\ultau^2/2} \left(\frac{\ultau}{\sqrt{2}}\right)^{k-2}
\leq \frac{k!}{2}{\sigma^2_{\ultau^2/2}}\left( \frac{\ultau}{3\sqrt{2}}\right)^{k-2}, ~\text{for all integers}~ k\geq 3. 
\$

Using Lemma \ref{lemma:bernstein.ine} with $v=n {\sigma^2_{\ultau^2/2}}$ and $c=\ultau/(3\sqrt{2})$, we have for any $t> 0$
\$
\PP\left( \sum_{i=1}^n |\varepsilon_i|  1\left(\varepsilon_i^2\leq \frac{\ultau^2}{2} - {r^2} \right) -\sum_{i=1}^n\EE |\varepsilon_i|  1\left(\varepsilon_i^2\leq \frac{\ultau^2}{2}- {r^2} \right) \geq \sqrt{{2n\sigma^2_{\ultau^2/2}}{t}}+ \frac{\ultau t}{3\sqrt{2}}\right)\leq \exp\left(-t\right).
\$
Taking $t= \log (1/\delta)$ acquires  the desired result. 
\end{proof}

\section{Proofs for Section \ref{sec:3.5}}
This section collects proofs for results in Section \ref{sec:3.5}.  

\subsection{Proof of Theorem \ref{thm:asym_mom}}

\begin{proof}[Proof of Theorem \ref{thm:asym_mom}]
First,  the MoM estimator $\hat\mu^{\mom}=M(z_1, \ldots, z_k)$ is equivalent to
\$
 \argmin \sum_{j=1}^k \left|z_j -\mu\right|. 
\$
For any $x\in \RR$, let $\ell(x)= |x|$ and define 
$
L(x)= \EE \ell'(x+Z) 
$
where $Z\sim \cN(0,1)$ and 
\$
\ell'(x)=
\begin{cases}
1, & \text{if $x>0$},\\
0, & \text{if $x=0$},  \\
-1, & \text{otherwise}. 
\end{cases}
\$
If the assumptions of Theorem 4 of \cite{minsker2019distributed} are satisfied, we obtain, after some algebra, that
\$
\sqrt{n} \left(\hat\mu^\mom - \mu^*\right) \rightsquigarrow \cN\left(0, \frac{\EE(\ell'(Z))^2}{(L'(0))^2}\right). 
\$
Some algebra derives  that
\$
\frac{\EE(\ell'(Z))^2}{(L'(0))^2}= \frac{\pi \sigma^2}{2}. 
\$
It remains to check the assumptions there. Assumptions (1), (4),  and (5) trivially hold. Assumption (2) can be verified  by  using the following Berry-Esseen bound. 
\begin{fact}
Let $y_1,\ldots, y_m$ be i.i.d. random copies of $y$ with mean $\mu$, variance $\sigma^2$ and $\EE|y-\mu|^{2+\iota}<\infty$ for some $\iota\in(0,1]$. Then there exists an absolute constant $C$ such that
\$
\sup_{t\in\RR} \left|\PP\left( \sqrt{m} \frac{\bar y -\mu}{\sigma} \leq t\right) -  \Phi(t) \right| \leq C \, \frac{\EE|y-\mu|^{2+\iota}}{\sigma^{2+\iota}m^{\iota/2}}. 
\$
\end{fact}
It remains to check Assumption (3).   
Because $g(m)\lesssim m^{-\iota/2}$, $\sqrt{k}g(m)\lesssim \sqrt{k} m ^{-\iota/2}\rightarrow 0$ if $k = o(n^{\iota/(1+\iota)})$ as $n\rightarrow \infty.$ Thus Assumption (3) holds if $k = o(n^{\iota/(1+\iota)})$ and $k\rightarrow \infty$. This completes the proof. 
\end{proof}

\subsection{Proof of Theorem \ref{thm:asym}}

In this subsection, we state and prove a stronger result of Theorem \ref{thm:asym}, aka Theorem \ref{thm:asym_strong}.  Theorem \ref{thm:asym} can then be proved following the same proof under the assumption that $\EE|\varepsilon_i|^{2+\iota} < \infty$ for any prefixed $0<\iota\leq 1$.  



\begin{theorem}\label{thm:asym_strong}
Assume the same assumptions as in Theorem \ref{thm:v}. 
Take $z^2 \geq 2\log(n)$. If $\EE \varepsilon_i^4 <\infty$, then
\$
\sqrt{n} \, 
\begin{bmatrix}
\hmu - \mus \\
\hv - \vs
\end{bmatrix}  
\rightsquigarrow \cN\left(0, \Sigma\right),
\text{ where } 
\Sigma 
=
\begin{bmatrix}
\sigma^2 & \sigma \, \EE\varepsilon_i^3/2 \\
\sigma  \, \EE\varepsilon_i^3/2  & (\sigma^2  \EE\varepsilon_i^4- \sigma^6)/4
\end{bmatrix}.   
\$
\end{theorem}


\begin{proof}[Proof of Theorem \ref{thm:asym_strong}]

Now we are ready to analyze the self-tuned mean estimator $\hmu = \hmu(\hv)$.  For any $\delta\in (0,1)$, following the proof of Theorem \ref{thm:v}, we obtain with probability at least $1-\delta$ that 
\$
|\widehat \mu(\hv)-\mu^*| 
\leq \sup_{v\in [v_0, V_0]}|\widehat \mu(v)-\mu^*|
\leq  2C\cdot \frac{V_0^2}{v_0} \cdot \frac{\log(n/\delta)}{z \sqrt n}.
\$
Taking $z^2 \geq \log(n/\delta)$ with $\delta = 1/n$ in the above inequality, we obtain  $\hmu \rightarrow \mus$ in probability. Theorem \ref{thm:v_consistency} implies that $\hv \, \rightarrow\, \sigma$ in probability. Thus we have $\|\htheta -\thetas \|_2 \rightarrow 0$ in probability, where 
\$
\widehat\theta = (\hmu, \hv)^\T, ~\text{and}~\theta^* = (\mus, \sigma)^\T. 
\$

Using the Taylor's theorem for vector-valued functions, we obtain
\$
\nabla L_n (\htheta)
= 0
=\nabla L_n (\thetas) + H_n(\thetas) (\htheta - \thetas) \, + \, \frac{R_2(\theta)}{2} \big(\htheta - \thetas \big)^{\otimes 2}, 
\$
where $\otimes$ indicates the tensor product. Let $\tsigma = \sigma \sqrt{n}/z$. We say that $X_n$ and $Y_n$ are asymptotically equivalent, denoted as $X_n \simeq Y_n$,  if both $X_n$ and $Y_n$ converge in distribution to some same random variable/vector $Z$.    Rearranging,  we obtain
\$
\sqrt{n} \, \big(\htheta - \thetas \big) 
&\simeq \left[ H_n(\thetas) \right]^{-1} \left( - \sqrt{n} \, \nabla L_n(\thetas)  \right ) \\
&=  \begin{bmatrix}
\frac{\sqrt{n}}{z} \cdot \frac{1}{n} \sum_{i=1}^n \frac{\tsigma^2}{(\tsigma^2 + \varepsilon_i^2)^{3/2}} & \frac{n}{z^2}\cdot \frac{1}{n} \sum_{i=1}^n \frac{\tsigma \varepsilon_i}{(\tsigma^2 + \varepsilon_i^2)^{3/2}} \\
\frac{n}{z^2}\cdot \frac{1}{n} \sum_{i=1}^n \frac{\tsigma \varepsilon_i}{(\tsigma^2 + \varepsilon_i^2)^{3/2}} & \frac{n^{3/2}}{z^3}\cdot \frac{1}{n} \sum_{i=1}^n \frac{\varepsilon_i^3}{(\tsigma^2 + \varepsilon_i^2)^{3/2}}\end{bmatrix}^{-1} 
 \begin{bmatrix}
\sqrt{n}  \cdot \frac{1}{n} \sum_{i=1}^n \frac{\tsigma \varepsilon_i}{\sigma \sqrt{\tsigma^2 + \varepsilon_i^2}} \\
\sqrt{n}  \cdot \frac{n}{z^2} \frac{1}{n} \sum_{i=1}^n \frac{\sqrt{1+ \varepsilon_i^2/\tsigma^2} -  1}{\sqrt{1 + \varepsilon_i^2/\tsigma^2}}   - \sqrt{n} \cdot a
\end{bmatrix}\\
&\simeq 
\begin{bmatrix}
{\sigma} & 0 \\
0 & {\sigma^3} 
\end{bmatrix}   
\begin{bmatrix}
\sqrt{n}  \cdot \frac{1}{n} \sum_{i=1}^n \frac{\tsigma \varepsilon_i}{\sigma \sqrt{\tsigma^2 + \varepsilon_i^2}} \\
\sqrt{n}  \cdot \frac{n}{z^2} \frac{1}{n} \sum_{i=1}^n \frac{\sqrt{1+ \varepsilon_i^2/\tsigma^2} -  1}{\sqrt{1 + \varepsilon_i^2/\tsigma^2}}   - \sqrt{n} \cdot a
\end{bmatrix} \\
&= \begin{bmatrix}
{\sigma} & 0 \\
0 & {\sigma^3} 
\end{bmatrix} \, 
\begin{bmatrix}
\Rom{1} \\
\Rom{2}
\end{bmatrix}, 
\$
where  the second $\simeq$ uses the fact  that 
\$
H_n(\thetas) 
 \overset{\rm a.s.}{\longrightarrow}
\begin{bmatrix}
\frac{1}{\sigma} & 0 \\
0 & \frac{1}{\sigma^3} 
\end{bmatrix}. 
\$
We proceed  to derive the asymptotic property of $(\Rom{1}, \Rom{2})^\T$.  For \Rom{1}, we have
\$
\Rom{1} 
&= \sqrt{n}  \cdot \left(\frac{1}{n} \sum_{i=1}^n \frac{\tsigma \varepsilon_i}{\sigma \sqrt{\tsigma^2 + \varepsilon_i^2}} - \EE\left[ \frac{\tsigma\varepsilon_i}{\sigma \sqrt{\tsigma^2 + \varepsilon_i^2}}   \right] \right) +  \sqrt{n} \cdot \EE\left[ \frac{\tsigma\varepsilon_i}{\sigma \sqrt{\tsigma^2+ \varepsilon_i^2}}\right] \\
&\rightsquigarrow \cN \left(0, \lim_{n\rightarrow \infty} \var\left[\frac{\tsigma \varepsilon_i}{ \sigma \sqrt{\tsigma^2 + \varepsilon_i^2}}\right] \right) + \lim_{n\rightarrow\infty} \sqrt{n} \cdot \EE\left[ \frac{\tsigma\varepsilon_i}{\sigma \sqrt{\tsigma^2+ \varepsilon_i^2}}\right]. 
\$
It remains to calculate
\$
\lim_{n\rightarrow \infty} \EE\left(\frac{\sqrt{n}\tsigma{\varepsilon_i}}{\sqrt{\tsigma^2+{\varepsilon_i^2}}}\right) ~~\text{and}~~ \lim_{n\rightarrow \infty} \var \left[ \frac{\tau {\varepsilon_i}}{\sqrt{\tsigma^2+{\varepsilon_i^2}}}\right]. 
\$
For the former term, if there exists some $0< \iota \leq 1$ such that $\EE|\varepsilon_i|^{2+\iota}<\infty$, using the fact that $\EE\varepsilon_i=0$, we have 
\#
\left|\EE\left(\frac{\sqrt{n}\tsigma{\varepsilon_i}}{\sqrt{\tsigma^2+{\varepsilon_i^2}}}\right)\right|
&=   {\sqrt{n}\tsigma}\cdot\left| \EE \left\{ \frac{-{{\varepsilon_i}}/\tsigma}{\sqrt{1+{\varepsilon_i^2}/\tsigma^2}}\right\}\right|
=  {\sqrt{n}\tsigma}\cdot\left|\EE \left\{ \frac{{{\tsigma^{-1}}{{\varepsilon_i}}} \left(\sqrt{1+{\varepsilon_i^2}/\tsigma^2}-1\right)}{\sqrt{1+{\varepsilon_i^2}/\tsigma^2}}\right\}\right| \nn \\
&\leq  \frac{{\sqrt{n}\tsigma}}{2}\cdot  \EE \left| \frac{\varepsilon_i^3/\tsigma^3}{\sqrt{1+{\varepsilon_i^2}/\tsigma^2}}\right| 
\leq  \frac{\sqrt{n}\tsigma}{2}\cdot \frac{\EE|\varepsilon_i|^{2+\iota}}{\tsigma^{2+\iota}} \nn\\
&\leq \frac{\sqrt{n} \, \EE |\varepsilon_i|^{2+\iota}}{2\tsigma^{1+\iota}} \rightarrow 0,  \label{eq:exp_limit}
\#
where the first inequality uses Lemma \ref{lemma:ine} (ii) with $r=1/2$, that is,  $\sqrt{1+x}\leq 1+ x/2$ for $x\geq -1$. 
For the second term, we have
\$
\lim_{n\rightarrow\infty}\var \left[ \frac{\tsigma {\varepsilon_i}}{\sqrt{\tsigma^2+{\varepsilon_i^2}}}\right] 
&= \lim_{n\rightarrow\infty}\EE \left[ \frac{\tsigma^2 {\varepsilon_i^2}}{{\tsigma^2+{\varepsilon_i^2}}}\right]= \sigma^2,
\$
by the dominated convergence theorem. Thus
\$
\Rom{1} \rightsquigarrow \cN(0, 1). 
\$

 For \Rom{2}, recall $a=1/2$ and using the facts that
\$
&\lim_{n\rightarrow \infty} \frac{n}{z^2} \cdot \EE \left(\frac{\sqrt{1+\varepsilon_i^2/\tsigma^2} - 1}{\sqrt{1+\varepsilon_i^2/\tsigma^2}} \right) 
= \lim_{n\rightarrow \infty} \frac{ n}{2\tsigma^2 z^2} \cdot \EE \left(\frac{1}{\sqrt{1+\varepsilon_i^2/\tsigma^2}} \cdot \frac{\sqrt{1+\varepsilon_i^2/\tsigma^2} - 1}{1/(2\tsigma^2)} \right) 
= \frac{1}{2},  \\
&\lim_{n\rightarrow \infty} \sqrt{n} \cdot \left( \frac{n}{z^2} \cdot \EE \left(\frac{\sqrt{1+\varepsilon_i^2/\tsigma^2} - 1}{\sqrt{1+\varepsilon_i^2/\tsigma^2}} \right) -\frac{1}{2} \right)
= 0, 
\$
we have
\$
\Rom{2} 
&= \sqrt{n}  \cdot \frac{n}{z^2} \cdot \frac{1}{n} \sum_{i=1}^n \frac{\sqrt{1+ \varepsilon_i^2/\tsigma^2} -  1}{\sqrt{1 + \varepsilon_i^2/\tsigma^2}}   - \sqrt{n} \cdot \frac{1}{2}\\
&\simeq \sqrt{n}\cdot \frac{1}{n}\sum_{i=1}^n \left( \frac{n}{z^2}\cdot \frac{\sqrt{1+ \varepsilon_i^2/\tsigma^2} -  1}{\sqrt{1 + \varepsilon_i^2/\tsigma^2}} - \EE\left(\frac{n}{z^2}\cdot\frac{\sqrt{1+ \varepsilon_i^2/\tsigma^2} -  1}{\sqrt{1 + \varepsilon_i^2/\tsigma^2}}   \right)\right)\\
&\simeq \cN\left(0, \lim_{n\rightarrow \infty} \var\left(\frac{n}{z^2}\cdot \frac{\sqrt{1+ \varepsilon_i^2/\tsigma^2} -  1}{\sqrt{1 + \varepsilon_i^2/\tsigma^2}}    \right) \right).
\$
If $\EE \varepsilon_i^4 < \infty$, then 
\$
 \lim_{n\rightarrow \infty} \var\left(\frac{n}{z^2}\cdot \frac{\sqrt{1+ \varepsilon_i^2/\tsigma^2} -  1}{\sqrt{1 + \varepsilon_i^2/\tsigma^2}} \right)= \frac{\EE\varepsilon_i^4}{4\sigma^4} -\frac{1}{4},
\$
and thus $\Rom{2}\simeq \cN\left(0, (\EE\varepsilon_i^4/\sigma^4 -1)/4\right)$. For the cross covariance, we have 
\$
&\lim_{n\rightarrow \infty} \cov \left( \frac{\tsigma \varepsilon_i}{\sigma \sqrt{\tsigma^2 + \varepsilon_i^2}}   , \frac{n}{z^2}\cdot \frac{\sqrt{1+ \varepsilon_i^2/\tsigma^2} -  1}{\sqrt{1 + \varepsilon_i^2/\tsigma^2}} \right) \\
&= \lim_{n\rightarrow \infty} \EE\left( \frac{\tsigma \varepsilon_i}{\sigma \sqrt{\tsigma^2 + \varepsilon_i^2}} \cdot  \frac{n}{z^2}\cdot \frac{\sqrt{1+ \varepsilon_i^2/\tsigma^2} -  1}{\sqrt{1 + \varepsilon_i^2/\tsigma^2}} \right) \\
&= \frac{\EE \varepsilon_i^3}{2\sigma^3}. 
\$
Thus 
\$
\sqrt{n} \, (\htheta -\thetas) \rightsquigarrow \cN(0, \Sigma),
\$
where 
\$
\Sigma 
=
\begin{bmatrix}
\sigma & 0 \\
0 & \sigma^3 
\end{bmatrix}  
\begin{bmatrix}
1 & {\EE\varepsilon_i^3}/({2\sigma^3}) \\
 {\EE\varepsilon_i^3}/({2\sigma^3}) & (\EE\varepsilon_i^4/\sigma^4 - 1)/4
\end{bmatrix}
\begin{bmatrix}
\sigma & 0 \\
0 & \sigma^3 
\end{bmatrix}  
=
\begin{bmatrix}
\sigma^2 & \sigma \EE\varepsilon_i^3/2 \\
\sigma \EE\varepsilon_i^3/2  & (\sigma^2  \EE\varepsilon_i^4- \sigma^6)/4
\end{bmatrix}.   
\$
Therefore, for $\hmu$ only, we have
\$
\sqrt{n} \, (\hmu -\mus) \rightsquigarrow \cN(0, \sigma^2). 
\$

\end{proof}

\subsection{Consistency of $\hv$}

This subsection proves that $\hv$ is a consistent estimator of $\sigma$.
Recall that
\$
\nabla_v L_n(\mu,v)
&=  \frac{n}{z^2} \cdot \frac{1}{n}\sum_{i=1}^n \left(\frac{\tau}{\sqrt{\tau^2 + {(y_i-\mu)^2}}} - 1\right)  + a
\$
where $a=1/2$. We emphasize that the following proof only needs the second moment assumption $\sigma^2=\EE\varepsilon_i^2<\infty$. 

\begin{theorem}[Consistency of $\hv$]\label{thm:v_consistency}
Assume the same assumptions as in Theorem \ref{thm:v}. 
Take $z^2 \geq \log(n)$. 
Then
\$
\hv \longrightarrow \sigma ~~\text{in probability}. 
\$
\end{theorem}

\begin{proof}[Proof of Theorem \ref{thm:v_consistency}]

By the proof of Theorem \ref{thm:v}, we obtain with probability at least $1-\delta$ that the following two results hold simultaneously:  
\#
\sup_{v\in [v_0, V_0]}|\widehat \mu(v)-\mu^*| 
&\leq  2C\cdot \frac{V_0^2}{v_0} \cdot \frac{\log(n/\delta)}{z \sqrt n} =: r, \label{eq:v_cons_1}\\
v_0 
&< c_0\sigma_{\tauvm} \leq \hv \leq C_0\sigma < V_0,  \label{eq:v_cons_2}
\#
provided that  $z^2\geq \log(5/\delta)$ and $n$ is large enough. 
Therefore, the constraint in the optimization problem \eqref{eq:const} is not active, and thus
\$
\nabla_v L_n(\hmu, \hv)  = 0. 
\$
Using Lemma  \ref{lemma:v.sc} together with the equality above, we obtain with probability at least $1-\delta$ that
\$
\frac{c_0}{V_0^3} |\hv-\sigma|^2 
&\leq \frac{c_0}{\hv^3\vee \sigma^3} |\hv-\sigma|^2  \leq \rho_\ell |\hv-\sigma|^2   \\
&\leq { \llangle  \nabla_v L_n(\hmu, \hv) -  \nabla_v L_n(\hmu,\sigma),  \hv -\sigma \rrangle} \\
&\leq \left|   \nabla_v L_n(\hmu, \sigma)\right| \left|\hv -\sigma \right|  \\
&\leq \left|  \frac{n}{z^2} \cdot \frac{1}{n}\sum_{i=1}^n \left(\frac{\tau_\sigma}{\sqrt{\tau_\sigma^2 + {(y_i-\hmu)^2}}} - 1\right)  + a  \right| \left|\hv - \sigma\right|.
\$
Plugging \eqref{eq:v_cons_1} into the above inequality and canceling $|\hv - \sigma |$ on both sides, we obtain with probability at least $1-2\delta$ that 
\$
\frac{c_0}{V_0^3} |\hv-\sigma| 
&\leq \left|  \frac{n}{z^2} \cdot \frac{1}{n}\sum_{i=1}^n \left(\frac{\tau_\sigma}{\sqrt{\tau_\sigma^2 + {(y_i-\hmu)^2}}} - 1\right)  + a  \right| \\
&\leq \sup_{\mu \in \BB_r(\mus)}\left|  \frac{n}{z^2} \cdot \frac{1}{n}\sum_{i=1}^n \left(\frac{\tau_\sigma}{\sqrt{\tau_\sigma^2 + {(y_i - \mu)^2}}} - 1\right)  + a  \right| \\
&=  \frac{n}{z^2} \cdot \sup_{\mu \in \BB_r(\mus)}\left|\frac{1}{n}\sum_{i=1}^n \left(\frac{\tau_\sigma}{\sqrt{\tau_\sigma^2 + {(y_i - \mu)^2}}} - 1\right)  + \frac{az^2}{n}  \right|\\
&\leq     \frac{n}{z^2} \cdot \sup_{\mu \in \BB_r(\mus)}\left|\frac{1}{n}\sum_{i=1}^n \left( 1- \frac{\tau_\sigma}{\sqrt{\tau_\sigma^2 + {(y_i - \mu)^2}}}  \right)  - \EE \left(  1- \frac{\tau_\sigma}{\sqrt{\tau_\sigma^2 + {(y_i - \mu)^2}}}   \right)\right|   \\
&\qquad\qquad\qquad   + \frac{n}{z^2} \cdot \sup_{\mu \in \BB_r(\mus)}\left|  \EE \left(  1- \frac{\tau_\sigma}{\sqrt{\tau_\sigma^2 + {(y_i - \mu)^2}}}   \right) - \frac{az^2}{n}  \right|    \\
&=: \Rom{1} + \Rom{2}. 
\$

It remains to bound terms \Rom{1} and \Rom{2}. We start with term \Rom{2}. Let $r_i^2 = (y_i-\mu)^2$.  We have 
\$
\Rom{2} 
&= \frac{n}{z^2} \cdot  \sup_{\mu \in \BB_r(\mus)}\left|  \EE \left(  1- \frac{\tau_\sigma}{\sqrt{\tau_\sigma^2 + {(y_i - \mu)^2}}}   \right) - \frac{az^2}{n}  \right| \\
&= \max\left\{ \sup_{\mu \in \BB_r(\mus)} \left( \frac{n}{z^2} \cdot   \EE \frac{\sqrt{1+r_i^2/\tsigma^2} - 1}{\sqrt{1+r_i^2/\tsigma^2}}  - a\right), ~\sup_{\mu \in \BB_r(\mus)}\left(   a - \frac{n}{z^2} +   \EE \frac{1}{\sqrt{1+ r_i^2/\tsigma^2}}   \right)  \right\}\\
&=: \Rom{2}_1 \vee \Rom{2}_2. 
\$

In order to bound $\Rom{2}$, we  bound $\Rom{2}_1$ and  $\Rom{2}_2$ respectively. For term $\Rom{2}_1$, using Lemma \ref{lemma:ine} (ii), aka $(1+x)^r \leq 1+rx$ for $x\geq -1$ and $r\in(0,1)$, and $a=1/2$,  we have
\$
\Rom{2}_1
&=  \sup_{\mu \in \BB_r(\mus)} \left( \frac{n}{z^2} \cdot   \EE \frac{\sqrt{1+r_i^2/\tsigma^2} - 1}{\sqrt{1+r_i^2/\tsigma^2}}  - a\right) \\
&\leq \sup_{\mu \in \BB_r(\mus)}  \left\{\frac{n}{z^2} \cdot  \left( 1+ \EE \frac{r_i^2}{2\tsigma^2} -1 \right)- a\right\}\\
&\leq \frac{n}{z^2}\cdot \EE \frac{\varepsilon_i^2 + 2r|\varepsilon_i| + {r^2}|}{2\tsigma^2} -\frac{1}{2} \tag{$a=1/2$}\\
&\leq \frac{r}{\sigma}\left(1+ \frac{r}{2\sigma}   \right)\\
&\leq \frac{2r}{\sigma}
\$ 
if $n$ is large enough such that $r\leq 2\sigma$. 
To bound $\Rom{2}_2$, we need Lemma \ref{lemma:ine_2}. Specifically, for any $0\leq \gamma <1$, we have 
\$
(1+x)^{-1}\leq 1- (1-\gamma)x, ~\text{for any}~0\leq x\leq \frac{\gamma}{1-\gamma}.
\$  
Using this result,  we obtain  
\$
\EE \frac{1}{\sqrt{1+ r_i^2/\tsigma^2}} 
&\leq \sqrt{\EE \frac{1}{1+ r_i^2/\tsigma^2}} \tag{concavity of $\sqrt{x}$} \\
&\leq \sqrt{\EE\left\{\left(1-\frac{{(1-\gamma)r_i^2}}{\tsigma^2}\right)1\left(\frac{{r_i^2}}{\tsigma^2} \leq \frac{\gamma}{1-\gamma} \right)+\frac{1}{{1+ {r_i^2}/\tsigma^2}}1\left(\frac{{r_i^2}}{\tsigma^2}> \frac{\gamma}{1-\gamma} \right)\right\}} \\
&\leq \sqrt{ 1 - (1-\gamma) \,  \EE \left(\frac{r_i^2}{\tsigma^2} 1\left( \frac{{r_i^2}}{\tsigma^2} \leq \frac{\gamma}{1-\gamma}   \right)  \right)} \tag{Lemma \ref{lemma:ine_2}}\\
&\leq  \sqrt{ 1 - (1-\gamma) \,  \EE \left(\frac{r_i^2}{\tsigma^2} 1\left( \frac{{r_i^2}}{\tsigma^2} \leq \frac{\gamma}{1-\gamma}   \right)  \right)}  \\
&\leq  \sqrt{ 1 - (1-\gamma) \,  \EE \left(\frac{\varepsilon_i^2- 2r|\varepsilon_i| + {r^2}}{\tsigma^2} 1\left( \frac{{2(\varepsilon_i^2 + {r^2})}}{\tsigma^2} \leq \frac{\gamma}{1-\gamma}   \right)  \right)}  \tag{ $\forall~\mu \in \BB_r(\mus)$}\\
&\leq   1- \frac{1-\gamma}{2} \, \EE \left(\frac{\varepsilon_i^2- 2r|\varepsilon_i| + {r^2}}{\tsigma^2} 1\left( \frac{{2(\varepsilon_i^2 + {r^2})}}{\tsigma^2} \leq \frac{\gamma}{1-\gamma}   \right)  \right),
\$
where the first inequality uses the concavity of $\sqrt{x}$, the third inequality uses Lemma \ref{lemma:ine_2}, and the last inequality uses the inequality that $(1+x)^{-1} \leq  1- x/2$ for $x\in [0,1]$, aka Lemma \ref{lemma:ine} (iii) with $r=-1$, provided that 
\$
& (1-\gamma) \,  \EE \left(\frac{\varepsilon_i^2- 2r|\varepsilon_i| - {r^2}}{\tsigma^2} 1\left( \frac{{2(\varepsilon_i^2 + {r^2})}}{\tsigma^2} \leq \frac{\gamma}{1-\gamma}   \right)  \right)
\leq (1-\gamma) \frac{\sigma^2 - 2r\sigma - {r^2} }{\tsigma^2}\leq 1. 
 \$
Thus term $\Rom{2}_2$ can be bounded as 
\$
\Rom{2}_2 
&= \sup_{\mu \in \BB_r(\mus)}\left(   a - \frac{n}{z^2} +  \frac{n}{z^2}\cdot \EE \frac{1}{\sqrt{1+ r_i^2/\tsigma^2}}   \right)  \\
&\leq  a - \frac{n}{z^2} + \frac{n}{z^2}\cdot  \left\{    1- \frac{1-\gamma}{2} \, \EE \left(\frac{\varepsilon_i^2- 2r|\varepsilon_i| + {r^2}}{\tsigma^2} 1\left( \frac{{2(\varepsilon_i^2 + {r^2})}}{\tsigma^2} \leq \frac{\gamma}{1-\gamma}   \right)  \right)  \right\} \\
&\leq a - \frac{1-\gamma}{2\sigma^2} \cdot \EE \varepsilon_i^2  + \frac{1-\gamma}{2\sigma^2} \cdot 2r \cdot  \EE\left(  |\varepsilon_i|  \right)\\
&\leq a - \frac{1-\gamma}{2} +    \frac{r(1-\gamma)}{\sigma } \\
&= \frac{\gamma}{2} + \frac{r(1-\gamma)}{\sigma} \tag{$a=1/2$}. 
\$
Combining the upper bound for $\Rom{2}_1$ and $\Rom{2}_2$ and using the fact that, we obtain 
\$
\Rom{2}
&\leq \max\{\Rom{2}_1, \Rom{2}_2\} \leq  \frac{\gamma}{2} + \frac{2r}{\sigma} \rightarrow 0, 
\$
if $\gamma = \gamma(n) \rightarrow 0$. 

We proceed to bound $\Rom{1}$. Recall that
\$
\Rom{1}
=  \frac{n}{z^2} \cdot \sup_{\mu \in \BB_r(\mus)}\left|\frac{1}{n}\sum_{i=1}^n \left( 1- \frac{\tau_\sigma}{\sqrt{\tau_\sigma^2 + {(y_i - \mu)^2}}}  \right)  - \EE \left(  1- \frac{\tau_\sigma}{\sqrt{\tau_\sigma^2 + {(y_i - \mu)^2}}}   \right)\right| . 
\$
For any $0< \epsilon \leq 2r$, there exists an $\epsilon$-cover  $\cN\subseteq \BB_r(\mus)$ of $\BB_r(\mu^*)$ such that 
 $
 |\cN| \leq  {6r}/{\epsilon}. 
 $ 
Then for any $\mu \in \BB_r(\mu^*)$ there exists a  $\omega \in \cN$ such that
$
|\omega-\mu |\leq \gamma,
$
and 
\$
&\left|\frac{1}{n}\sum_{i=1}^n\left( 1 - \frac{\tsigma}{\sqrt{\tsigma^2+(y_i-\mu)^2}}\right)  - \EE \left( 1 - \frac{\tsigma}{\sqrt{\tsigma^2+(y_i-\mu)^2}}\right)\right|\\
&=  \left|\frac{1}{n}\sum_{i=1}^n\frac{\sqrt{1+ (y_i-\mu)^2/\tsigma^2}- 1}{\sqrt{1+(y_i-\mu)^2/\tsigma^2}} 
		- \EE  \frac{\sqrt{1 + (y_i - \mu)^2/\tsigma^2}  - 1}{\sqrt{1 + (y_i - \mu)^2/\tsigma^2}}\right|\\
&\leq  \left|\frac{1}{n}\sum_{i=1}^n\frac{\sqrt{1+ (y_i-\omega)^2/\tsigma^2}- 1}{\sqrt{1+(y_i-\omega)^2/\tsigma^2}} 
		- \EE  \frac{\sqrt{1 + (y_i- \omega)^2/\tsigma^2}  - 1}{\sqrt{1 + (y_i-\omega)^2/\tsigma^2}}\right| \\
 &\qquad +   \left|\frac{1}{n}\sum_{i=1}^n\frac{\sqrt{1+ (y_i - \mu )^2/\tsigma^2}- 1}{\sqrt{1+(y_i - \mu )^2/\tsigma^2}} 
		    -   \frac{1}{n}\sum_{i=1}^n \frac{\sqrt{1 + (y_i- \omega)^2/\tsigma^2}  - 1}{\sqrt{1 + (y_i-\omega)^2/\tsigma^2}}\right| \\
 & \qquad +  \left|\EE \frac{\sqrt{1+ (y_i- \mu  )^2/\tsigma^2}- 1}{\sqrt{1+(y_i - \mu )^2/\tsigma^2}}
			- \EE  \frac{\sqrt{1 + (y_i- \omega)^2/\tsigma^2}  - 1}{\sqrt{1 + (y_i-\omega)^2/\tsigma^2}}\right| \\
 &= \Rom{1}_1 + \Rom{1}_2 + \Rom{1}_3. 
\$
For $\Rom{1}_1$, using Lemma \ref{lemma:c6} acquires with probability at least $1-2\delta$ that 
\$
\Rom{1}_1 &\leq  \sqrt{\frac{ \EE (y_i -\omega)^2\, \log({1}/{\delta})}{n\tsigma^2}} + \frac{\log({1}/{\delta})}{3 n}\\
&\leq  \sqrt{\frac{2  (\sigma^2+ {r^2})  \log({1}/{\delta})}{n\tsigma^2}} + \frac{ \log({1}/{\delta})}{3n} \\
&\leq \frac{2z\sqrt{\log(1/\delta)}}{n} + \frac{\log(1/\delta)}{3 n}
\$
provided ${r^2}\leq \sigma^2$. 
Let 
\$
g(x)= - \frac{1}{n}\sum_{i=1}^n \frac{ \tau }{\sqrt{\tau^2 + (x + {\varepsilon_i})^2}}. 
\$
Using the mean value theorem and the inequality that $|x/(1+x^2)^{3/2}| \leq 1/2$, we obtain 
\$
\left | g(x) - g(y) \right| 
&=  \left|\frac{1}{n}\sum_{i=1}^n \frac{ (\tilde x + {\varepsilon_i})/\tsigma }{(1 + (\tilde x + {\varepsilon_i})^2/\tsigma^2)^{3/2}} \cdot \frac{x-y} {\tsigma }\right|\leq  \frac{|x-y|}{2\tsigma}, 
\$
where $\tilde x$ is some convex combination of $x$ and $y$.  
Then we have  
\$
\Rom{1}_2 
&=\left|\frac{1}{n}\sum_{i=1}^n \frac{ (\tilde \Delta + {\varepsilon_i})/\tsigma }{(1 + (\tilde \Delta + {\varepsilon_i})^2/\tsigma^2)^{3/2}} \cdot \frac{\Delta_\mu - \Delta_\omega}{\tsigma }\right|
\leq \frac{ \epsilon }{2\tsigma}
\$
where $\tilde \Delta $ is some convex combination of $\Delta_w = \mu^* - w$ and $\Delta_\mu = \mu^* - \mu$. 
For $\Rom{2}_3$, a similar argument for bounding $\Rom{2}_2$ yields 
\$
\Rom{1}_3
&=\left|\EE \left( \frac{ (\tilde \Delta + {\varepsilon_i})/\tsigma }{(1 + (\tilde \Delta + {\varepsilon_i})^2/\tsigma^2)^{3/2}}\right) \cdot \frac{\Delta_\mu - \Delta_\omega}{\tsigma } \right| \\
&\leq  \EE|\tilde\Delta +{\varepsilon_i}| \cdot \frac{\epsilon}{\tsigma^2}  \\
&\leq \frac{\epsilon \sqrt{2({r^2} +\sigma^2)}}{\tsigma^2},  
\$
where the last inequality uses Jensen's inequality, i.e. $\EE|\tilde\Delta +{\varepsilon_i}| \leq \sqrt{\EE (\tilde\Delta + \varepsilon_i^2)}\leq \sqrt{2({r^2} +\sigma^2)}$. Putting the above pieces together and using the union bound, we obtain with probability at least $1- 12\epsilon^{-1}r \delta$ 
\$
\Rom{1}
&\leq \frac{n}{z^2}\cdot \sup_{\omega \in \cN} \left|\frac{1}{n}\sum_{i=1}^n\frac{\sqrt{1+ (y_i-\omega)^2/\tsigma^2}- 1}{\sqrt{1+(y_i-				\omega)^2/\tsigma^2}} - \EE  \frac{\sqrt{1 + (y_i- \omega)^2/\tsigma^2}  - 1}{\sqrt{1 + (y_i-\omega)^2/\tsigma^2}}\right| \\
	&\qquad\qquad 	   +  \frac{n}{z^2}\cdot \frac{\epsilon}{2\tsigma} \left(1 + \frac{2\sqrt{2({r^2}+\sigma^2)}}{\tsigma}\right) \\
&\leq   \frac{2\sqrt{\log(1/\delta)}}{z} + \frac{\log(1/\delta)}{3 z^2}    +   \frac{\epsilon \sqrt{n}}{\sigma z},
\$
provided that 
\$
{2\sqrt{2({r^2}+\sigma^2)}}  \leq  {\tsigma}.
\$

Putting above results together, we obtain with probability at least $1- (12r/\epsilon + 2)\delta$ that
\$
|\hv - \sigma| 
&\lesssim \Rom{1} + \Rom{2}\\
&\leq  \frac{2\sqrt{\log(1/\delta)}}{z} + \frac{\log(1/\delta)}{3 z^2}    +   \frac{\epsilon \sqrt{n}}{\sigma z} + \frac{\gamma}{2} + \frac{2r}{\sigma}. 
\$
Let $C'=24CV_0^2/v_0 $. Therefore, taking $\epsilon = 1/\sqrt{n}$, $\delta = {1}/{\log n}$, and $z^2\geq \log(n)$, we obtain with probability at least 
\$
1- \frac{C'\big(\sqrt{\log n} + {\log\log n}/{\sqrt{\log n}}\big) + 2}{\log n}
\$ 
that 
\$
|\hv -\sigma| \lesssim \sqrt{\frac{\log \log n}{\log n }} + \frac{\log \log n}{\log n} + \frac{1}{\sqrt{\log n}} + \gamma + r 
\rightarrow 0. 
\$
Therefore  $\hv \rightarrow \sigma$ in probability.  This finishes the proof. 

\end{proof}

\subsection{Local strong convexity in $v$}

In this section, we first present the local  strong convexity of the empirical loss function with respect to $v$ uniformly over a neighborhood of $\mus$. 


\begin{lemma}[Local strong convexity in $v$]\label{lemma:v.sc}
Let $\BB_r(\mu^*) = \{\mu: |\mu-\mu^*|\leq r\}$.   Assume $r= r(n)=o(1)$. 
Let $0<\delta <1$ and $n$ is sufficiently large.  Take   $\varpi$ such that $\max\{\varpi r\sqrt{n}, \varpi \} \rightarrow 0 ~\text{and}~\varpi \sqrt{n}\rightarrow \infty$. 
Then, with probability at least $1-\delta$, we have  
\$
\inf_{\mu \in \BB_r(\mus)}    \frac{ \llangle  \nabla_v L_n(\mu, v) -  \nabla_v L_n(\mu,\vs),  v -\sigma  \rrangle}{|v -\sigma|^2}           
 \geq  \rho_\ell  =  \frac{\sigma^2_{{c\varpi^2 n}/{(4z^2)}}}{2(v^3\vee \sigma^3)}\geq \frac{c_0}{v^3\vee \sigma^3},
\$ 
where $c$ and $c_0$ are some constants. 
\end{lemma}

\begin{proof}[Proof of Lemma \ref{lemma:v.sc}]

Recall $\tau = v\sqrt{n}/z$.  For notational simplicity, write $\tau_\sigma= \sigma\sqrt{n}/z$, $\tau_{v_0} = v_0 \sqrt{n}/z$, $\tauvp= \varpi \sqrt{n}/z$,  and $\Delta = \mu^* - \mu$. It follows that  
\$
 \langle \nabla_v  L_n(\mu, v) -  \nabla_v L_n(\mu, \sigma)  , v -  \sigma \rangle
 &= \frac{n}{z^2}\llangle \frac{1}{n}\sum_{i=1}^n \frac{\tau}{\sqrt{\tau^2+(y_i-\mu)^2}}  -\frac{1}{n}\sum_{i=1}^n \frac{\tau_\sigma}{\sqrt{\tau_\sigma^2+(y_i-\mu)^2}}, v - \sigma \rrangle \\
 &= \frac{n^{3/2}}{z^3}\cdot \frac{1}{n}\sum_{i=1}^n \frac{(y_i - \mu)^2}{(\ttau^2+(y_i-\mu)^2)^{3/2}} \, |v-\sigma|^2 \\
 &\geq \frac{n^{3/2}}{z^3}\cdot \frac{1}{n}\sum_{i=1}^n \frac{(y_i - \mu)^2}{( (\tau\vee \tau_\sigma)^2+(y_i-\mu)^2)^{3/2}} \, |v-\sigma|^2
 \$
 where $\ttau$ is some convex combination of $\tau$ and $\tau_\sigma$, that is $\ttau =(1- \lambda)\tau_\sigma + \lambda\tau$ for some $\lambda \in [0,1]$. 
Because $\tau^3x^2/(\tau^2+ x^2)^{3/2}$ is an increasing function of $\tau$, if $\tauvp \leq \tau\vee \tau_\sigma$, we have 
 \$
 \frac{\langle \nabla_v  L_n(\mu, v) -  \nabla_v L_n(\mu, \sigma)  , v -  \vs \rangle}{| v - \sigma|^2}&\geq \frac{n^{3/2}}{z^3(\tau\vee\tau_\sigma)^3}\cdot\frac{1}{n}\sum_{i=1}^n \frac{(\tau\vee\tau_\sigma)^3(y_i - \mu)^2}{(\tau^2\vee \tau_\sigma^2+(y_i-\mu)^2)^{3/2}}\\
&\geq \frac{n^{3/2}}{z^3(\tau\vee\tau_\sigma)^3}\cdot\frac{1}{n}\sum_{i=1}^n \frac{\tauvp^3(y_i - \mu)^2}{(\tauvp^2+(y_i-\mu)^2)^{3/2}}. 
 \$
Thus
 \$
&\inf_ {\mu\in  \BB_r (\mu^*)} \frac{\langle \nabla_v  L_n(\mu, v) -  \nabla_v L_n(\mu, \sigma)  , v -  \vs \rangle}{|v - \sigma|^2} \\
&\geq \frac{n^{3/2}}{z^3(\tau\vee\taus)^3}\cdot   \inf_ {\mu\in  \BB_r (\mu^*)} \frac{1}{n}\sum_{i=1}^n \frac{\tauvp^3(y_i - \mu)^2}{(\tauvp^2+(y_i-\mu)^2)^{3/2}} \\
&= \frac{n^{3/2}}{z^3(\tau\vee\tau_\sigma)^3}\cdot \Bigg(\inf_ {\mu\in  \BB_r (\mu^*)} \left(\EE  \frac{\tauvp^3(y_i - \mu)^2}{(\tauvp^2+(y_i-\mu)^2)^{3/2}}\right)  \\
&\qquad\qquad\qquad\quad\quad  - \sup_ {\mu\in  \BB_r (\mu^*)} \left|\frac{1}{n}\sum_{i=1}^n \frac{\tauvp^3(y_i - \mu)^2}{(\tauvp^2+(y_i-\mu)^2)^{3/2}}  - \EE  \frac{\tauvp^3(y_i - \mu)^2}{(\tauvp^2+(y_i-\mu)^2)^{3/2}}\right|\Bigg)\\
& = \frac{n^{3/2}}{z^3(\tau\vee\tau_\sigma)^3}\cdot \left( \Rom{1} -\Rom{2}\right). 
 \$

 It remains to lower bound  \Rom{1} and upper bound \Rom{2}. We start with \Rom{1}. Let $f(x)= x/(1+x)^{3/2}$ which satisfies
\$
f(x)
 \geq 
\begin{cases}
  \epsilon x & x\leq c_{\epsilon} \\
 0 & x > c_{\epsilon},
\end{cases}
\$
and  $Z={(y-\mu)^2}/{\tauvp^2}$ in which $y\sim y_i$. Suppose 
$
{r^2} \leq c_\epsilon \tauvp^2/4, 
$
then we have
 \$
 \inf_{\mu \in\BB_r(\mus)}\left(\EE \frac{\tauvp^3 (y_i-\mu)^2}{(\tauvp^2+(y_i-\mu)^2)^{3/2}} \right)
 &=   \inf_{\mu \in\BB_r(\mus)}\EE\left(\frac{ \tauvp^2 Z}{\left(1+Z\right)^{3/2}} \right)  \\
 &\geq \epsilon \cdot  \inf_{\mu \in\BB_r(\mus)}\EE\left[ (y-\mu)^2 1((y-\mu)^2\leq c_\epsilon \tauvp^2)\right]\\
 &\geq \epsilon \cdot  \inf_{\mu \in\BB_r(\mus)}\EE\left[ (y-\mu)^2 1({\varepsilon^2} \leq c_\epsilon\tauvp^2/2 -  {r^2}) \right]\\
 &\geq \epsilon \cdot  \inf_{\mu \in\BB_r(\mus)}\left(\EE\left[ (\Delta^2 + {\varepsilon^2}) 1\left({\varepsilon^2} \leq \frac{c_\epsilon\tauvp^2}{4}\right) \right]  -  \frac{8\Delta \sigma^2}{c_\epsilon \tauvp^2}\right)\\
 &\geq   \epsilon \cdot  \left(\EE\left[ {\varepsilon^2} 1\left({\varepsilon^2} \leq \frac{c_\epsilon\tauvp^2}{4}\right) \right]  -  \frac{8 r \sigma^2}{c_\epsilon \tauvp^2}\right). 
 \$

 We then proceed with \Rom{2}.  For any $0< \gamma \leq 2r $, there exists an $\gamma$-cover  $\cN$ of $\BB_r(\mu^*)$ such that 
 $
 |\cN| \leq  {6r}/{\gamma}. 
 $ 
Then for any $\mu \in \BB_r(\mu^*)$ there exists an $\omega \in \cN$ such that
$
|\omega-\mu |\leq \gamma,
$
and thus by Lemma \ref{lemma:c5} we have 
\$
&\left|\frac{1}{n}\sum_{i=1}^n \frac{\tauvp^3(y_i - \mu)^2}{(\tauvp^2+(y_i-\mu)^2)^{3/2}}  - \EE  \frac{\tauvp^3(y_i - \mu)^2}{(\tauvp^2+(y_i-\mu)^2)^{3/2}}\right|\\
&\leq 
\left|\frac{1}{n}\sum_{i=1}^n \frac{\tauvp^3(y_i - \omega)^2}{(\tauvp^2+(y_i- \omega)^2)^{3/2}}  - \EE  \frac{\tauvp^3(y_i - \omega)^2}{(\tauvp^2+(y_i- \omega)^2)^{3/2}}\right| \\
 &\qquad +   \left|\frac{1}{n}\sum_{i=1}^n \frac{\tauvp^3(y_i - \omega)^2}{(\tauvp^2+(y_i- \omega)^2)^{3/2}}  - \frac{1}{n}\sum_{i=1}^n \frac{\tauvp^3(y_i - \mu)^2}{(\tauvp^2+(y_i- \mu)^2)^{3/2}}  \right|\\
 & \qquad +  \left| \EE  \frac{\tauvp^3(y_i - \omega)^2}{(\tauvp^2+(y_i- \omega)^2)^{3/2}} -  \EE  \frac{\tauvp^3(y_i - \mu)^2}{(\tauvp^2+(y_i- \mu)^2)^{3/2}}\right| \\
 &= \Rom{2}_1 + \Rom{2}_2 + \Rom{2}_3. 
\$
For $\Rom{2}_1$, Lemma \ref{lemma:c5} implies with probability at least $1-2\delta$
\$
\Rom{2}_1 
&\leq  \sqrt{\frac{2 \tauvp^2 \EE (y_i -\omega)^2  \log({1}/{\delta})}{3n}} + \frac{\tauvp^2 \log({1}/{\delta})}{3\sqrt{3}n}
\leq  \sqrt{\frac{2 \tauvp^2 (\sigma^2+ {r^2})  \log({1}/{\delta})}{3n}} + \frac{\tauvp^2 \log({1}/{\delta})}{3\sqrt{3}n}. 
\$
Let 
\$
g(x)= \frac{1}{n}\sum_{i=1}^n \frac{\tau^3 (x + {\varepsilon_i})^2}{(\tau^2 + (x + {\varepsilon_i})^2)^{3/2}}. 
\$
Using the mean value theorem and the inequality that $|\tau^2x/(\tau^2+ x^2)^{3/2}|\leq 1/\sqrt{3}$, we obtain 
\$
\left | g(x) - g(y) \right| 
&=  \left|\frac{1}{n}\sum_{i=1}^n \frac{\tau^3 (\tilde x + {\varepsilon_i})\left(\tau^2 - (\tilde x + {\varepsilon_i})^2\right)}{(\tau^2 + (\tilde x + {\varepsilon_i})^2)^{5/2}} (x-y )\right|\leq  \frac{\tau}{\sqrt{3}}|x-y|. 
\$
{
Then we have 
\$
\Rom{2}_2 
&=\left|\frac{1}{n}\sum_{i=1}^n \frac{\tauvp^3 (\tilde \Delta + {\varepsilon_i})\left(\tauvp^2 - (\tilde \Delta + {\varepsilon_i})^2\right)}{(\tauvp^2 + (\tilde \Delta + {\varepsilon_i})^2)^{5/2}} (\Delta_w - \Delta_\mu )\right|
\leq \frac{\tauvp \gamma }{\sqrt{3}}
\$
where $\tilde \Delta $ is some convex combination of $\Delta_w = \mu^* - w$ and $\Delta_\mu = \mu^* - \mu$. 
}
For $\Rom{2}_3$, we have 
\$
\Rom{2}_3
&=\left|\EE\left( \frac{\tauvp^3 (\tilde \Delta + {\varepsilon_i})\left(\tauvp^2 - (\tilde \Delta + {\varepsilon_i})^2\right)}{(\tauvp^2 + (\tilde \Delta + {\varepsilon_i})^2)^{5/2}}\right) (\Delta_w - \Delta_\mu )\right|
\leq {\gamma \EE|\tilde\Delta +{\varepsilon_i}|} 
\leq {\gamma \sqrt{\EE\left(\tilde\Delta + {\varepsilon_i} \right)^2}},
\$
where the last inequality uses Jensen's inequality. Putting the above pieces together and using the union bound, we obtain with probability at least $1- 12\gamma^{-1}r \delta$ 
\$
\Rom{2}
&\leq \sup_{\omega \in \cN} \left|\frac{1}{n}\sum_{i=1}^n \frac{\tauvp^3(y_i - \omega)^2}{(\tauvp^2+(y_i- \omega)^2)^{3/2}}  - \EE  \frac{\tauvp^3(y_i - \omega)^2}{(\tauvp^2+(y_i- \omega)^2)^{3/2}}\right| +  \frac{\tauvp \gamma }{\sqrt{3}} + \gamma \sqrt{{r^2} + \sigma^2}  \\
&\leq  \sqrt{\frac{2 \tauvp^2 ( {r^2} + \sigma^2)  \log({1}/{\delta})}{3n}} + \frac{\tauvp^2 \log({1}/{\delta})}{3\sqrt{3}n} 
      + \frac{\tauvp \gamma }{\sqrt{3}} + \gamma \sqrt{{r^2} + \sigma^2}\\
&= \sqrt{{r^2}+\sigma^2}\left(\sqrt{\frac{2\varpi^2 \log(1/\delta)}{3z^2}} + \gamma\right) + \frac{\varpi^2 \log(1/\delta)}{3\sqrt{3}z^2} + \frac{\varpi \gamma \sqrt{n}}{\sqrt{3}}. 
\$

Combining the bounds for \Rom{1} and \Rom{2} yields with probability at least $1-\delta$
\$
&\inf_ {\mu\in  \BB_r (\mu^*)} \frac{\langle \nabla_v  L_n(\mu, v) -  \nabla_v L_n(\mu, \sigma)  , v - \sigma \rangle}{|v - \sigma|^2}\\
&\geq \frac{n^{3/2}}{z^3(\tau\vee\tau_\sigma)^3} \Bigg\{ \epsilon \left(\EE\left[ {\varepsilon^2} 1\left({\varepsilon^2} \leq \frac{c_\epsilon\tauvp^2}{4}\right) \right]   -  \frac{8 r \sigma^2}{c_\epsilon \tauvp^2}\right)\\ 
& \qquad \qquad \qquad \qquad   -   \sqrt{{r^2}+\sigma^2}\left(\sqrt{\frac{2\varpi^2 \log(1/\delta)}{3z^2}} + \gamma\right)  - \frac{\varpi^2 \log(1/\delta)}{3\sqrt{3}z^2} - \frac{\varpi \gamma \sqrt{n}}{\sqrt{3}}  \Bigg\}\\
&\geq  \frac{1}{2(v\vee\sigma)^3} \EE\left[{\varepsilon^2} 1\left({\varepsilon^2} \leq \frac{c_\epsilon\tauvp^2}{4}  \right)\right]
\$
where $\epsilon, \varpi, \gamma, n$ are picked such that $\epsilon=3/4$, $\gamma = 12 r $, and 
\$
&\epsilon \left(\EE\left[ {\varepsilon^2} 1\left({\varepsilon^2} \leq \frac{c_\epsilon\tauvp^2}{4}\right) \right]   -  \frac{8 r \sigma^2 z^2}{c_\epsilon \varpi ^2 n }\right)  -   \sqrt{{r^2}+\sigma^2}\left(\sqrt{\frac{2\varpi^2 \log(1/\delta)}{3z^2}} + \gamma\right)  - \frac{\varpi^2 \log(1/\delta)}{3\sqrt{3}z^2} - \frac{\varpi \gamma \sqrt{n}}{\sqrt{3}} \\
&\geq \frac{1}{2}\EE\left[ {\varepsilon^2} 1\left({\varepsilon^2} \leq \frac{c_\epsilon\tauvp^2}{4}\right) \right]\geq \frac{1}{4}\sigma. 
\$
For example, we can pick  $\varpi$ such that 
\$
\max\{\varpi r\sqrt{n}, \varpi \} \rightarrow 0 ~\text{and}~\varpi \sqrt{n}\rightarrow \infty
\$
as $n\rightarrow \infty$. 
This completes the proof.  

\end{proof}


\subsection{Supporting lemmas}


This subsection proves  a supporting lemma that is used prove Lemma \ref{lemma:v.sc}.   

\begin{lemma}\label{lemma:c5}
Let $w_i$ be i.i.d. copies of $w$.  For any $0< \delta < 1$, we have 
\$
\frac{1}{n}\sum_{i=1}^n \frac{\tau^3 w_i^2}{(\tau^2+w_i^2)^{3/2}}   -  \EE\frac{\tau^3 w_i^2}{({\tau^2+ w_i ^2})^{3/2}}
&\geq   - \sqrt{\frac{2 \tau^2 \EE w_i^2  \log({1}/{\delta})}{3n}} - \frac{\tau^2 \log({1}/{\delta})}{3\sqrt{3}n}, ~\text{with prob. } 1-\delta, \\
\left|\frac{1}{n}\sum_{i=1}^n \frac{\tau^3 w_i^2}{(\tau^2+w_i^2)^{3/2}}   -  \EE\frac{\tau^3 w_i^2}{({\tau^2+ w_i ^2})^{3/2}}\right|
&\leq    \sqrt{\frac{2 \tau^2 \EE w_i^2  \log({1}/{\delta})}{3n}} + \frac{\tau^2 \log({1}/{\delta})}{3\sqrt{3}n}, ~\text{with prob. } 1-2\delta.
\$
\end{lemma}
\begin{proof}[Proof of Lemma \ref{lemma:c5}]
We only prove the first result and the second result follows similarly. 
The random  variables $Z_i = Z_i(\tau):=\tau^3 w_i^2/(\tau^2+w_i^2)^{3/2}$ with $\mu_z=\EE Z_i$ and $\sigma_z^2=\var (Z_i)$ are bounded i.i.d. random variables such that 
\$
0\leq Z_i&={\tau^3 w_i^2}/({\tau^2 + w_i^2})^{3/2}\leq w_i^2\wedge \frac{\tau^2}{\sqrt{3}} \wedge\frac{\tau |w_i|}{\sqrt{3}}. 
\$
Moreover we have
\$
\EE Z_i^2 &=\EE\left(\frac{\tau^6w_i^4}{(\tau^2+{\varepsilon_i^2})^3}\right) \leq \frac{\tau^2 \EE w_i^2}{3},~\sigma^2_z:=\var (Z_i)\leq \frac{\tau^2 \EE w_i^2}{3}.
\$
For third and higher order absolute moments, we have 
\$
\EE |Z_i|^k&=\EE\left|\frac{\tau^3w_i^2}{(\tau^2+{\varepsilon_i^2})^{3/2}}\right|^k\leq   \frac{\tau^2 \EE w_i^2}{3} \cdot  \left(\frac{\tau^2}{\sqrt{3}} \right)^{k-2} \leq \frac{k!}{2}\cdot  \frac{\tau^2 \EE w_i^2}{3} \cdot \left(\frac{\tau^2}{3\sqrt{3}}\right)^{k-2}, ~\text{for all integers}~ k\geq 3. 
\$
Therefore, using Lemma \ref{lemma:bernstein.ine} with $v=n\tau^2 \, \EE w_i^2 /3$ and $c=\tau^2/\big(3\sqrt{3}\big)$ acquires that  for any $t\geq 0$
\$
\PP\left(\sum_{i=1}^n \frac{\tau^3 w_i^2 }{(\tau^2+{\varepsilon_i^2})^{3/2}}- \sum_{i=1}^n  \EE  \left( \frac{\tau^3 w_i^2}{(\tau^2+{\varepsilon_i^2})^{3/2}} \right)\geq -\sqrt{\frac{2 n\tau^2 \EE w_i^2 t}{3}} - \frac{\tau^2 t}{3\sqrt{3}} \right)\leq  \exp(-t). 
\$
Taking $t= \log (1/\delta)$ acquires  that for any $0< \delta < 1$
\$
\PP\left( \frac{1}{n}\sum_{i=1}^n  \frac{\tau^3 w_i^2}{(\tau^2+w_i^2)^{3/2}} - \frac{1}{n}\sum_{i=1}^n  \EE  \left(  \frac{\tau^3 w_i^2}{(\tau^2+{\varepsilon_i^2})^{3/2}} \right) 
> - \sqrt{\frac{2 \tau^2 \EE w_i^2  \log({1}/{\delta})}{3n}} - \frac{\tau^2 \log({1}/{\delta})}{3\sqrt{3}n}  \right)
>  1-\delta. 
\$
This finishes the proof. 
\end{proof}

\section{Proofs for Section \ref{sec:6}}
We first prove Proposition \ref{prop:multi}. 

\begin{proof}[Proof of Proposition \ref{prop:multi}]
The proof directly follows from Theorem \ref{thm:main_random} and the union bound. 
\end{proof}

Next, we prove Proposition \ref{prop:asym_multi}. 

\begin{proof}[Proof of Proposition \ref{prop:asym_multi}]
We only sketch the proof, as most of the proof follows from that of Theorem \ref{thm:asym_strong}. By Proposition \ref{prop:multi} and taking $z^2\geq 2\log{n}$, we obtain
\$
\|\hmu- \mus\|_2 \rightarrow 0 ~~\text{in probability}.
\$
Similarly, following the proof of Theorem \ref{thm:v_consistency}, we obtain
\$
\|\hv - \sigma\|_2 \rightarrow 0 ~~\text{in probability},
\$
where $\hv=(\hv_1, \ldots, \hv_d)^\T$ and $\sigma=(\sigma_{11}, \ldots, \sigma_{dd})^\T$.

With a slight overload of notation, let 
$
L_n(\mu)= L_n(\mu, \sigma). 
$
Let $\tau_{\sigma_k} = \sigma_{kk} \sqrt{n}/z$.
Then following the proof of Theorem \ref{thm:asym_strong}, we obtain
\$
\sqrt{n} \, \big(\hmu - \mus \big) 
&\simeq \left[ H_n(\mus) \right]^{-1} \left( - \sqrt{n} \, \nabla L_n(\mus)  \right ) \\
&=  
\begin{bmatrix}
\frac{\sqrt{n}}{z} \cdot \frac{1}{n} \sum_{i=1}^n \frac{\tau^2_{\sigma_{11}}}{(\tau^2_{\sigma_{11}} + \varepsilon_{i1}^2)^{3/2}} & 0 &  \dots\\
\vdots &\ddots & \ \\
0 &  & \frac{\sqrt{n}}{z} \cdot \frac{1}{n} \sum_{i=1}^n \frac{\tau^2_{\sigma_{dd}}}{(\tau^2_{\sigma_{dd}} + \varepsilon_{id}^2)^{3/2}} 
\end{bmatrix}^{-1} 
\begin{bmatrix}
\sqrt{n}  \cdot \frac{1}{n} \sum_{i=1}^n \frac{\tau_{\sigma_{11}} \varepsilon_{i1}}{\sigma_{11} \sqrt{\tau^2_{\sigma_{11}} + \varepsilon_{i1}^2}} \\
\vdots\\
\sqrt{n}  \cdot \frac{1}{n} \sum_{i=1}^n \frac{\tau_{\sigma_{dd}} \varepsilon_{id}}{\sigma_{dd} \sqrt{\tau^2_{\sigma_{dd}} + \varepsilon_{id}^2}} 
\end{bmatrix}
\\
&\simeq 
\begin{bmatrix}
{\sigma_{11}} & 0      &\dots \\
\vdots      & \ddots & \\
0           &        &  {\sigma_{dd}} 
\end{bmatrix}   
\begin{bmatrix}
\sqrt{n}  \cdot \frac{1}{n} \sum_{i=1}^n \frac{\tau_{\sigma_{11}} \varepsilon_{i1}}{\sigma_{11} \sqrt{\tau^2_{\sigma_{11}} + \varepsilon_{i1}^2}} \\
\vdots\\
\sqrt{n}  \cdot \frac{1}{n} \sum_{i=1}^n \frac{\tau_{\sigma_{dd}} \varepsilon_{id}}{\sigma_{dd} \sqrt{\tau^2_{\sigma_{dd}} + \varepsilon_{id}^2}} 
\end{bmatrix}
\\
&=: \Lambda \, \Rom{1}, 
\$
where $\Lambda=\diag(\sigma_{11},\dots, \sigma_{dd} )$.

We only to derive the asymptotic distribution of the term \Rom{1}:
\$
\Rom{1}
&= \sqrt{n}\cdot 
\left(
\begin{bmatrix}
 \frac{1}{n} \sum_{i=1}^n \frac{\tau^2_{\sigma_{11}} \varepsilon_{i1}}{\sigma \sqrt{\tau^2_{\sigma_{11}} + \varepsilon_{i1}^2}} \\
\vdots\\
 \frac{1}{n} \sum_{i=1}^n \frac{\tau^2_{\sigma_{dd}}  \varepsilon_{id}}{\sigma \sqrt{\tau^2_{\sigma_{dd}} + \varepsilon_{id}^2}}
\end{bmatrix} 
-
\EE
\begin{bmatrix}
\frac{1}{n} \sum_{i=1}^n \frac{\tau_{\sigma_{11}} \varepsilon_{i1}}{\sigma_{11} \sqrt{\tau^2_{\sigma_{11}} + \varepsilon_{i1}^2}} \\
\vdots\\
 \frac{1}{n} \sum_{i=1}^n \frac{\tau_{\sigma_{dd}} \varepsilon_{id}}{\sigma_{dd} \sqrt{\tau^2_{\sigma_{dd}} + \varepsilon_{id}^2}} 
\end{bmatrix} \right)
+ 
\sqrt{n}\cdot \EE
\begin{bmatrix}
\frac{1}{n} \sum_{i=1}^n \frac{\tau_{\sigma_{11}} \varepsilon_{i1}}{\sigma_{11} \sqrt{\tau^2_{\sigma_{11}} + \varepsilon_{i1}^2}} \\
\vdots\\
 \frac{1}{n} \sum_{i=1}^n \frac{\tau_{\sigma_{dd}} \varepsilon_{id}}{\sigma_{dd} \sqrt{\tau^2_{\sigma_{dd}} + \varepsilon_{id}^2}} 
\end{bmatrix} \\
&= \Rom{1}_1 + \Rom{1}_2. 
\$

Again, following the proof of Theorem \ref{thm:asym_strong}, the $\ell_2$ norm of the second term goes to 0. For the first term \Rom{1}, we have
\$
\Rom{1}_1 
&\rightsquigarrow
\cN\left(0, \lim_{n\rightarrow \infty} 
\cov\left(
\begin{bmatrix}
  \frac{\tau^2_{\sigma_{11}} \varepsilon_{i1}}{\sigma \sqrt{\tau^2_{\sigma_{11}} + \varepsilon_{i1}^2}} \\
\vdots\\
  \frac{\tau^2_{\sigma_{dd}}  \varepsilon_{id}}{\sigma \sqrt{\tau^2_{\sigma_{dd}} + \varepsilon_{id}^2}}
\end{bmatrix}\right)    
\right)\\
&= \cN\left(0, 
\Lambda^{-1}\Sigma \Lambda^{-1}  
\right).  
\$
Thus we have
\$
\sqrt{n} \, \big(\hmu - \mus \big)\rightsquigarrow \cN(0, \Sigma). 
\$
This finishes the proof. 

\end{proof}


\section{Preliminary lemmas}

This section collects preliminary lemmas that  are frequently used in the proofs for the main results and supporting lemmas. We first collect the Hoeffding's inequality and then present  a form of Bernstein's inequality. We omit their proofs and refer interested readers  to \cite{boucheron2013concentration}. 

\begin{lemma}[Hoeffding's inequality]\label{lemma:hoeffding}
Let $Z_1,\ldots, Z_n$ be independent real-valued random variables such that 
$
a\leq Z_i\leq b
$
almost surely. 
Let $S_n=\sum_{i=1}^n (Z_i-\EE Z_i)$ and $v=n(b-a)^2$. Then for all $t\geq0$,
\$
\PP\left( S_n\geq \sqrt{vt/2}\right) \leq e^{-t}, ~ \PP\left( S_n\leq - \sqrt{vt/2}\right) \leq e^{-t}, ~\PP\left( |S_n|\geq \sqrt{vt/2}\right) \leq 2e^{-t}. 
\$
\end{lemma}

\begin{lemma}[Bernstein's inequality]\label{lemma:bernstein.ine}
Let $Z_1,\ldots, Z_n$ be independent real-valued random variables such that 
\$
\sum_{i=1}^n \EE Z_i^2\leq v, \, \sum_{i=1}^n \EE |Z_i|^k\leq \frac{k!}{2}v c^{k-2} ~ \textnormal{for all}~ k\geq 3. 
\$
If $S_n=\sum_{i=1}^n (Z_i-\EE Z_i)$, then for all $t\geq0$,
\$
\PP\left( S_n\geq \sqrt{2vt}+ct\right) \leq e^{-t}, ~ \PP\left( S_n\leq - (\sqrt{2vt}+ct)\right) \leq e^{-t}, ~\PP\left( |S_n|\geq \sqrt{2vt}+ct\right) \leq 2e^{-t}. 
\$
\end{lemma}
\begin{proof}[Proof of Lemma \ref{lemma:bernstein.ine}]
This lemma involves a two-sided extension of Theorem  2.10 by \cite{boucheron2013concentration}. The proof follows from a similar argument used in the proof of Theorem 2.10, and thus is omitted. 
\end{proof}

Our third lemma concerns  the localized Bregman divergence for convex functions. It was first   established in \cite{fan2018lamm}.
For any loss function ${L}$, define the Bregman divergence and the  symmetric Bregman divergence as 
\$
D_{L}(\beta_1,\beta_2)&={L}(\beta_1)-{L}(\beta_2)- \langle \nabla {L}(\beta_2),\beta_1-\beta_2  \rangle,\\
D_{L}^s(\beta_1,\beta_2)&=D_{L}(\beta_1,\beta_2)+D_{L}(\beta_2,\beta_1).
\$
\begin{lemma}\label{lemma:l2g}
 For any $\beta_\eta  =\beta^*+ \eta (\beta-\beta^*)$ with $\eta \in (0,1]$ and any convex loss function ${L}$, we have
	\$
	D_{L}^s(\beta_{\eta} ,\beta^*)\leq \eta D_{L}^s(\beta,\beta^*).
	\$
\end{lemma}

Our forth lemma in this  section concerns three basic inequalities that are frequently used in the  proofs. 

\begin{lemma}\label{lemma:ine}
The following inequalities hold:
\begin{itemize}
\item[(i)] $(1+x)^r\geq 1+rx$ for $x\geq -1$ and  $r\in \RR\setminus (0,1)$; 
\item[(ii)] $(1+x)^r\leq 1+rx$ for $x\geq -1$ and $r\in (0,1)$; 
\item[(iii)] $(1+x)^r\leq 1+(2^r-1)x$ for $x\in[0,1]$ and $r\in \RR\setminus(0,1)$. 
\end{itemize}
\end{lemma}

\end{document}